\documentclass[11pt,letter]{article}
\pdfoutput=1 

\usepackage{jheppub}
\usepackage{amssymb,amsmath,amsthm,graphicx}
\usepackage{graphics, color}
\usepackage{subfigure}
\usepackage{latexsym}
\usepackage{bm}
\usepackage[hang,flushmargin]{footmisc}
\usepackage{epsfig}
\usepackage{textcomp}
\hypersetup{
 pdfauthor={Gary T. Horowitz, Jorge  E. Santos},
 citecolor=black,linkcolor=black,urlcolor=black}
\usepackage[utf8]{inputenc}
\allowdisplaybreaks[1]

\begin{document}
\title{\centering{Smooth extremal horizons are the exception,\\not the rule}}
\author[1]{Gary~T.~Horowitz}
\author[2]{and Jorge~E.~Santos}

\affiliation[1]{Department of Physics, University of California, Santa Barbara, CA 93106, U.S.A.}
\affiliation[2]{DAMTP, Centre for Mathematical Sciences, University of Cambridge, Wilberforce Road, \\ Cambridge CB3 0WA, UK}
\emailAdd{horowitz@ucsb.edu}
\emailAdd{jss55@cam.ac.uk}

\newcommand{\blue}{\color{blue}}
\newcommand{\red}{\color{red}}

\abstract{We show that the general charged, rotating black hole in five-dimensional Einstein-Maxwell theory has a singular extremal limit. Only the known analytic solutions with exactly zero charge or zero angular momenta have smooth extremal horizons.
We also consider general black holes in five-dimensional Einstein-Maxwell-Chern-Simons theory, and show that they also have singular extremal limits except for one special value of the coefficient of the Chern-Simons term (the one  fixed by supergravity). Combining this with earlier results showing that 
extremal black holes have singular horizons in four-dimensional general relativity with small higher derivative corrections, and in anti-de Sitter space with perturbed boundary conditions, one sees that smooth extremal horizons are indeed the exception and not the rule.  }

\maketitle

\section{Introduction}

The five-dimensional (5D) generalization of the Kerr-Newman solution, describing charged rotating black holes, is not known. The 5D solution should contain  four parameters: a mass $M$, charge $Q$, and two angular momenta, $J_\phi, J_\psi$, corresponding to rotations in two orthogonal planes. Exact solutions to the 5D Einstein-Maxwell equations are known only when some of these parameters vanish. When $J_\phi= J_\psi =0$, one has the static 5D generalization of the Reissner-Nordstr\"om solution \cite{Tangherlini:1963bw}, and when $Q=0$, one has the vacuum Myers-Perry solutions \cite{Myers:1986un}. Both of these solutions have smooth extremal limits (when $J_\phi, J_\psi$ are both nonzero). If one adds a Chern-Simons term to the action with its coefficient fixed by supergravity, the general four-parameter family of black hole solutions is known \cite{Chong:2005hr}, and again has  smooth extremal limits. Given these results, one is tempted to conclude that the generic 5D (spherical\footnote{We will consider only (asymptotically flat) black holes with horizon topology $S^3$.}) rotating charged black hole will have a smooth extremal limit.

The goal of this paper is to show that this intuition is completely wrong. We will show that generic 5D black holes have singular extremal limits. More precisely, black hole solutions to the 5D Einstein-Maxwell equations 
with nonzero $Q,J_\phi, J_\psi$ develop curvature singularities on their horizon in the extremal limit. This is probably why, despite considerable effort over the decades since the Myers-Perry solution was discovered (see \cite{Deshpande:2024vbn} for a recent discussion), no one has found its charged generalization analytically. If one adds a Chern-Simons term with a coefficient $\lambda$ normalised so that $\lambda = 1$ is the supergravity value, then extremal black holes remain singular as one increases $\lambda$, except for the precise value $\lambda =1$. We find it remarkable that by studying black holes in a theory without fermions, one can select the value of $\lambda$ required by supersymmetry. General relativity seems to know about supergravity.

There are two different types of curvature singularities that can arise depending on the parameters. For $|\lambda| < 1$, generic extremal black holes have diverging tidal forces and diverging electric fields for infalling observers as they approach the horizon. However, all scalars constructed from the curvature and Maxwell field do not blow up and remain finite. This is directly analogous to the singularities that are found for extremal black holes in four dimensions when small higher curvature corrections are added to general relativity \cite{Horowitz:2023xyl,Horowitz:2024dch}. Similar singularities are found for extremal black holes in 4D anti-de Sitter space with perturbed boundary conditions \cite{Horowitz:2022mly}. The remarkable thing about five dimensions is that the generic extremal black hole is singular even in the simplest Einstein-Maxwell theory, without higher derivative corrections or perturbed boundary conditions.

When $|\lambda| > 1$, the singularity becomes stronger, and curvature scalars do blow up.
We also find that for all $\lambda$, if one of the angular momenta vanish, then the extremal limit is singular in this strong sense. This is a well known property of 5D 
Myers-Perry black holes, but we will see that it continues to hold when any charge is added. In particular, if one adds an arbitrarily small amount of angular momentum in one plane to a 5D Reissner-Nordstr\"om black hole, the extremal horizon has diverging curvature scalars.

When the curvature scalars remain finite, the metric
 and vector potential are $C^0$ but not  $C^1$ at the extremal horizon. Since the horizon becomes infinitely far away in spacelike directions, one can define a  near-horizon geometry. 
 When $Q\to 0$, it  approaches the Myers-Perry near-horizon geometry. It is natural to  expect that when the angular momenta vanish, it will approach the 5D  Reissner-Nordstr\"om near-horizon geometry. However this is incorrect.   We find that the solution approaches the more general static $U(1) \times  U(1)$ symmetric near-horizon geometries \cite{Kunduri:2009ud} when $J_\phi$ and  $J_\psi$ both vanish. These metrics involve a squashed $S^3$, and were originally interpreted as the near-horizon geometries of a nonrotating extremal black hole in a background electric field. We will show that the same geometries arise (without a background electric field) for an extremal  black hole in the  limit $J_\phi, J_\psi \to 0$, with the amount of squashing determined by the ratio $J_\phi / J_\psi$.

 These near-horizon geometries are smooth, but the exact metrics differ from them by a term than can be written $\rho^\gamma h$  where $\rho$ is an affine distance to the horizon, $h$ is a smooth tensor field, and the scaling dimension, $\gamma$, is determined by the field equations. We will show that for $|\lambda | < 1$, $\gamma$ is not an integer and lies in the range $0 < \gamma < 1/2$, except for a couple of special cases (corresponding to the known analytic solutions). Similarly, the vector potential has power law behavior near the horizon with the same exponent. This is the origin of the diverging tidal forces and diverging electric fields for observers crossing the extremal horizon.  Even though curvature scalars remain finite, this singularity is strong enough that Einstein's equation cannot be defined even in a distributional sense. For larger coefficients of the Chern-Simons term ($|\lambda | > 1$),  curvature scalars diverge and there is no near-horizon geometry.\footnote{Strictly speaking, there is a solution with the symmetries of a near-horizon geometry, but it does not arise in the asymptotically flat solution.}

We do not obtain these results by finding the general 5D rotating charged black hole solution explicitly. Even the general near-horizon geometry is not known analytically. Instead, we use a combination of numerical and perturbative arguments to establish our results. For example, in the case of pure Einstein-Maxwell theory, we start with the Myers-Perry near-horizon geometry and add charge perturbatively. We find that the black hole entropy does not depend on the charge and is given by\footnote{We set $G_5 = 1$ throughout this paper.}  
\begin{equation}
S_{1}=2\pi \sqrt{|J_{\phi} J_{\psi}|}\,.
\label{eq:MPS}
\end{equation}
We also start with the general static $U(1)\times U(1)$ symmetric near-horizon geometry and add angular momenta perturbatively. The entropy is then given by
\begin{equation}
{S_2}=\frac{4 \sqrt{\pi }}{3^{3/4}}|Q|^{3/2} +\frac{3^{3/4} \pi ^{3/2}}{4} \frac{\left|J_{\phi }J_{\psi}\right|}{|Q|^{3/2}}
\label{eq:JsS}
\end{equation}
Although this is initially derived just for small angular momenta, when we numerically compute the general near-horizon geometry, we find that it holds exactly for all angular momenta until $S_1 = S_2$. This occurs when
\begin{equation}\label{eq:meet}
\left|J_{\phi} J_{\psi}\right| =\frac{16\ |Q|^3}{3 \sqrt{3} \pi}   
\end{equation}
When $|J_{\phi} J_{\psi}|$ exceeds this bound, the entropy is given by \eqref{eq:MPS} exactly. So there are two branches of near-horizon geometries (when $\lambda = 0$) which meet at \eqref{eq:meet}. The fact that these two branches have simple entropy formulas strongly suggests that there is an analytic solution for the general near-horizon geometry in this case. Finding this solution explicitly remains an open problem. In the special case $J_{\phi} = J_{\psi}$, there is enhanced symmetry and the analytic solution is known \cite{Blazquez-Salcedo:2013yba}. Our entropy formulas \eqref{eq:MPS} and \eqref{eq:JsS} reduce to the known expressions in this case.

After finding the near-horizon geometries of the extremal solutions, we compute the scaling dimensions $\gamma$. We again first do this perturbatively for small charge and small angular momenta, and then compute $\gamma$ numerically for the general case.
 Finally, we numerically construct the full asymptotically flat, near extremal solutions and show that the horizon becomes singular as one approaches extremality.

 In the next section we introduce the one parameter family of 5D theories that we will consider, labelled by the coefficient of a Chern-Simons term. In section 3 we solve the field equations to determine the near-horizon geometries. The scaling dimensions are computed in Section 4, and the full asymptotically flat solutions are constructed in Section 5. Section 6 contains some discussion of these results, and the Appendices compute some of the higher scaling dimensions of the near-horizon geometries with zero charge or zero angular momenta.

\section{5D Einstein-Maxwell theory with possible Chern-Simons term}
We start with  the following action
\begin{equation}
S=\frac{1}{16\pi }\int_{\mathcal{M}}\mathrm{d}^5 x\sqrt{-g}\left(R-\frac{1}{4}F^{ab}F_{ab}-\frac{\lambda}{12\sqrt{3}}\varepsilon^{abcde}F_{ab}F_{cd}A_e\right)\,,
\label{eq:action}
\end{equation}
where $R$ is the Ricci scalar associated with $g$, $F={\rm d}A$ is the Maxwell field strength, $A$ is its gauge potential, and $\lambda$ is a parameter that we take to be real. For $\lambda=1$, the action above admits a supersymmetric completion, becoming the bosonic sector of minimal supergravity in five dimensions.

The equations of motion derived from the action (\ref{eq:action}) are:
\begin{equation}
R_{ab}-\frac{R}{2}g_{ab}=\frac{1}{2}\left(F_{a}^{\phantom{a}c}F_{bc}-\frac{g_{ab}}{4}F^{cd}F_{cd}\right)\quad\text{and}\quad \nabla_b F^{ba}=\frac{\lambda}{4\sqrt{3}}\varepsilon^{abcde}F_{bc}F_{de}\,.
\label{eqs:EOM}
\end{equation}
We are interested in constructing stationary and asymptotically flat black hole solutions to the above equations of motion. Furthermore, we focus on solutions that have a spatial $U(1)^2$ symmetry in addition to the timelike symmetry $\mathbb{R}$, with each of the $U(1)$ factors parametrising rotations around a two-dimensional axis.

Let $k=\partial/\partial t$ be the stationary Killing vector field, normalised so that $k^2 = -1$ near null infinity. Additionally, let $m_{\phi} = \partial/\partial \phi$ generate a one-parameter group of isometries isomorphic to one of the $U(1)$ factors, and similarly for $m_{\psi}=\partial/\partial \psi$, with both $\phi\sim\phi+2\pi$ and $\psi\sim\psi+2\pi$.

For such solutions, one can compute their energy $M$, angular momenta $J_{\phi}$ and $J_{\psi}$ and charge $Q$ via Komar integrals \cite{Komar:1958wp} evaluated at spatial infinity as
\begin{subequations}
\label{eq:charges}
\begin{equation}
M=\frac{3}{32\pi }\lim_{r\to+\infty}\int_{S^3_r} \star \mathrm{d}k\,,
\end{equation}
\begin{equation}
J_{\phi}=-\frac{1}{16\pi }\lim_{r\to+\infty}\int_{S^3_r} \star \mathrm{d}m_{\phi}\,,
\end{equation}
\begin{equation}
J_{\psi}=-\frac{1}{16\pi }\lim_{r\to+\infty}\int_{S^3_r} \star \mathrm{d}m_{\psi}\,,
\end{equation}
and
\begin{equation}
Q=\frac{1}{16\pi } \int_{\Sigma}\left(\star F+\frac{\lambda}{\sqrt{3}}F\wedge A\right)\,.
\end{equation}
\end{subequations}%
where $r$ is the standard radial coordinate near spatial infinity,  and $\Sigma$ is any (topological) three-sphere surrounding the black hole.

Note that there are several possible definitions of charge, due to the presence of the Chern-Simons term. Here we use the so called Page charge \cite{Page:1983mke}, which satisfies a Gauss law, making the result independent of the choice of integration surface $\Sigma$. One might worry that the Page charge is gauge dependent (since it explicitly depends on $A$). However, the gauge  ambiguity can be fixed by demanding regularity of the vector field at the horizon, which we will assume henceforth.

Similarly, the angular momenta Komar integrals can be recast as integrals over a surface $\Sigma$, provided that $\Sigma$ is a (topological) three-sphere surrounding the black hole. For vacuum spacetimes this can be done trivially, since $R_{ab}=0$ implies that the integral does not change when one deforms the surface of integration. However this is not the case for the theory at hand. Indeed, there are matter contributions that must be included. After some algebra, one finds \cite{Hanaki:2007mb,Kunduri:2013gce}
\begin{equation}
J_{\phi}=-\frac{1}{16\pi }\int_{\Sigma}\left[\star{\rm d}m_{\phi}+(m_{\phi}\cdot A) \star F+\frac{2\lambda}{3\sqrt{3}}(m_{\phi}\cdot A)A\wedge F\right]\,,
\label{eq:angH}
\end{equation}
and similarly for $J_{\psi}$. This expression can be used to read off the angular momenta of any black hole solution directly at the horizon and it will be of great help to interpret the integration constants when studying near-horizon geometries. Indeed, a similar expression also exists for the energy $M$.

We focus on black hole solutions satisfying Hawking's rigidity theorem \cite{Bardeen:1973gs} for which the event horizon $H$ is a Killing horizon with the horizon generator being
\begin{equation}
K=k+\Omega_{\phi} m_{\phi}+\Omega_{\psi} m_{\psi}\,
\end{equation}
with $\Omega_{\phi}$ and $\Omega_{\psi}$ being constant \cite{Carter:1969zz} and interpreted as the black hole angular velocity along the $\phi$ and $\psi$ angular directions. Furthermore, throughout, we will assume that spatial horizon cross sections have topology $S^3$ and that a bifurcating Killing three-sphere $\mathcal{B}^+$ exists where $K=0$.

To the Killing horizon we can associate a constant \cite{Bardeen:1973gs} Hawking temperature \cite{Hawking:1974rv},
\begin{equation}
T=\frac{1}{2\pi}\sqrt{-\frac{1}{4}\left.\frac{\nabla_a(K^cK_c)\nabla^a (K^d K_d)}{K^e K_e}\right|_{H}}\,,
\label{eq:temperature}
\end{equation}
from which we note that extremal black holes have $T=0$. Finally, we define the chemical potential of the black hole event horizon, by defining
\begin{equation}
\mu = \left.K^a A_a\right|_{H}-\lim_{r\to+\infty}K^a A_a\,.
\end{equation}
Note that we are assuming a gauge choice where the Maxwell potential is regular on 
$H$. This is achieved by choosing a gauge where $\left.K^a A_a\right|_{H}=0$, ensuring that only the last term contributes. This gauge can always be attained because the results of \cite{Bardeen:1973gs} can be used to show that $\mu$ is necessarily constant.

One can show that, with the definitions above, the solutions we are seeking to construct must satisfy the first law of black hole mechanics
\begin{subequations}
\begin{equation}
\delta M=T\,\delta S+\Omega_{\phi}\,\delta J_{\phi}+\Omega_{\psi}\,\delta J_{\psi}+\mu\,\delta Q\,
\end{equation}
where
\begin{equation}
S=\frac{A}{4 }\,,
\end{equation}
\end{subequations}
with $A$ being the area of the bifurcating Killing three-sphere $\mathcal{B}^+$.
\section{Near-horizon geometries}

In this section
we construct rotating and electrically charged near-horizon geometries. It has been shown that spherical, near-horizon geometries with a $U(1)\times U(1)$ symmetry must also have a $SO(2,1)$ symmetry \cite{Kunduri:2007vf}, so we will impose this below. We can choose an angular coordinate $x\in[-1,1]$ so that the most general line element compatible with these symmetries is
\begin{subequations}
\begin{multline}
{\rm d}s_{\rm NH}^2=Q_1(x)\left(-\rho^2 \mathrm{d}t^2+\frac{\mathrm{d}\rho^2}{\rho^2}\right)+\frac{Q_2(x)\mathrm{d}x^2}{1-x^2}+ (1-x^2)Q_3(x)\left(\mathrm{d}\phi+\omega_{\phi}\,\rho\,\mathrm{d}t\right)^2
\\
+x^2Q_4(x)\left[{\rm d}\psi+\omega_{\psi}\,\rho\,{\rm d}t+(1-x^2)Q_5(x)\left(\mathrm{d}\phi+\omega_{\phi}\,\rho\,\mathrm{d}t\right)\right]^2
\end{multline}
and
\begin{multline}
A_{\rm NH}=-Q_{\rm NH}\,\rho\,\mathrm{d}t+(1-x^2)Q_6(x)\left(\mathrm{d}\phi+\omega_{\phi}\,\rho\,\mathrm{d}t\right)
\\
+x^2Q_7(x)\left[{\rm d}\psi+\omega_{\psi}\,\rho\,{\rm d}t+(1-x^2)Q_5(x)\left(\mathrm{d}\phi+\omega_{\phi}\,\rho\,\mathrm{d}t\right)\right]
\end{multline}
\label{eqs:all}%
\end{subequations}%
Note that the first expression in parenthesis is just two-dimensional anti-de Sitter space. There are a total of seven functions of $x$ to solve for, denoted $Q_I(x)$ where $I=1,\ldots,7$. There are also three constants,
 $\omega_{\phi}$, $\omega_{\psi}$ and $Q_{\rm NH}$, which will determine the angular momenta and charge of the solution. At this point, we have not yet chosen a specific gauge and are keeping the ansatz as general as possible, but consistent with a $O(2, 1) \times U(1)_{\phi}\times U(1)_{\psi}$ symmetry.

Note that regularity at $x=1$, the axis of the $\phi$ rotation, demands that
\begin{subequations}
\begin{equation}
Q_2(1)=Q_3(1)
\end{equation}
while regularity at $x=0$, the axis of the $\psi$ rotation, demands that
\begin{equation}
Q_2(0)=Q_4(0)
\end{equation}
\end{subequations}

For $\lambda=1$, analytic solutions are known to exist, with the most general being the near-horizon limit of a particular extremal family of solutions found in \cite{Cvetic:1996xz}\footnote{This general solution contains extra fields coming from string theory. We restrict our attention to the particular solution presented in \cite{Chong:2005hr}, and set the cosmological constant to zero.}. These have
\begin{subequations}
\begin{multline}
\omega_{\phi}=\frac{a_{\psi } \left(a_{\phi }+\hat{\delta}\,a_{\psi }\right)^2+\left(a_{\phi }+2 \,\hat{\delta}\,a_{\psi }\right) q}{2 \sqrt{|a_{\phi } a_{\psi }+q|} \left[\left(a_{\phi }+\hat{\delta}\,a_{\psi }\right)^2+\hat{\delta}\,q\right]}\hat{\delta}\,,\quad
\omega_{\psi}=\frac{a_{\phi } \left(\hat{\delta}\,a_{\phi }+a_{\psi }\right)^2+\left(2 \hat{\delta}\,a_{\phi }+a_{\psi }\right) q}{2 \sqrt{|a_{\phi } a_{\psi }+q|} \left[\left(a_{\phi }+\hat{\delta}\,a_{\psi }\right)^2+\hat{\delta}\,q\right]}\hat{\delta}\,,\\
\text{and}\quad Q_{\rm NH}=\frac{\sqrt{3} q^2\hat{\delta}}{2 \sqrt{|a_{\phi } a_{\psi }+q|} \left[\left(a_{\phi }+\hat{\delta}\,a_{\psi }\right)^2+\hat{\delta}\,q\right]}\,,
\end{multline}
with $q$, $a_{\phi}$ and $a_{\psi}$ being constants that parametrise charge and rotations along the $\phi$ and $\psi$ axis and
\begin{equation}
\hat{\delta}=\left\{
\begin{array}{c}
\phantom{+}1\quad \text{if}\quad a_{\phi}a_{\psi}+q>0
\\
-1\quad \text{if}\quad a_{\phi}a_{\psi}+q<0
\end{array}\right.\,.
\end{equation}
Furthermore,
\begin{align}
&Q_2(x)=4 Q_1(x)=|a_{\phi } a_{\psi }+q|+a_{\phi }^2 x^2+a_{\psi }^2\left(1-x^2\right)
\\
&Q_3(x)=\frac{4 |a_{\phi } a_{\psi }+q| \left[\left(a_{\phi }+\hat{\delta}\,a_{\psi }\right)^2+\hat{\delta}\,q\right]^2 Q_1(x)}{Z(x)}
\\
&Q_4(x)=\frac{Z(x)}{Q_2(x)^2}
\\
&Q_5(x)=\frac{4 \left(a_{\phi }+\hat{\delta}\,a_{\psi }\right)^2 \left(a_{\phi } a_{\psi }+q\right) Q_1(x)-a_{\phi } a_{\psi } q^2}{Z(x)}
\\
&Q_6(x)=-\frac{\sqrt{3} q}{Z(x)} \left\{4 \left(a_{\phi }+\hat{\delta}\,a_{\psi }\right) |a_{\phi } a_{\psi }+q| Q_1(x)-a_{\psi } q\,\hat{\delta}\,\left[a_{\psi } \left(a_{\phi }+\hat{\delta}\,a_{\psi }\right)+q\right]\right\}
\\
&Q_7(x)=-\frac{\sqrt{3} a_{\psi } q}{Q_2(x)}\,,
\end{align}
where
\begin{multline}
Z(x)=\left(a_{\psi }^2+|a_{\phi}a_{\psi}+q|\right)^3+\left(a_{\phi}^2-a_{\psi}^2\right) \left(a_{\phi }+\hat{\delta}\,a_{\psi }\right)^2 |a_{\phi }
   a_{\psi }+q|x^4
   \\
   +\Big[\left(2 a_{\phi }-\hat{\delta}\,a_{\psi }\right) a_{\psi }^2 \left(a_{\phi }+\hat{\delta}\,a_{\psi }\right)^3+\left(4 a_{\phi }-\hat{\delta}\,a_{\psi }\right) a_{\psi }
   \left(a_{\phi }+\hat{\delta}\,a_{\psi }\right)^2 q
   \\
   +\left(2 a_{\phi }^2+2 a_{\phi } a_{\psi }\hat{\delta}-a_{\psi }^2\right) q^2\Big] x^2\,.
\end{multline}%
\label{eqs:crazyNH}%
\end{subequations}%
For $a_{\phi} = a_{\psi} = 0$, we recover the standard near-horizon geometry of a Reissner-Nordstr\"om black hole, while for $q = 0$, we have the near-horizon geometry of an extremal Myers-Perry black hole in five spacetime dimensions \cite{Myers:1986un}. For both of these particular cases, the above solution is valid for \emph{any} value of $\lambda$.

We aim to investigate whether \emph{smooth} solutions exist for $\lambda \neq 1$ and generic values of the charge and angular momenta. This will be done in three steps. First, we will take the limit of small angular momenta, where perturbation theory can be applied to proceed analytically. Similarly, we will study the small charge limit analytically. Finally, we will numerically construct typical near-horizon geometries for generic values of the angular momenta and charge. 

\subsection{Perturbative in the angular momenta}

To find the near-horizon geometry with slow rotation, it is natural to try to perturb the near-horizon geometry of 5D Reissner-Nordstr\"om. However, one soon encounters an obstruction. To obtain a perturbative solution, one must start with
 the most general static near-horizon geometry in Einstein-Maxwell-Chern-Simons theory (with   vanishing magnetic field), possessing $O(2,1)\times U(1)_{\phi}\times U(1)_{\psi}$ isometry, as found in \cite{Kunduri:2009ud}:
\begin{subequations}
\begin{equation}
{\mathrm{d}s^2_{\rm NH}}=\Gamma(x)\left(-\rho^2 \mathrm{d}T^2+\frac{\mathrm{d}\rho^2}{\rho^2}\right)+\frac{\sigma(x)^2{\rm d}x^2}{1-x^2}+\frac{c_2^3}{\sigma(x)}(1-x^2)\mathrm{d}\phi^2+\frac{c_1^3}{\sigma(x)}x^2\mathrm{d}\psi^2
\end{equation}
and
\begin{equation}
A=-\frac{\sqrt{3c_1 c_2}}{2}\,\rho\,\mathrm{d}T
\end{equation}
\label{eq:U1U1}%
\end{subequations}%
with
\begin{equation}\label{eq:cidef}
   \sigma(x) = c_1 (1 - x^2) + c_2 x^2, \qquad  \Gamma(x) = \sigma(x)^2 / 4
\end{equation}
 where both $c_1$ and $c_2$ are positive real constants that describe the size and squashing of an $S^3$. When $c_1 = c_2$ the sphere is round and the solution reduces to the near-horizon geometry of an extremal 5D Reissner-Nordström black hole.

To account for the effects of rotation, we modify the above expression accordingly and set:
\begin{subequations}\label{eq:Jansatz}
\begin{multline}
{\mathrm{d}s^2_{\rm NH}}=\Gamma(x)\left(-\rho^2 \mathrm{d}T^2+\frac{\mathrm{d}\rho^2}{\rho^2}\right)+\frac{\sigma_1(x)\sigma_2(x){\rm d}x^2}{1-x^2}+\frac{B_2}{\sigma_1(x)}(1-x^2)\left(\mathrm{d}\phi+\Omega_{\phi}\,\rho\,\mathrm{d}T\right)^2
\\
+\frac{B_1}{\sigma_2(x)}x^2\left[\mathrm{d}\psi+\Omega_{\psi}\,\rho\,\mathrm{d}T+(1-x^2)\alpha(x)\left(\mathrm{d}\phi+\Omega_{\phi}\,\rho\,\mathrm{d}T\right)\right]^2
\end{multline}
and
\begin{multline}
A=-Q_H\,\rho\,\mathrm{d}T+\beta(x)(1-x^2)\left(\mathrm{d}\phi+\Omega_{\phi}\,\rho \mathrm{d}T\right)
\\
+\delta(x)x^2\left[\mathrm{d}\psi+\Omega_{\psi}\,\rho\,\mathrm{d}T+(1-x^2)\alpha(x)\left(\mathrm{d}\phi+\Omega_{\phi}\,\rho\,\mathrm{d}T\right)\right]\,.
\end{multline}
\label{eqs:U1U1}%
\end{subequations}%
Here, $B_1$ and $B_2$ are constants that will be adjusted within perturbation theory to eliminate any conical singularities, while $\Omega_{\phi}$ and $\Omega_{\psi}$ are real constants that parameterize rotation along the $\phi$ and $\psi$ directions, respectively. There are now six functions of $x$ to solve for: $\Gamma(x)$, $\sigma_1(x)$, $\sigma_2(x)$, $\alpha(x)$, $\beta(x)$, and $\delta(x)$. Note that, since the equations of motion are invariant under the rescalings $g \to \hat{\lambda}^2 g$ and $A \to \hat{\lambda} A$, with $\hat{\lambda}$ an arbitrary constant, we can always choose $Q_H$ such that
\begin{equation}
Q_H=\frac{\sqrt{3c_1 c_2}}{2}\,.
\end{equation}
We have fixed the gauge such that the determinant of a spatial cross-section of the extremal horizon is simply given by $x \sqrt{B_1 B_2}$. In these coordinates, we are left with six coupled differential equations for $\Gamma$, $\sigma_1$, $\sigma_2$, $\alpha$, $\beta$, and $\delta$, five of which are second-order in $x$, and one is first-order.

We are now ready to establish our perturbative scheme. Pick two constants, $\omega_{\phi}$ and $\omega_{\psi}$, to describe the rotations we are adding, and define
\begin{equation}
\Omega_{\phi}=\varepsilon\,\omega_{\phi}\,,\quad \Omega_{\psi}=\varepsilon\,\omega_{\psi}
\end{equation}
and set
\begin{align}
&\Gamma(x)=\frac{\sigma(x)^2}{4}\left[1+\sum_{i=1}^{+\infty}\varepsilon^{2i} \Gamma^{(2i)}(x)\right]\,,\qquad \alpha(x)=\sum_{i=1}^{+\infty}\varepsilon^{2i} \alpha^{(2i)}(x)\,\nonumber
\\
&B_{I}=c_{I}^3\left[1+\sum_{i=1}^{+\infty}\varepsilon^{2i} B_I^{(2i)}\right]\,,\qquad \sigma_I(x)=\sigma(x)\left[1+\sum_{i=1}^{+\infty}\varepsilon^{2i} \sigma_I^{(2i)}(x)\right]\quad\text{with}\quad I=1,2\,,
\\
&\beta(x)=\sum_{i=0}^{+\infty}\varepsilon^{2i+1} \beta^{(2i+1)}(x)\,\quad \text{and}\quad \delta(x)=\sum_{i=0}^{+\infty}\varepsilon^{2i+1} \delta^{(2i+1)}(x)\,,\nonumber
\label{eqs:expanq}
\end{align}
while taking $\varepsilon$ to be  arbitrarily small. Note that the introduction of $\Omega_{\phi}$ and $\Omega_{\phi}$ sources first order perturbations to the vector potential but only second order perturbations to the metric functions.

At zeroth order we recover Eqs.~(\ref{eq:U1U1}),  
At linear order we find 
\begin{align}
&\beta^{(1)}(x)=-\frac{2  \sqrt{3c_2} \left(c_2^2 \omega_{\phi }-2 \lambda  c_1^2 \omega _{\psi }\right)}{\left(1-4 \lambda^2\right) \sqrt{c_1} \,\sigma(x)}+\frac{\Omega_1}{1-x^2}+\frac{\lambda  Z(x)}{1-x^2}\,,\nonumber
\\
&\delta^{(1)}(x)=-\frac{2  \sqrt{3c_1} \left(c_1^2 \omega_{\psi }-2 \lambda  c_2^2 \omega_{\phi }\right)}{\left(1-4 \lambda ^2\right) \sqrt{c_2}\,\sigma(x)}+\frac{\Omega_2}{x^2}+\frac{c_1 \sigma (x) Z'(x)}{4 x c _2^2}\,,\nonumber
\\
&Z(x)=x^{4 \lambda } \sigma (x)^{-2 \lambda } \Omega _3 \, _2F_1\left(-2 \lambda ,-2 \lambda ;1-4 \lambda ;\frac{\sigma (x)}{x^2 c _2}\right)\nonumber
\\
&\qquad \qquad \qquad \qquad\qquad +x^{-4 \lambda } \sigma (x)^{2 \lambda } \Omega _4 \,
   _2F_1\left(2 \lambda ,2 \lambda ;1+4 \lambda ;\frac{\sigma (x)}{x^2 c _2}\right)\,,
\end{align}
where $_2F_1\left(a ,b;c ;z\right)$ is a Gauss hypergeometric function, and $\Omega_i$, $i = 1,\cdots, 4$ are constant coefficients. Regularity requires $\Omega_1 =\Omega_2=0 $. In addition, if $\lambda \ge 0$, then $\Omega_4 = 0$ and if  $\lambda \le 0$, then $\Omega_3 = 0$. Finally, if $2\lambda$ is not an integer, regularity requires both   $\Omega_3 = \Omega_4 = 0$ so $Z(x)$ vanishes.
 For $\lambda=1/2$ or $\lambda=1$, $Z(x)$ can be nontrivial, and in particular
\begin{subequations}
\begin{align}
&Z(x)=-\frac{c _1\left(1-x^2\right) }{c _2 \sigma (x)} \Omega _3\quad\text{for}\quad \lambda=1/2
\\
&Z(x)=\frac{c _1 \left(1-x^2\right)}{3 c _2^2 \sigma(x)^2} \left[c _1 \left(1-x^2\right)-2 c _2 x^2\right]\Omega_3\quad\text{for}\quad \lambda=1\,.
\end{align}
\end{subequations}
Our conclusions remain the same if we take $2\lambda$ to be a non-zero integer, but for clarity, we will assume otherwise in this presentation. Without loss of generality, we will also assume $\lambda \ge 0$. 

We now move to second order perturbation theory in $\varepsilon$. At second order we find that $\alpha^{(2)}$ obeys a second order differential equation that we can readily solve while imposing the relevant boundary conditions
\begin{equation}
\alpha^{(2)}(x)=\frac{4 c_1}{\left(1-4 \lambda^2\right)^2 \sigma (x)} \left[3 \lambda  \left(\hat{\alpha}^4 \omega_{\phi }^2+\omega _{\psi }^2\right)-\hat{\alpha}^2 \left(1+10 \lambda ^2-8 \lambda ^4\right) \omega _{\psi } \omega _{\phi }\right]
\end{equation}
with $\hat{\alpha}\equiv c_2/c_1$. Finally, $\Gamma^{(2)}$, $\sigma_1^{(2)}$, and $\sigma_2^{(2)}$ satisfy three coupled ordinary differential equations. To solve these, we introduce two auxiliary variables, $\mathfrak{a}_1$ and $\mathfrak{a}_2$, defined as follows:
\begin{equation}
\Gamma^{(2)}(x)=\mathfrak{a}_1(x)\,\quad\text{and}\quad\sigma_1^{(2)}(x)=\mathfrak{a}_2(x)-\sigma_2^{(2)}(x)\,.
\end{equation}
As a result of this substitution, $\mathfrak{a}_1(x)$ and $\mathfrak{a}_2(x)$ decouple from $\sigma_2^{(2)}(x)$, which now satisfies a first-order differential equation driven by $\mathfrak{a}_1(x)$, $\mathfrak{a}_2(x)$, and their first derivatives. We are left with two second-order differential equations for $\mathfrak{a}_1(x)$ and $\mathfrak{a}_2(x)$. It turns out that these last two equations can be solved in full generality. We will refrain from presenting these solutions explicitly. The general solutions for $\mathfrak{a}_1(x)$ and $\mathfrak{a}_2(x)$ depend on four integration constants, and we have not yet imposed any regularity conditions. Therefore, we can solve for $\sigma_2^{(2)}(x)$ in full generality, though this introduces an additional integration constant. In total, the general solution for $\Gamma^{(2)}(x)$, $\sigma_1^{(2)}$, and $\sigma_2^{(2)}$ depends on five integration constants. The final step involves requiring these functions to remain regular at the poles $x=0$ and $x=1$. This procedure fixes three of the integration constants  \emph{and} demands that
\begin{equation}
\omega_{\phi}=\pm\frac{c_1^2}{c_2^2}\omega_{\psi}\, \qquad {\rm for\ all}\ \lambda^2 \ne 1 \, .
\label{eq:matchu1u1}
\end{equation}
We will consider only the upper sign, as we assume that both angular momenta are positive. The analysis for opposite sign angular momenta can be repeated \emph{mutatis mutandis}. This last expression justifies why we had to take Eqs.~(\ref{eq:U1U1}) as our background. Indeed, to ensure that the angular momenta can be different from each other, we need to consider cases where $c_1 \neq c_2$ and the $S^3$ is squashed! This also shows that if we try to add rotation in only one plane and set the other to zero, the sphere becomes infinitely squashed and hence singular.

Finally, we could now adjust $B_1^{(2)}$ and $B_2^{(2)}$   to remove any hypothetical conical singularity at $x=0$ and $x=1$, respectively. Once the procedure outlined above is done, it is a simple matter to compute the entropy, charge and angular momenta of the new near-horizon geometries. These combine in the following simple result
\begin{equation}
\frac{S}{|Q|^{3/2}}=\frac{4 \sqrt{\pi }}{3^{3/4}}+\frac{3^{3/4} \pi ^{3/2} (1+2\lambda)}{2 (2+\lambda)} \frac{J_\phi J_\psi}{|Q|^{3}} +\mathcal{O}(\epsilon^3)
\label{eq:smallJsS}
\end{equation}
The above expression was derived using perturbation theory. However, for $\lambda = 0$, it turns out to yield the full nonlinear result (with no $\mathcal{O}(\epsilon^3)$ correction). At this point, we have no justification for why this is the case, as we only know the geometry perturbatively. However, when comparing with the full numerical results in the next sections, we find that the above expression for $\lambda = 0$ agrees with the full nonlinear results to the first twenty decimal places, even when $J_{\phi}$ and $J_{\psi}$ are of the same order as $Q^{3/2}$. This agreement is exact only for $\lambda = 0$. Indeed, using the known analytic solution with $\lambda = 1$ (see Eqs.~(\ref{eqs:crazyNH})), it is a straightforward exercise to show that the expression for $\frac{S}{|Q|^{3/2}}$ given above is valid only to leading order in $J_{\phi}$ and $J_{\psi}$. The above result assumed $J_\phi J_\psi \ge 0.$ When $J_\phi J_\psi < 0$, we get instead that
\begin{equation}
\frac{S}{|Q|^{3/2}}=\frac{4 \sqrt{\pi }}{3^{3/4}}-\frac{3^{3/4} \pi ^{3/2} (1-2\lambda)}{2 (2-\lambda)} \frac{J_\phi J_\psi}{|Q|^{3}} +\mathcal{O}(\epsilon^3)\,.
\label{eq:smallJsSn}
\end{equation}
\subsection{Perturbative in the charge}
Introducing charge perturbatively is considerably more cumbersome than introducing angular momenta perturbatively. The reason is that the starting point, namely the five-dimensional extreme Myers-Perry solution, is more intricate than the general $U(1) \times U(1)$ static near-horizon geometry given in Eqs.~(\ref{eq:U1U1}).

We change gauge and set
\begin{subequations}
\begin{multline}
\mathrm{d}s^2=\Gamma(x)\Bigg\{-\rho^2 \mathrm{d}T^2+\frac{\mathrm{d}\rho^2}{\rho^2}+\frac{\Gamma_{\rm NH}^2{\rm d}x^2}{1-x^2}+h_{11}(x)(1-x^2)\left(\mathrm{d}\phi+\Omega_{\phi}\,\rho\,\mathrm{d}T\right)^2
\\
+h_{22}(x)x^2\left[\mathrm{d}\psi+\Omega_{\psi}\,\rho\,\mathrm{d}T+(1-x^2)h_{12}(x)\left(\mathrm{d}\phi+\Omega_{\phi}\,\rho\,\mathrm{d}T\right)\right]^2\Bigg\}\,,
\end{multline}
\begin{multline}
A=-Q_{\rm NH}\,\mathrm{d}T+b_1(x)(1-x^2)\left(\mathrm{d}\phi+\Omega_{\phi}\,\rho\,\mathrm{d}T\right)
\\
+b_2(x)x^2\left[\mathrm{d}\psi+\Omega_{\psi}\,\rho\,\mathrm{d}T+(1-x^2)h_{12}(x)\left(\mathrm{d}\phi+\Omega_{\phi}\,\rho\,\mathrm{d}T\right)\right]\,,
\end{multline}
\end{subequations}%
with $\Gamma_{\rm NH}$ and $Q_{\rm NH}$ real constants to be determined in what follows. Note that in these coordinates, the coefficients of $\mathrm{d}x^2$ and $\mathrm{d}\rho^2$ are related in a simpler manner than in our previous ansatz Eq~\eqref{eq:Jansatz}. At the nonlinear level we are left with six second order differential equations in $x$ for $h_{11}$, $h_{12}$, $h_{22}$, $\Gamma$, $b_1$ and $b_2$ and one nonlinear constraint quadratic in the first derivatives of $h_{11}$, $h_{12}$, $h_{22}$, $\Gamma$, $b_1$ and $b_2$. The absence of conical singularities at $x=0$ and $x=1$ demand
\begin{equation}
\Gamma_{\rm NH}^2=h_{22}(0)\quad\text{and}\quad \Gamma_{\rm NH}^2=h_{11}(1)\,
\end{equation}
respectively.

Our background is simply a five-dimensional extremal Myers-Perry black hole, for which
\begin{subequations}
\begin{align}
&\Gamma(x)=\Gamma^{(0)}(x)=\frac{1}{4}(a+b)\left[a\,x^2+b(1-x^2)\right]
\\
&h_{11}(x)=h_{11}^{(0)}(x)=\frac{4 a(a+b)}{(b+a x^2)\left[a\,x^2+b(1-x^2)\right]}
\\
&h_{12}(x)=h_{12}^{(0)}=\frac{a}{b+a x^2}
\\
&h_{22}(x)=h_{22}^{(0)}=\frac{4b(b+a x^2)}{\left[a\,x^2+b(1-x^2)\right]^2}
\\
&\Gamma_{\rm NH}=2
\\
&\Omega_{\phi}=\frac{1}{2}\sqrt{\frac{b}{a}}\quad\text{and}\quad \Omega_{\psi}=\Omega_{\psi}^{(0)}=\frac{1}{2}\sqrt{\frac{a}{b}}=\frac{1}{4\Omega_{\phi}}
\end{align}
\end{subequations}%
while $b_1$, $b_2$ and $Q_{\rm NH}$ vanish. Rather than label the solutions by the rotation parameters $(a,b)$, perturbation theory will be easier to present if we label them by $(a, \Omega_\phi)$ using $b = 4a\Omega_{\phi}^2$. In the near-horizon geometry of an extremal Myers-Perry black hole, $\Omega_\psi$ is then fixed by $4\Omega_{\psi}\Omega_{\phi} = 1$.

Our perturbation scheme reads
\begin{align}
&\Omega_{\psi}=\sum_{i=0}^{+\infty} q^i \Omega_{\psi}^{(i)}\,,\quad Q_{\rm NH}=\sum_{i=1}^{+\infty}q^i Q^{(i)}_{\rm NH}\,,\quad \Gamma(x)=\sum_{i=0}^{+\infty} q^i \Gamma^{(i)}(x)\,,\nonumber
\\
& h_{IJ}(x)=\sum_{i=0}^{+\infty}q^i h_{IJ}^{(i)}(x)\,,\quad\text{and}\quad b_{I}(x)=\sum_{i=1}^{+\infty}q^i b_{I}^{(i)}(x)\quad \text{with}\quad I,J=1,2\,,
\label{eq:expq}
\end{align}
while maintaining $\Gamma_{\rm NH}=2$.

To zeroth order, we recover the near-horizon geometry of an extremal Myers-Perry black hole by construction. At linear order, the equations for the $b_I(x)$ decouple from the remainder and we find
\begin{equation}
b_1(x)=\frac{\varphi_0+4\Omega_{\phi}^2\varphi_1}{x^2+4\Omega_{\phi}^2}\quad\text{and}\quad b_2(x)=\varphi_1+\frac{4  \Omega _{\phi }^2\varphi_0}{x^2+4\Omega_{\phi}^2 (1-x^2)}\,,
\end{equation}
with $\varphi_0$ and $\varphi_1$ integration constants. The equations for $\Gamma^{(1)}(x)$ and the $h_{IJ}^{(1)}(x)$ are coupled, but can be solved after the following field redefinition:
\begin{multline}
\Gamma^{(1)}(x)=\tilde{\Gamma}^{(1)}(x)-\frac{a^2}{48} \left(x^2+4 \Omega _{\phi }^2\right) \left[x^2+4 \Omega _{\phi }^2 \left(1-x^2\right)\right]^2 h_{11}^{(1)}(x)
\\
-\frac{a^2 \left(1+4 \Omega _{\phi }^2\right) \left[x^2+4 \Omega
   _{\phi }^2 \left(1-x^2\right)\right]^3}{192 \Omega _{\phi }^2 \left(x^2+4 \Omega _{\phi }^2\right)}h_{22}^{(1)}(x)\,.
\end{multline}
We are left with a second order equation for $\tilde{\Gamma}^{(1)}(x)$ (which can be readily solved) and three equations for the $h_{IJ}^{(1)}(x)$, with two being second order and one being first order. These equations, in turn, can be solved analytically. The full solution at this order, and prior to imposing any boundary condition, depends on seven integration constants. Regularity conditions, including the absence of conical singularities, determine five of these integration constants along with: 
\begin{equation}
Q_{\rm NH}^{(1)}=0\,.
\end{equation}

Once we have determined the solution up to this order, we can readily compute all the corresponding charges, as well as the entropy. In fact, we carried out the above calculation to second order in $q$. Once the dust settles, we find
\begin{equation}
S=2\pi \sqrt{|J_{\phi} J_{\psi}|}+\mathcal{O}(q^3)\,.
\label{eq:smallQS}
\end{equation}
Note that $S$, $J_{\phi}$, and $J_{\psi}$ are complicated functions of $a$, $\varphi_0$, $\varphi_1$, $\Omega_{\phi}$, and $\lambda$, yet the above expression still holds. Since the Chern-Simons term is cubic in the vector potential, it will only enter the analysis at $\mathcal{O}(q^3)$. However, if $\lambda = 0$,  the exact numerical data described in the next section agrees with the above expression  exactly (with no $\mathcal{O}(q^3)$ corrections)! In other words, the extremal entropy is independent of the charge. At the moment we have no understanding of why that is the case. Our results at small angular momenta, and small charge strongly suggest that, at least for $\lambda=0$, an analytic solution interpolating between the static $U(1)\times U(1)$ near-horizon geometry and the near-horizon geometry of an extreme Myers-Perry black hole ought to exist.
\subsection{Fully nonlinear results\label{subsec:fullback}}
To find the near-horizon geometry  for generic values of the angular momenta, charge and $\lambda$ we proceed numerically. We use the DeTurck trick, first proposed in \cite{Headrick:2009pv} and reviewed in \cite{Wiseman:2011by,Dias:2015nua}. We start with a generic line element and gauge field that are compatible with $O(2,1) \times U(1)_{\phi} \times U(1)_{\psi}$, as in Eqs.~(\ref{eqs:all}). Instead of solving the Einstein equation (\ref{eqs:EOM}), we solve the Einstein-DeTurck equation:
\begin{equation} R_{ab} - \nabla_{(a}\xi_{b)} = \frac{1}{2}\left(F_{a}^{\phantom{a}c}F_{bc} - \frac{g_{ab}}{6}F^{cd}F_{cd}\right)\,,
\label{eq:deturck}
\end{equation}
while leaving Maxwell's equation unchanged. If $\xi = 0$, the above reduces to the trace-reversed form of (\ref{eqs:EOM}). The vector $\xi$ is called the DeTurck vector and is defined as $\xi^a =\left[\Gamma^{a}_{bc}(g) - \Gamma^{a}_{bc}(\bar{g})\right]g^{bc}$, where $\Gamma^{a}_{bc}(\mathfrak{g})$ are the components of the Levi-Civita connection associated with a metric $\mathfrak{g}$. $\bar{g}$ is the so-called reference metric, and essentially controls the gauge choice, as we will shortly see.

When $\xi = 0$, solutions of (\ref{eq:deturck}) coincide with those of the original Einstein equation (\ref{eqs:EOM}). However, this is almost certainly not the case when $\xi \neq 0$. Solutions with $\xi \neq 0$ are called Ricci solitons, and when $F = 0$, they were shown not to exist when searching for certain stationary solutions \cite{Figueras:2016nmo}. In our case, such unphysical solutions can, in principle, exist. However, for stationary solutions like the ones we aim to construct, Eq.~(\ref{eq:deturck}) leads to a system of elliptic partial differential equations, once appropriate boundary conditions are specified. Crucially, elliptic partial differential equations yield locally unique solutions. This implies that solutions with $\xi \neq 0$ cannot be arbitrarily close to those with $\xi = 0$. Therefore, once we find a solution, we can monitor $\xi^a \xi_a \geq 0$ and verify whether it approaches zero as we refine the resolution of our numerical scheme. Since we are solving the Einstein equation when $\xi^a = 0$, one might wonder which gauge this procedure naturally induces. We recall that $\xi^a = \left[\Gamma^{a}_{bc}(g) - \Gamma^{a}_{bc}(\bar{g})\right]g^{bc} = \Box x^a - \tilde{S}^a$, where $\tilde{S}^a \equiv \Gamma^{a}_{bc}(\bar{g})g^{bc}$. This indicates that the DeTurck gauge is essentially a stationary version of the usual generalized harmonic gauge.

We now turn our attention to boundary conditions for the functions in \eqref{eqs:all}. At $x=0$, to avoid a conical singularity, we require that $Q_2(0)=Q_4(0)$, with the other functions being even across $x=0$. At $x=1$, however, we have $Q_2(1)=Q_3(1)$, while the remaining functions satisfy Robin boundary conditions, which we will refrain from presenting here.

It remains to specify what reference metric we take in our construction. For the problem at hand, we picked $\bar{g}$ to be given by $4Q_1=Q_2=Q_3=Q_4=4$ and $Q_5=0$ in Eqs.~(\ref{eqs:all}).  To solve the Einstein-DeTurck and Maxwell's equations, we used spectral collocation methods on a Gauss-Lobatto grid with $N$ collocation points. Once we have a given solution, we checked that as $N$ increases, $\xi^a \xi_a$ approached zero exponentially fast in $N$. For all the plots shown in this section we used $N=200$.

Our moduli space of solutions depends on the Chern-Simons coefficient $\lambda$, the angular momenta $J_{\phi}$ and $J_{\psi}$, and the electric charge $Q$. Using the scaling symmetry $g \to \hat{\lambda}^2 g$ and $F \to \hat{\lambda} F$, with $\hat{\lambda}$ a real constant, we can always rescale one of these quantities so that only the appropriately normalised ratios are meaningful. We will work with $J_{\phi}/|Q|^{3/2}$ and $J_{\psi}/|Q|^{3/2}$, demonstrating that for fixed $\lambda$, we have a continuous two-dimensional moduli space of solutions. In our numerical scheme we have access to $Q_{\rm NH}$, $\omega_{\phi}$ and $\omega_{\psi}$,  which we dial to move across the two-dimensional phase space of solutions

\begin{figure}
\centering
\includegraphics[width=0.48\textwidth]{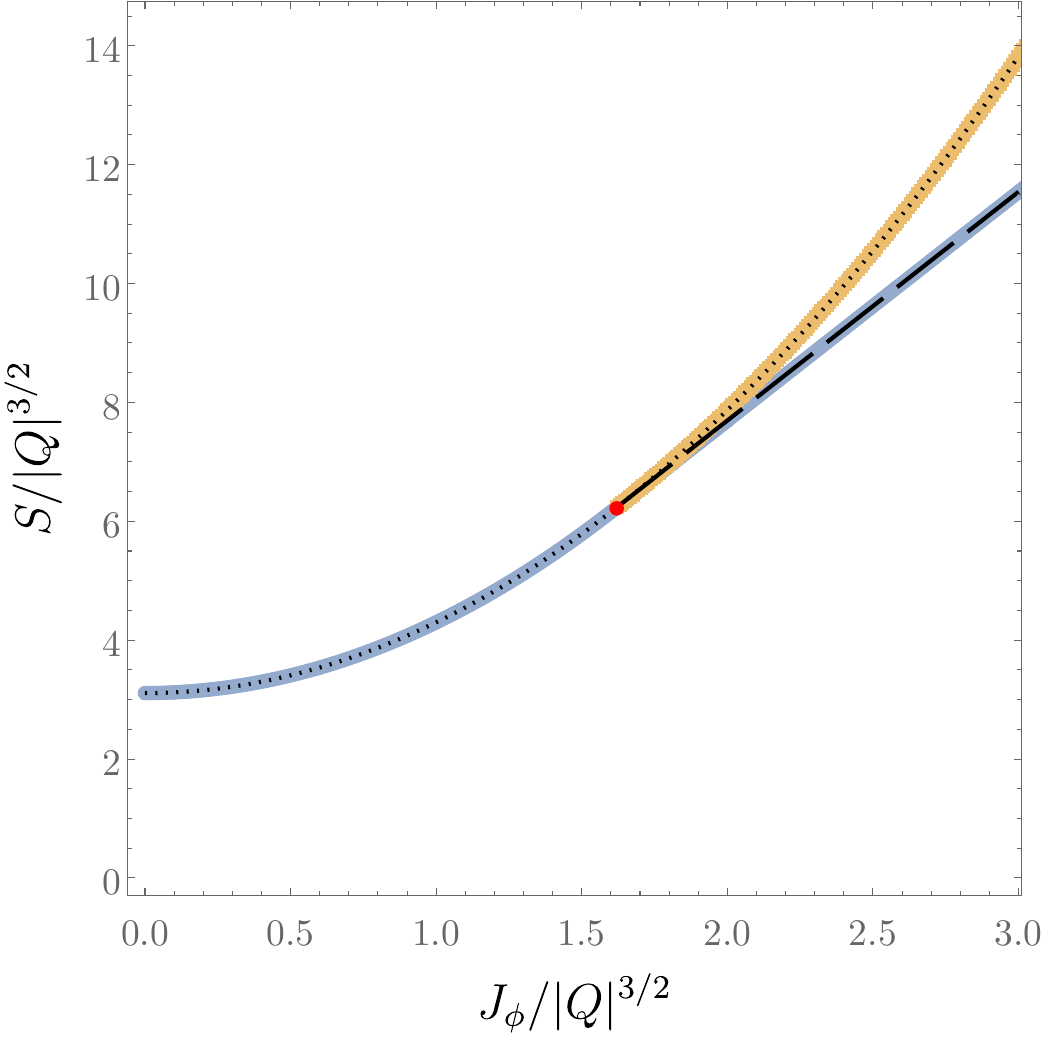}
\caption{Normalised entropy $S/|Q|^{3/2}$ as a function of the normalised angular momentum $J_{\phi}/|Q|^{3/2}$ for fixed $J_{\psi}=3 J_{\phi}/8$ in 5D Einstein-Maxwell theory. To the left of the red dot, the blue curve satisfies Eq.~(\ref{eq:JsS}), and to the right it satisfies Eq.~\eqref{eq:MPS}. The orange branch corresponds to 
near-horizon geometries which do not extend to asymptotically flat solutions.}
\label{fig:entropyla0}
\end{figure}

In Fig.~\ref{fig:entropyla0}, we plot $S/|Q|^{3/2}$ as a function of $J_{\phi}/|Q|^{3/2}$ for $J_{\psi}/J_{\phi} = 3/8$, with $\lambda$ fixed at 0. (Other ratios of $J_{\psi}/J_{\phi}$ display qualitatively similar behavior.)
 The blue and  orange curves describe our numerical solutions. For small angular momenta, we find a single family of solutions and the blue curve appears to follow Eq.~(\ref{eq:JsS}) (denoted by a black dashed line) exactly. This indicates that our perturbative result for small angular momenta (\ref{eq:smallJsS}) does not receive any higher order corrections, and actually holds beyond the perturbative regime. For large angular momenta, we find two different families of solutions. The blue curve appears to follow Eq.~\eqref{eq:MPS} (denoted by a long-dashed gray line) exactly, indicating that our perturbative result for small charge (\ref{eq:smallQS}) also does not receive corrections. The second branch of solutions, denoted by the orange curve, appears to follow a continuation of Eq.~(\ref{eq:JsS}). It starts at the point where the two entropy formulas \eqref{eq:MPS} and (\ref{eq:JsS}) agree, which occurs when 
\begin{equation}\label{eq:matching}
J_{\phi} J_{\psi}= \frac{16}{3\sqrt{3}\pi} Q^{3}
\end{equation}
This point is denoted by the red dot in Fig.~\ref{fig:entropyla0}. We will show in Sec. 4 that the orange branch of near-horizon geometries cannot be extended to asymptotically flat solutions. So only the blue curve represents extremal charged rotating black holes in 5D Einstein-Maxwell theory.

The picture above changes when we turn on $\lambda$.  For any $\lambda \neq 0$, we find that Eq.~(\ref{eq:smallJsS}) and Eq.~(\ref{eq:smallQS}) are only valid in the regimes where they were derived, i.e., for small angular momenta and small charge, respectively. Interestingly, we no longer observe a merger between distinct families of solutions. The two families of solutions represented by the blue and orange curves in Fig.~\ref{fig:entropyla0} separate, and the kink in the  blue curve at \eqref{eq:matching} smooths out. The orange curve stops above the blue curve, and connects to yet another family of solutions. However we will see that only the blue curve describes near-horizon geometries that connect to asymptotically flat regions. 

\begin{figure}
\centering
\includegraphics[width=\textwidth]{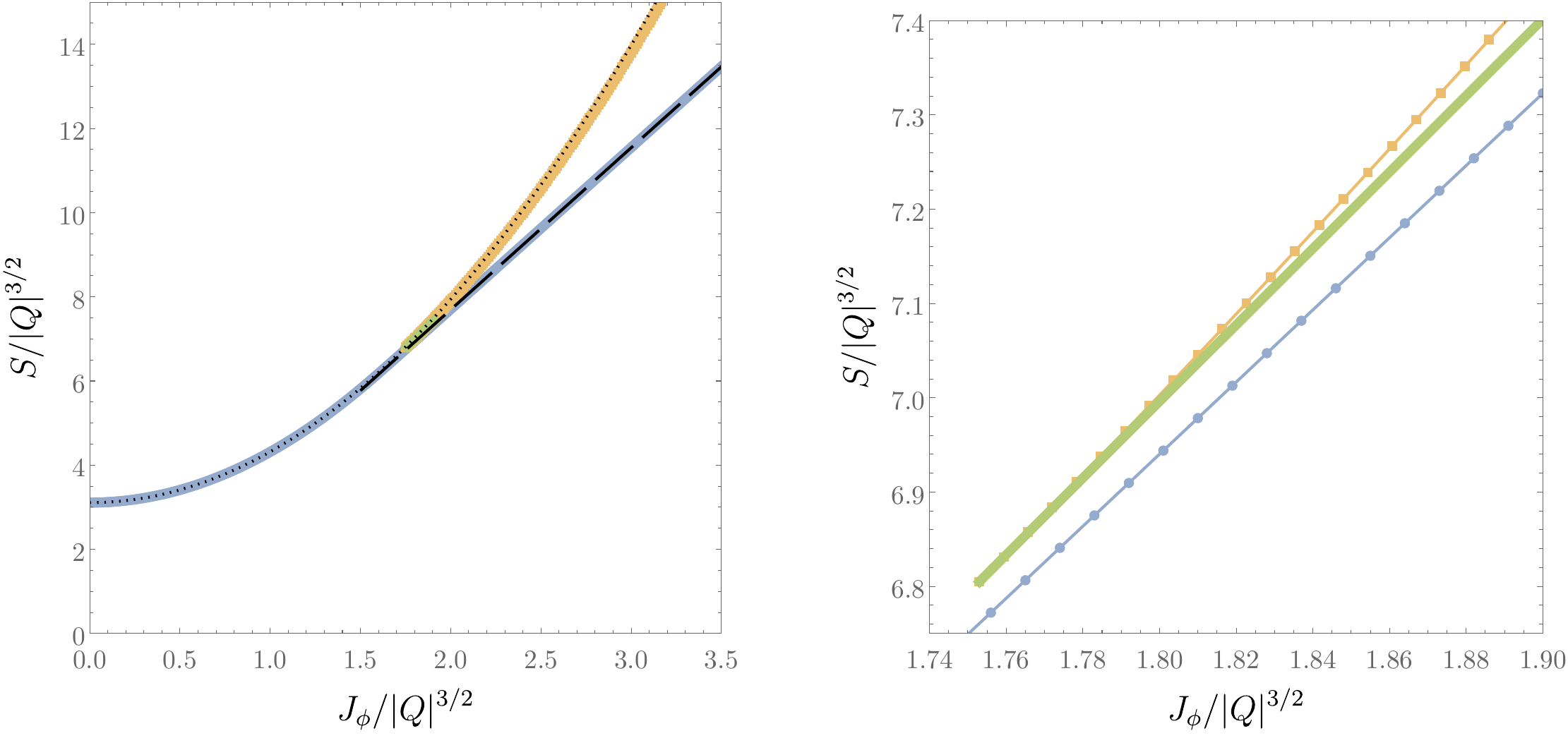}
\caption{{\bf Left Panel}: Normalised entropy $S/|Q|^{3/2}$ as a function of the normalised angular momentum $J_{\phi}/|Q|^{3/2}$ for fixed $J_{\psi}=3 J_{\phi}/8$ and $\lambda=10^{-2}$. {\bf Right Panel}: A blow-up of the center region of the left panel. Three distinct families of near-horizon geometries are visible, but only the one denoted by blue dots can be extended to asymptotically flat solutions. }
\label{fig:entropyla0.01}
\end{figure}

 We illustrate this behaviour in Fig.~\ref{fig:entropyla0.01}, which was constructed for fixed $\lambda=0.01$. On the left panel we show the normalised entropy $S/|Q|^{3/2}$ as a function of the normalised angular momentum $J_{\phi}/|Q|^{3/2}$ for all the different branches found, while on the right panel we plot a blow-up of the region of angular momenta close to \eqref{eq:matching}  (the red dot in 
 Fig.~\ref{fig:entropyla0}).
 In both panels, the new branch of solutions is represented in green.
  This new branch does not appear to extend to arbitrarily large values of $J_{\phi}/|Q|^{3/2}$. The gap between the blue branch of solutions and the rest is clear on the right. The point where the orange and green branches of solutions merge moves to larger angular momenta as we increase $\lambda$. For instance, for $\lambda\approx 0.12$ this starting point is already above $J_{\phi}/|Q|^{3/2}=3.7$ for fixed $J_{\psi}=3J_{\phi}/8$, so the diagram on the left would show only the blue curve.
 
As we continue to increase $\lambda$, we eventually reach $\lambda=1$, and the numerical curve coincides with
\begin{equation}
S=\sqrt{\left|\frac{16 \pi}{3 \sqrt{3}}Q^3+4  \pi ^2 J_{\phi } J_{\psi }\right|}\,,
\end{equation}
which was computed directly from Eqs.~(\ref{eqs:crazyNH}), giving reassurance that our numerical code is correct.
\section{Scaling dimensions\label{sec:scalings}}
We now turn our attention to computing the scaling dimensions associated with our near-horizon geometries. Following the same strategy as in the previous section, we first perform the calculations analytically at small angular momenta and small charge, before proceeding to the fully nonlinear regime, which we once again explore using numerical methods.
\subsection{Perturbative in the angular momenta\label{sec:smalljs}}
 In this section we describe how to compute the scaling dimensions $\gamma$ using perturbation theory about the static $U(1)\times U(1)$ symmetric near-horizon geometries described in Sec. 3.1 above. Our perturbatively small parameters will be the angular momenta $J_{\phi}$ and $J_{\psi}$ as in that section. It was noted in \cite{Horowitz:2022mly} that the lowest scaling dimension associated with the near-horizon geometry of an extremal Reissner-Nordstr\"om black hole in 5D is $\gamma = 0$. This occurs when deformations in the near-horizon region are generated by an $\ell = 2$ harmonic mode on the three-sphere. Since adding rotation to a spherically symmetric spacetime involves vector spherical harmonics, we expect the backreaction of angular momenta on the geometry to alter the corresponding scaling dimensions at second order. We will show that there is a $\gamma = 0$ perturbation even when we deform the $S^3$ in the near-horizon geometry, and adding angular momenta indeed shifts $\gamma$ above zero. When this happens, the horizon becomes singular since the metric is  continuous but no longer differentiable there. 

We will chose our gauge so that
\begin{subequations}
\begin{multline}
\mathrm{d}s^2=Q_1(x,\rho)\left(-\rho^2 \mathrm{d}T^2+\frac{\mathrm{d}\rho^2}{\rho^2}\right)+\frac{Q_2(x,\rho){\rm d}x^2}{1-x^2}+\frac{1-x^2}{Q_3(x,\rho)}\left(\mathrm{d}\phi+\rho\,Q_6(x,\rho)\,\mathrm{d}T\right)^2
\\
+\frac{x^2}{Q_4(x,\rho)}\left[\mathrm{d}\psi+\rho\,Q_7(x,\rho)\,\mathrm{d}T+(1-x^2)Q_5(x,\rho)\left(\mathrm{d}\phi+\rho\,Q_6(x,\rho)\mathrm{d}T\right)\right]^2
\end{multline}
and
\begin{multline}
A=-\rho\,Q_8(x,\rho)\,\mathrm{d}T+Q_9(x,\rho)(1-x^2)\left(\mathrm{d}\phi+\rho\,Q_6(x,\rho)\,\mathrm{d}T\right)
\\
+Q_{10}(x,\rho)x^2\left[\mathrm{d}\psi+\rho\,Q_7(x,\rho)\,\mathrm{d}T+(1-x^2)Q_5(x,\rho)\left(\mathrm{d}\phi+\rho\,Q_6(x,\rho)\,\mathrm{d}T\right)\right]\,,
\end{multline}
\label{eqs:U1U1g}%
\end{subequations}%
resulting in a total of ten functions $Q_I$ of $\rho$ and $x$ to solve for. 
We  compute the scaling dimensions by considering linearized deformations of the near-horizon geometries proportional to $\rho^\gamma$. A linearized approximation is sufficient since (for $\gamma > 0$) the deformation becomes very small near the horizon. Note that, in effect, we have a double expansion. We know the near-horizon geometries in an expansion in angular momenta, and for each order in that expansion, we compute $\gamma$ by solving the equations for a linear deformation.

To compute the scaling dimensions we set
\begin{align}
&Q_1(x,\rho)=\Gamma(x)\left(1+\epsilon\,\rho^\gamma\,q_1(x)\right)\nonumber
\\
&Q_2(x,\rho)=\sigma_1(x)\sigma_2(x)\left(1+\epsilon\,\rho^\gamma\,q_2(x)\right)\nonumber
\\
&Q_3(x,\rho)=\frac{\sigma_1(x)}{B_2}\left(1+\epsilon\,\rho^\gamma\,q_3(x)\right)\nonumber
\\
&Q_4(x,\rho)=\frac{\sigma_2(x)}{B_1}\left(1+\epsilon\,\rho^\gamma\,q_4(x)\right)\nonumber
\\
&Q_5(x,\rho)=\alpha(x)\left(1+\epsilon\,\rho^\gamma\,q_5(x)\right)
\\
&Q_6(x,\rho)=\Omega_{\phi}\left(1+\epsilon\,\rho^\gamma\,q_6(x)\right)\nonumber
\\
&Q_7(x,\rho)=\Omega_{\psi}\left(1+\epsilon\,\rho^\gamma\,q_7(x)\right)\nonumber
\\
&Q_8(x,\rho)=Q_{\rm NH}\left(1+\epsilon\,\rho^\gamma\,q_8(x)\right)\nonumber
\\
&Q_9(x,\rho)=\beta(x)\left(1+\epsilon\,\rho^\gamma\,q_9(x)\right)\nonumber
\\
&Q_{10}(x,\rho)=\delta(x)\left(1+\epsilon\,\rho^\gamma\,q_{10}(x)\right)\,,\nonumber
\end{align}
with $\epsilon$ a bookkeeping parameter that we take to be infinitesimally small. (Note that this is different from the expansion parameter $\varepsilon$ we used to add angular momenta.) When $\epsilon=0$ we get Eq.~(\ref{eqs:U1U1}), showing that we are perturbing the appropriate near-horizon geometry.  For any $\gamma\neq0,1$ the Einstein equation implies that
\begin{equation}
q_2(x)=q_3(x)+q_4(x)\,.
\end{equation}
For exactly $\gamma=0,1$ we still impose the above, but as a gauge condition instead, which can be achieved via a residual gauge symmetry similar to the one found in \cite{Horowitz:2024dch}. The resulting perturbed Einstein and Maxwell equations yield first order differential equations for $q_1$ and $q_3$ and seven second order equations for the remaning variables. These are the equations that one wishes to solve in perturbation theory about $\varepsilon\ll1$, i.e. small angular momenta. Note that $\Gamma$, $\sigma_I$, $\alpha$, $\beta$, $\delta$ and $A_I$ admit themselves an expansion in terms of $\varepsilon$ (see Eqs.~(\ref{eqs:expanq})), and we would like to determine $\gamma$ using said expansion.

Let $q_i^{(0)}(x)$ and $\gamma^{(0)}$ denote a solution to these equations with $\varepsilon=0$, that is before we add angular momenta. So $\gamma^{(0)}$ is a scaling dimension of the static $U(1)\times U(1)$ symmetric near-horizon geometry found in \cite{Kunduri:2009ud}.  With $\varepsilon=0$, the equations decouple into three groups:
\begin{equation}
{\rm G}_{1}\equiv \left\{q_1^{(0)},q_3^{(0)},q_4^{(0)},q_8^{(0)}\right\}\,,\quad {\rm G}_{2}\equiv \left\{q_5^{(0)}\right\}\,,\quad \text{and}\quad {\rm G}_3\equiv \left\{q_6^{(0)},q_7^{(0)},q_9^{(0)},q_{10}^{(0)}\right\}\, .\label{eq:groups}
\end{equation} 
We first study the modes with the lowest scaling dimension since they will dominate near the horizon.   This turns out to be $\gamma^{(0)}=0$.  Using this fact and 
imposing regularity at $x=0$ and $x=1$ we find\footnote{Throughout these intermediate calculations we are assuming that $2\lambda$ is not an integer. However, our final result can be analytically continued to these semi-integer values.}
\begin{gather}
 q^{(0)}_1(x)=q^{(0)}_2(x)=\left[\frac{2 c_1 c_2}{\sigma (x)}-c_1-c_2\right]a_0\,,\quad q^{(0)}_3(x)=\left[\frac{c_1 c_2}{\sigma (x)}-2c_1+c_2\right]a_0\,,\nonumber
\\
 q^{(0)}_4(x)=\left[\frac{c_1 c_2}{\sigma (x)}-2c_2+c_1\right]a_0\,,\quad q^{(0)}_5(x)=a_2+a_3+\left[\frac{c_1 c_2}{\sigma (x)}-2c_1+c_2\right]a_0\,,\nonumber
\\
 q^{(0)}_6(x)=a_2\,,\quad q^{(0)}_7(x)=a_3\,,\quad q^{(0)}_8(x)=a_1+(1-2x^2)(c_1-c_2)\frac{a_0}{2}\,,
\\
 q^{(0)}_9(x)=\frac{1}{1+2\lambda}\left\{a_2+2a_3 \lambda-\left[\frac{c_1 c_2}{\sigma(x)}-2c_1+(1-2\lambda)c_2\right]a_0\right\}\,,\nonumber
\\
 q^{(0)}_{10}(x)=\frac{1}{1+2\lambda}\left\{a_3+2a_2 \lambda-\left[\frac{c_1 c_2}{\sigma(x)}-2c_2+(1-2\lambda)c_1\right]a_0\right\}\,.\nonumber
\end{gather}
The solution depends on four undetermined parameters,  $a_0$, $a_1$, $a_2$, and $a_3$, so there is a 4-fold degeneracy of modes with $\gamma^{(0)}=0$.  To see how $\gamma$ changes when we add angular momenta, we  use 
 degenerate perturbation theory in $\varepsilon$.

We set the perturbation theory in $\varepsilon$ as follows:
\begin{equation}
q_{I}(x)=\sum_{i=0}^{+\infty}\varepsilon^i q^{(i)}_I(x)\quad\text{with}\quad I=1,\ldots,10\quad\text{and}\quad \gamma=\sum_{i=1}^{+\infty}\varepsilon^i \gamma^{(i)}\,,
\end{equation}
with all the $q_I^{(0)}(x)$ given above. To order $\varepsilon$, we find no constraints on $a_0$, $a_1$, $a_2$ and $a_3$, but we find that $\gamma^{(1)}=0$. This implies that $\gamma$ does not change sign when we change the sign of $J_i$.

The equations at quadratic order become more complicated to solve, but nevertheless can be integrated in full generality. In particular, we can still decouple the  $q_I^{(2)}$ according to the same groups as in Eq.~(\ref{eq:groups}), with the upperscript ${}^{(0)}$ replaced by ${}^{(2)}$. After some lengthy algebra we find that the smallest scaling dimension becomes
\begin{equation}
\gamma=2 \sqrt{3} \pi  \frac{1-\lambda ^2}{4-\lambda ^2}\frac{J_\phi J_\psi}{ Q^{3}}+\mathcal{O}(\varepsilon^3)\,.
\label{eq:anasmalljs}
\end{equation}
and that the relation between $a_0$, $a_1$, $a_2$, and $a_3$ becomes  determined, in accordance to general expectations from degenerate perturbation theory.  We have assumed that the angular momenta and charge are positive. If not, the $J_i$'s and $Q$ should be replaced by their absolute value, so $\gamma$ is positive for $|\lambda| < 1$. The fact that $\gamma$ becomes negative for $|\lambda| > 1$ strongly suggests that the near-horizon geometries will cease to smoothly connect to an asymptotically flat end when $|\lambda| > 1$.

We now discuss some of the modes of the zeroth order static geometries with larger scaling dimensions. Although they are subdominant at small angular momenta, we will see that without the Chern-Simons term ($\lambda = 0$), there is ``level crossing" as one increases $J_i$, and these modes can have smaller scaling dimension than the one starting at $\gamma^{(0)}=0$.
Recall that when $\varepsilon =0$, the linearized equations decouple into three groups \eqref{eq:groups}.
The corresponding scaling dimensions for $G_1$ and $G_2$ can be determined analytically for any values of the squashing parameters $c_1$ and $c_2$ in Eq.~\eqref{eq:cidef}, and they are found to be integers. The details of this calculation can be found in appendix \ref{app:U1U1}. However, for $G_3$, we were only able to determine the scaling dimensions analytically in the special case $c_1 = c_2$, which corresponds to the near-horizon geometry of an extremal 5D Reissner-Nordstr\"om black hole. For $c_1\neq c_2$ we will proceed numerically.

When $c_1 = c_2$, we can exploit the background spherical symmetry to decompose the functions in $G_3$ using vector spherical harmonics on $S^3$. This reduces the task of determining the scaling dimensions to solving an algebraic problem. Within our symmetry class, vector spherical harmonics on $S^3$ are simply given by
\begin{equation}
S^{\phi}_{\hat{a}}{\rm d}x^{\hat{a}}=\frac{S(x)}{1-x^2} \mathrm{d}\phi\quad\text{and}\quad S^{\psi}_{\hat{a}}{\rm d}x^{\hat{a}}=\frac{S^\prime(x)}{x} \mathrm{d}\psi
\end{equation}
where $\hat{a}=\{x,\phi,\psi\}$ and 
\begin{equation}
S(x)={}_2F_1\left(-p-1,p+1;1;x^2\right)
\end{equation}
with $p=0,1,2\ldots$ and ${}_2F_1(a,b;c;z)$ the standard Gauss hypergeometric function.  Using the above, we set
\begin{equation}
q_6^{(0)}(x)=\frac{S(x)}{1-x^2} a_0\,,\quad q_7^{(0)}(x)=\frac{S^\prime(x)}{x} b_0\,,\quad q_9^{(0)}(x)=\frac{S(x)}{1-x^2} c_0\quad\text{and}\quad q_{10}^{(0)}(x)=\frac{S^\prime(x)}{x} d_0
\end{equation}
where $a_0$, $b_0$, $c_0$ and $d_0$ are constants to be determined. Plugging in the above into the ${\rm G}_3$ equations with $c_1=c_2$, yields an eigenvalue equation for $\gamma$, with the $a_0$, $b_0$, $c_0$ and $d_0$ regarded as the components of an eigenvector. Once the dust settles, we find four families of modes. Modes with $p=0$ are special, so we discuss these first.

When $p=0$ the four families of modes are given by
\begin{equation}
\gamma^{(0)}=0\,,\quad \gamma^{(0)}_{\pm}=\frac{1}{2}\left(\sqrt{17\pm8\lambda}-1\right)
\end{equation}
where the mode with $\gamma^{(0)}=0$ has degeneracy two\footnote{We showed above that $\gamma^{(0)}=0$ has four-fold degeneracy, but here we are only considering the equations in group $G_3$ which correspond mainly to perturbations of the Maxwell field.},  i.e. one can express $c_0$ and $d_0$ as a function of \emph{both} $a_0$ and $b_0$.  This special mode persists even when $c_1\neq c_2$ and was discussed above. Note that one can map $\gamma_+$ into $\gamma_-$ by flipping the sign of $\lambda$, so we take $\lambda>0$ without loss of generality. $\gamma_+$ is positive and real for all values of $\lambda>0$, however, $\gamma_-$ exhibits a more interesting behaviour. Namely, it becomes negative for $\lambda>2$, and complex for $\lambda>17/8$.  A negative $\gamma$ signals a breakdown of perturbation theory and indicates that the full asymptotically flat solution does not approach the near-horizon geometry. A complex $\gamma$ means that the mode has an effective AdS$_2$ mass below the 2D Breitenlohner-Friedman bound. This implies that there is a  dynamical instability of the near extremal asymptotically flat black hole when $\lambda>17/8$ \cite{Durkee:2010ea}.

For $p\geq1$ we find the following four families of modes (dropping the $0$ superscript)
\begin{subequations}
\begin{equation}
\gamma^{+}_{\pm}=\sqrt{\frac{1}{4}+[2+\lambda  (p+1)]+p (p+2)\pm\sqrt{[2+\lambda  (p+1)]^2+3 p (p+2)}}-\frac{1}{2}\,.
\end{equation}
and
\begin{equation}
\gamma^{-}_{\pm}=\sqrt{\frac{1}{4}+[2-\lambda  (p+1)]+p (p+2)\pm\sqrt{[2-\lambda  (p+1)]^2+3 p (p+2)}}-\frac{1}{2}\,.
\end{equation}
\end{subequations}
Since we can map the $\gamma^{-}_{\pm}$ modes to the $\gamma^{+}_{\pm}$ modes by flipping the sign of $\lambda$, we will again take $\lambda>0$ without loss of generality. The $\gamma^+_{\pm}$, $\gamma^-_+$ modes are always real and positive, but $\gamma^-_-$ becomes negative when $\lambda>(1+p)/2$. In fact, for $p=1$, this happens precisely when $\lambda>1$, suggesting that the region $\lambda>1$ does not admit a smooth zero temperature limit if modes with $p\geq1$ are excited in the near-horizon region. What is more, if we increase $\lambda$ further, we find that $\gamma^-_-$ becomes complex for
\begin{equation}
\lambda>\frac{16 p^4+64 p^3+88 p^2+48 p+17}{32 p^3+96 p^2+72 p+8}
\end{equation}
signaling a dynamical instability of the full extremal geometry for any value of $p$.

To calculate these scaling dimensions for the nonspherical near-horizon geometries,  we proceed numerically. All physical results depend only on the ratio of the squashing parameters $c_2/c_1\equiv \alpha$. Furthermore, since sending $\alpha$ to $1/\alpha$ merely exchanges $x=0$ with $x=1$, which is an isometry that relabels $\phi$ as $\psi$ and vice versa, we can restrict $\alpha$ to the range $0 < \alpha < 1$.

To proceed we define
\begin{equation}
q_{6}^{(0)}(x)=\gamma\,r_1(x)\,,\quad q_{7}^{(0)}(x)=\gamma\,r_2(x)\,,\quad q_9^{(0)}(x)=\sqrt{c_1 c_2}\,r_3(x)\quad\text{and}\quad q_{10}^{(0)}(x)=\sqrt{c_1 c_2}\,r_4(x)\,
\end{equation}
and solve for $\{r_i,\gamma(\gamma+1)\}$ numerically as a function of $\alpha$ and $\lambda$. Note that since we are solving for $\gamma(\gamma+1)$, $\gamma$ will become complex if $\gamma(\gamma+1)<-1/4$. When $\alpha=1$, we recover the analytic results discussed above. At $x=1$ we impose regularity, which yields Robin-type boundary conditions for all eigenfunctions $r_i$, and similarly at $x=0$. The resulting second order equations are solved numerically using a spectral collocation grid in the $x$ direction, on Gauss-Lobatto nodes. 

\begin{figure}
\centering
\includegraphics[width=0.78\textwidth]{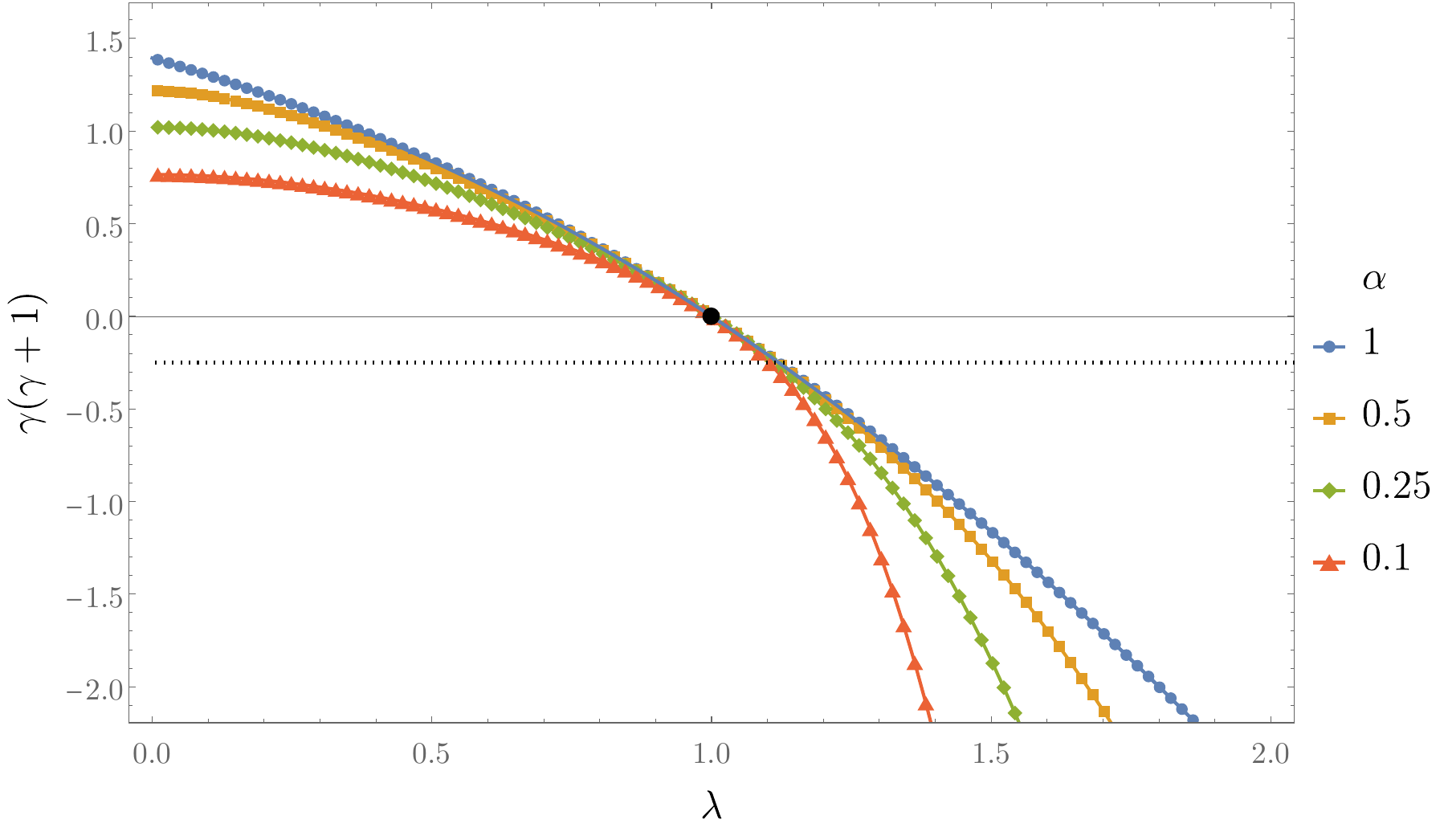}
\caption{$\gamma(\gamma+1)$ as a function of the Chern-Simons coefficient $\lambda$, for several fixed values of the squashing parameter $\alpha \equiv c_2/c_1$ \eqref{eq:cidef}, indicated on the right. Note that $\gamma < 0$ for $\lambda > 1$ indicating that the extremal black hole has a singular horizon with diverging curvature scalars. The horizontal black dashed line reprensents $\gamma(\gamma+1) = -1/4$, and any points below this line will necessarily have complex values of $\gamma$ indicating an instability of the near extremal black hole.}
\label{fig:u1u1}
\end{figure}

In Fig.~\ref{fig:u1u1}, we plot $\gamma(\gamma+1)$ as a function of $\lambda$ for several fixed values of $\alpha$, indicated on the right. We focused on extending the modes $\gamma^-_-$ with $p=1$ when $c_1 = c_2$ (blue disks in Fig.~\ref{fig:u1u1}), as these impose the strongest constraints on the physical range of $\lambda$. The horizontal black dashed line represents $\gamma(\gamma+1) = -1/4$, and any points below this line correspond to complex values of $\gamma$. We see that this occurs for all $\alpha$ as we increase $\lambda$ indicating that near extremal black holes become unstable \cite{Durkee:2010ea}.\footnote{The fact that black holes can be unstable when $\lambda>1$ was discussed in \cite{Kunz:2005ei}.}  For $\lambda>1$ all these scaling dimensions become negative, for any value of $\alpha$, signaling that if modes with $p=1$ are turned on in the near-horizon region, perturbation theory breaks down.  We will show in section 5 that these modes are sourced just by adding an asymptotically flat region, with the result that curvature scalars now diverge on the horizon in the extremal limit.

\subsection{Perturbative in the charge\label{sec:pcharge}}
In this section, we compute the scaling dimensions perturbatively in the charge. It is helpful to make a slight gauge adjustment to match the known form of the near-horizon geometry of an extremal Myers-Perry black hole. We now take the following metric and gauge field ansatz
\begin{subequations}
\begin{multline}
{\rm d}s^2=Z_1(x,\rho)\Bigg\{-\rho^2 \mathrm{d}T^2+\frac{\mathrm{d}\rho^2}{\rho^2}+\frac{Z_7(x,\rho){\rm d}x^2}{1-x^2}+Z_{2}(x,\rho)(1-x^2)\left(\mathrm{d}\phi+\rho\,Z_{5}(x,\rho)\,\mathrm{d}T\right)^2
\\
+Z_{4}(x,\rho)x^2\left[\mathrm{d}\psi+\rho\,Z_6(x,\rho)\,\mathrm{d}T+(1-x^2)Z_{3}(x,\rho)\left(\mathrm{d}\phi+\rho\,Z_5(x,\rho)\mathrm{d}T\right)\right]^2\Bigg\}\,,
\end{multline}
and
\begin{multline}
A=-Z_8(x,\rho){\rm d}T+Z_9(x,\rho)(1-x^2)\left(\mathrm{d}\phi+\rho\,Z_5(x,\rho)\,\mathrm{d}T\right)+
\\
Z_{10}(x,\rho)x^2\left[\mathrm{d}\psi+\rho\,Z_6(x,\rho)\,\mathrm{d}T+(1-x^2)h_{12}(x,\rho)\left(\mathrm{d}\phi+\rho\,Z_5(x,\rho)\mathrm{d}T\right)\right]\,,
\end{multline}
\end{subequations}%
with a total of ten functions $Z_{I}(x,\rho)$ to solve for. Since we are interested in computing the scaling dimensions $\gamma$, we set
\begin{equation}
\begin{aligned}
&Z_1(x,\rho)=\Gamma(x)\left[1+\epsilon\,\rho^\gamma z_1(x)\right]\,,
\\
&Z_2(x,\rho)=h_{11}(x)\left[1+\epsilon\,\rho^\gamma z_2(x)\right]\,,
\\
&Z_3(x,\rho)=h_{12}(x)\left[1+\epsilon\,\rho^\gamma z_3(x)\right]\,,
\\
&Z_4(x,\rho)=h_{22}(x)\left[1+\epsilon\,\rho^\gamma z_4(x)\right]\,,
\\
&Z_5(x,\rho)=\Omega_{\phi}\left[1+\epsilon\,\rho^\gamma z_5(x)\right]\,,
\\
&Z_6(x,\rho)=\Omega_{\psi}\left[1+\epsilon\,\rho^\gamma z_6(x)\right]\,,
\\
&Z_7(x,\rho)=\Gamma_{\rm NH}\left[1+\epsilon\,\rho^\gamma z_7(x)\right]\,,
\\
&Z_8(x,\rho)=Q_{\rm NH}+\epsilon\,\rho^\gamma z_8(x)\,,
\\
&Z_9(x,\rho)=b_1(x)\left[1+\epsilon\,\rho^\gamma z_9(x)\right]\,,
\\
&Z_{10}(x,\rho)=b_1(x)\left[1+\epsilon\,\rho^\gamma z_{10}(x)\right]\,,
\end{aligned}
\end{equation}%
where $\epsilon$ is a bookkeeping parameter assumed to be infinitesimally small. Linearising around $\epsilon=0$, yields a linear set of equations coupling all  ten functions $z_I$ and the corresponding scaling dimension $\gamma$.

For $\gamma\neq0,1$, the Einstein equation implies that
\begin{equation}
z_7=-\frac{3}{2}z_1-\frac{1}{2}z_2-\frac{1}{2}z_4\,,
\end{equation}
while for $\gamma=0,1$ we can impose the above relation as a gauge choice. The remaining equations yield two first order constraints on the $z_I$, and seven second order ordinary differential equations in total. The two first order constraints are then combined to give a second order order differential equation. Once the dust settles, these can be combined into a set of eight second order differential equations for $z_1$, $z_2$, $z_4$, $z_5$, $z_6$, $z_8$, $z_9$ and $z_{10}$ which are all coupled. 

We  solve these equations perturbatively in the black hole charge. 
 Using the same expansion we used in Eqs.~(\ref{eq:expq}), we set
\begin{equation}
z_{I}(x)=\sum_{i=0}^{+\infty}q^i z_{I}^{(i)}(x)\quad \text{and}\quad \gamma=\sum_{i=0}^{+\infty}q^i \gamma^{(i)}\,,
\end{equation}
so that at zeroth order in $q$, we are simply computing the scaling dimensions $\gamma^{(0)}$ of the near-horizon geometry of an extremal Myers-Perry black hole.
These can be computed in full generality, as shown in Appendix \ref{app:MP}, and turn out to be nonnegative integers, consistent with earlier results in the literature \cite{Durkee:2010ea,Murata:2012ct}. 
We are primarily interested in the lowest scaling dimension, which is $\gamma^{(0)}=0$.

Even for $\gamma^{(0)}=0$, the explicit expressions are somewhat cumbersome for several of the functions. However, to facilitate the reproduction of the calculations in this paper, we present a few examples here, highlighting the main features:
\begin{equation}
\begin{aligned}
&z^{(0)}_3(x)=\frac{1}{x^2+4 \Omega _{\phi }^2}\left[\mu_0\,x^2+2\,\mu_1 \,\Omega_{\phi}^2 \left(4-\frac{137+965 \Omega_{\phi }^2-4745 \Omega _{\phi }^4+3003 \Omega_{\phi }^6}{1384-4337 \Omega_{\phi }^2-4061 \Omega_{\phi }^4+13809\Omega_{\phi }^6+3573 \Omega_{\phi }^8}x^2\right)\right]
\\
&z^{(0)}_5(x)=-\mu_1\,,
\\
&z^{(0)}_6(x)=\mu_1\,,
\\
&z^{(0)}_8(x)=0\,,
\\
&z^{(0)}_9(x)=\frac{\nu_0+4\Omega_{\phi}^2\nu_1}{x^2+4\Omega_{\phi}^2}\,,
\\
&z^{(0)}_{10}(x)=\nu_1+\frac{4  \Omega _{\phi }^2\nu_0}{x^2+4\Omega_{\phi}^2 (1-x^2)}\,.
\end{aligned}
\end{equation}
where $\mu_0,\mu_1, \nu_0,\nu_1$ are arbitrary constants.
The remaining functions not shown, i.e., $z^{(0)}_1(x)$, $z^{(0)}_2(x)$, $z^{(0)}_4(x)$, and $z^{(0)}_7(x)$, depend only on $x$, $\mu_0$, and $\mu_1$. The discussion above demonstrates that the near-horizon geometry of an extreme Myers-Perry black hole has a fourfold degeneracy for modes with vanishing scaling dimension. We then proceed to higher orders in $q$, applying degenerate perturbation theory. At first order in $q$, we find $\gamma^{(1)} = 0$, which is consistent with our expectation that the scaling dimensions should not depend on the sign of the electric charge. At first order, we also find that $\nu_0 = -\nu_1$, a constraint that ensures our deformations do not alter the background electric charge. {At this stage, we find ourselves with a three-fold degeneracy which we can choose to parametrise by $\mu_0$, $\mu_1$ and $\nu_1$.}

The equations at second order become increasingly difficult to solve, but can still be solved in full generality.  This is possible because the perturbations decouple into three groups at each $i$-th order in perturbation theory,  and each group can be integrated separately. The groups in this case are:
\begin{equation}
{\rm G}_1=\{z^{(i)}_1,z^{(i)}_2,z^{(i)}_4\}\,,\quad {\rm G}_2=\{z^{(i)}_5,z^{(i)}_6\}\quad \text{and}\quad {\rm G}_3=\{z^{(i)}_8,z^{(i)}_9,z^{(i)}_{10}\}\,.
\end{equation}
While this occurs for modes that smoothly connect to $\gamma^{(0)} = 0$ in the $q \to 0$ limit, in general, the functions in ${\rm G}_1$ and ${\rm G}_2$ mix for modes with $\gamma^{(0)} \neq 0$. At second order we finally determine $\gamma^{(2)}q^2$ after imposing regularity at $x=0$ and $x=1$. The expression appears as a function of $a$, $\Omega_{\phi}$ and $q$, but we can readily rewrite these in terms of $J_{\phi}$, $J_{\psi}$ and $Q$, respectively. {We find two possible solutions. Either $\gamma^{(2)}q^2=0$ and we need to proceed to fourth order to determine $\gamma^{(4)}q^4$ and the remaining degeneracy, or:} 
\begin{equation}
\begin{aligned}
\gamma& = \gamma^{(2)}_+q^2+\mathcal{O}(q^3)
\\
&=\frac{16^{1/3}}{3 \pi ^{2/3}}\frac{1}{J_{\phi } J_{\psi }}\frac{J_+^{2/3}}{J_-^2}\left[J_{\phi }^2-6 J_{\phi } J_{\psi }+J_{\psi}^2+2 J_{\phi } J_{\psi }\frac{J_+}{J_-}\log \left(\frac{J_{\phi }}{J_{\psi }}\right)\right] \left(1-\lambda ^2\right)Q^2+\mathcal{O}(Q^3)\,,
\end{aligned}
\label{eq:anasmallq}
\end{equation}
with $J_{\pm}\equiv J_{\phi }\pm J_{\psi }$. {If we fix $J_{\psi}=\beta J_{\phi}$ with $\beta>0$, the above expression predicts $\gamma\sim \left(|Q|^{3/2}/J_{\phi}\right)^{4/3}$ for large values of $J_{\phi}/|Q|^{3/2}$.}
This shows that the addition of charge indeed lifts $\gamma$ to a noninteger value causing the horizon to become singular. Notice that
 at least for small values of $|Q|^{3/2}/J_{\phi}$, where we expect perturbation theory to be valid, $\gamma < 0$ for $\lambda > 1$. This indicates that we do not expect these particular near-horizon geometries to smoothly connect to an asymptotically flat region.

 Even though there appears to be a divergence in \eqref{eq:anasmallq} when $J_{\phi}=J_{\psi}$, this is not the case. 
 Indeed, in the limit $J_{\phi}\to J_{\psi}$ the above expression collapses to
\begin{equation}
\gamma=\frac{16}{9 \pi ^{2/3}} \left(1-\lambda ^2\right)\frac{Q^2}{J_{\psi}^{4/3}}+\mathcal{O}(Q^3)\,.
\end{equation}
\subsection{Fully nonlinear results}
In this section, we avoid approximations and solve for the scaling dimensions $\gamma$ numerically. Since we want to use the backgrounds constructed in section \ref{subsec:fullback}, we again use the De Turck trick. Our metric and gauge field ansatz read
\begin{subequations}
\begin{multline}
{\rm d}s^2=-\rho^2 Q_1(x,\rho){\rm d}T^2+\frac{Q_8(x,\rho)\left({\rm d}\rho+\rho\,Q_{9}(x,\rho){\rm d}x\right)^2}{\rho^2}
\\
+4\Bigg\{\frac{Q_2(x,\rho){\rm d}x}{1-x^2}+(1-x^2)Q_3(x,\rho)\left({\rm d}\phi+\omega_{\phi}\,\rho\,Q_{10}(x,\rho){\rm d}T\right)^2
\\
+x^2 Q_{4}(x,\rho)\left[{\rm d}\psi+\omega_{\psi}\,\rho\,Q_{11}(x,\rho){\rm d}T+(1-x^2)Q_5(x,\rho)\left({\rm d}\phi+\omega_{\phi}\,\rho\,Q_{10}(x,\rho){\rm d}T\right)\right]^2\Bigg\}\,,
\end{multline}
\begin{multline}
A=-Q_{\rm NH}\,\rho\,Q_{12}(x,\rho)+(1-x^2)Q_6(x,\rho)\left({\rm d}\phi+\omega_{\phi}\,\rho\,Q_{10}(x,\rho){\rm d}T\right)+
\\
x^2Q_{7}(x,\rho)\left[{\rm d}\psi+\omega_{\psi}\,\rho\,Q_{11}(x,\rho){\rm d}T+(1-x^2)Q_5(x,\rho)\left({\rm d}\phi+\omega_{\phi}\,\rho\,Q_{10}(x,\rho){\rm d}T\right)\right]\,,
\end{multline}
\end{subequations}%
with a total of twelve functions $Q_I(x,\rho)$ to solve for. For our reference metric, we take
\begin{equation}
\begin{aligned}
&Q_1(x,\rho)=Q_2(x,\rho)=Q_3(x,\rho)=Q_4(x,\rho)=Q_{8}(x,\rho)=Q_{10}(x,\rho)=Q_{11}(x,\rho)=1\,,
\\
&Q_5(x,\rho)=Q_9(x,\rho)=0\,,
\end{aligned}
\end{equation}
which was precisely the reference metric used in section \ref{subsec:fullback}. To proceed, we define
\begin{equation}
\begin{aligned}
&Q_I(x,\rho)=Q_I(x)\left[1+\epsilon\,\rho^\gamma\,q_I(x)\right]\,,\quad \text{for}\quad I=1,\ldots,7
\\
&Q_8(x,\rho)=Q_1(x)\left[1+\epsilon\,\rho^\gamma\,q_8(x)\right]\,,
\\
&Q_9(x,\rho)=\epsilon\,\rho^\gamma\,q_9(x)\,,
\\
&Q_I(x,\rho)=1+\epsilon\,\rho^\gamma\,q_I(x)\quad \text{for}\quad I=10,11,12\,.
\end{aligned}
\end{equation}
As expected, the dependence on $\rho$ cancels out, leaving us with twelve ordinary differential equations of the St\"urm-Liouville type for the $q_I(x)$, with $\gamma$ appearing as an eigenvalue.

The boundary conditions at $x = 0$ and $x = 1$ are determined by regularity. At $x = 0$, Neumann boundary conditions apply to all variables except for $q_9$, which satisfies a Dirichlet condition, $q_9(0) = 0$. Additionally, there is a Dirichlet condition that imposes $q_2(0) = q_4(0)$. For $x = 1$, we apply Robin-type boundary conditions to all variables, along with a Dirichlet condition $q_2(1) = q_3(1)$.

The resulting St\"urm-Liouville problem is solved using spectral collocation methods, discretized on a Gauss-Lobatto grid with $N$ collocation points. As expected, we observe exponential convergence as $N$ increases, which is characteristic of spectral collocation methods. Additionally, we monitor $\xi_a \xi^a$ and find that it approaches zero in the continuum limit, i.e., as $N$ increases, with exponential speed. All the plots presented in this section were generated with $N=200$.

\begin{figure}
\centering
\includegraphics[width=0.78\textwidth]{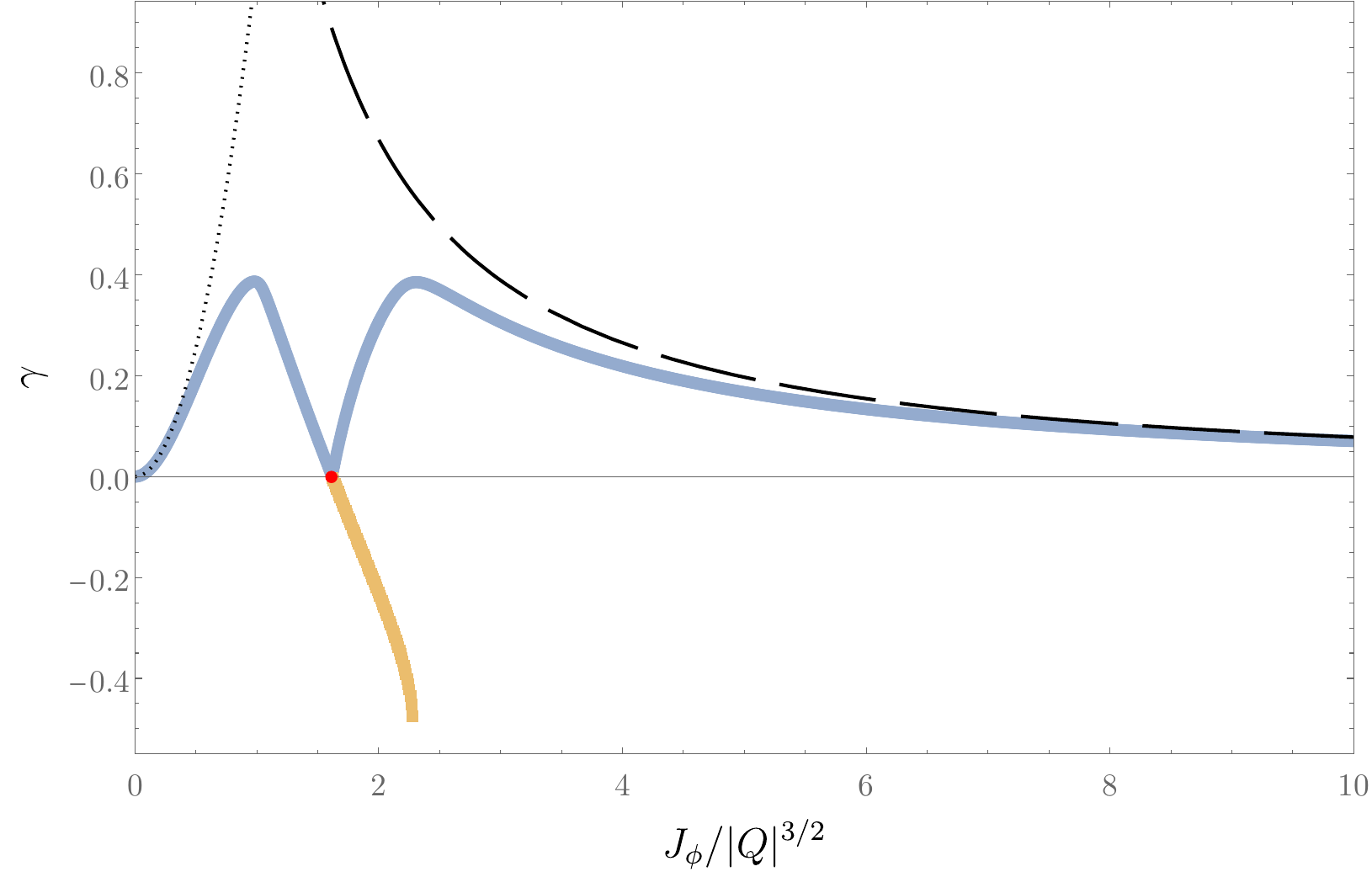}
\caption{The scaling dimension of a mode on the general near-horizon geometry,
as a function of the dimensionless angular momentum $J_{\phi}/|Q|^{3/2}$. The same color coding as in Fig.~\ref{fig:entropyla0} is used. The perturbative calculation, valid at small angular momenta and given in Eq.~(\ref{eq:anasmalljs}) (black dotted line), as well as the perturbative prediction valid at small charge, provided in Eq.~(\ref{eq:anasmallq}) (black dashed line), are shown. This figure was generated for fixed values of $\lambda=0$ and $J_{\psi}=3J_{\phi}/8$.}
\label{fig:ananum}
\end{figure}

In Fig.~\ref{fig:ananum} we plot the scaling dimension amenable to the perturbative analysis discussed in Sections \ref{sec:smalljs} and \ref{sec:pcharge} as a function of the normalised angular momentum $J_{\phi}/|Q|^{3/2}$. This figure was generated for fixed values of $\lambda=0$ and $J_{\psi}=3J_{\phi}/8$. We also plot the perturbative calculation valid at small angular momenta given in Eq.~(\ref{eq:anasmalljs}) (black dotted line) and the perturbative prediction valid at small charge and given in Eq.~(\ref{eq:anasmallq}) (black dashed line). In both cases, the agreement is excellent. The same color coding as in Fig.~\ref{fig:entropyla0} is used. The family of near-horizon geometries represented in orange in Fig.~\ref{fig:entropyla0} is observed to have negative scaling dimensions, supporting the earlier claim that it cannot be connected to an asymptotically flat end. The family of near-horizon geometries represented in blue in Fig.~\ref{fig:entropyla0} always has small \emph{positive} scaling dimensions,  with $0<\gamma<1/2$. Although we have chosen $J_{\psi}/J_{\phi} = 3/8$, we have observed this to be the case for a large range of values of $J_{\psi}/J_{\phi}$.

\begin{figure}
\centering
\includegraphics[width=0.78\textwidth]{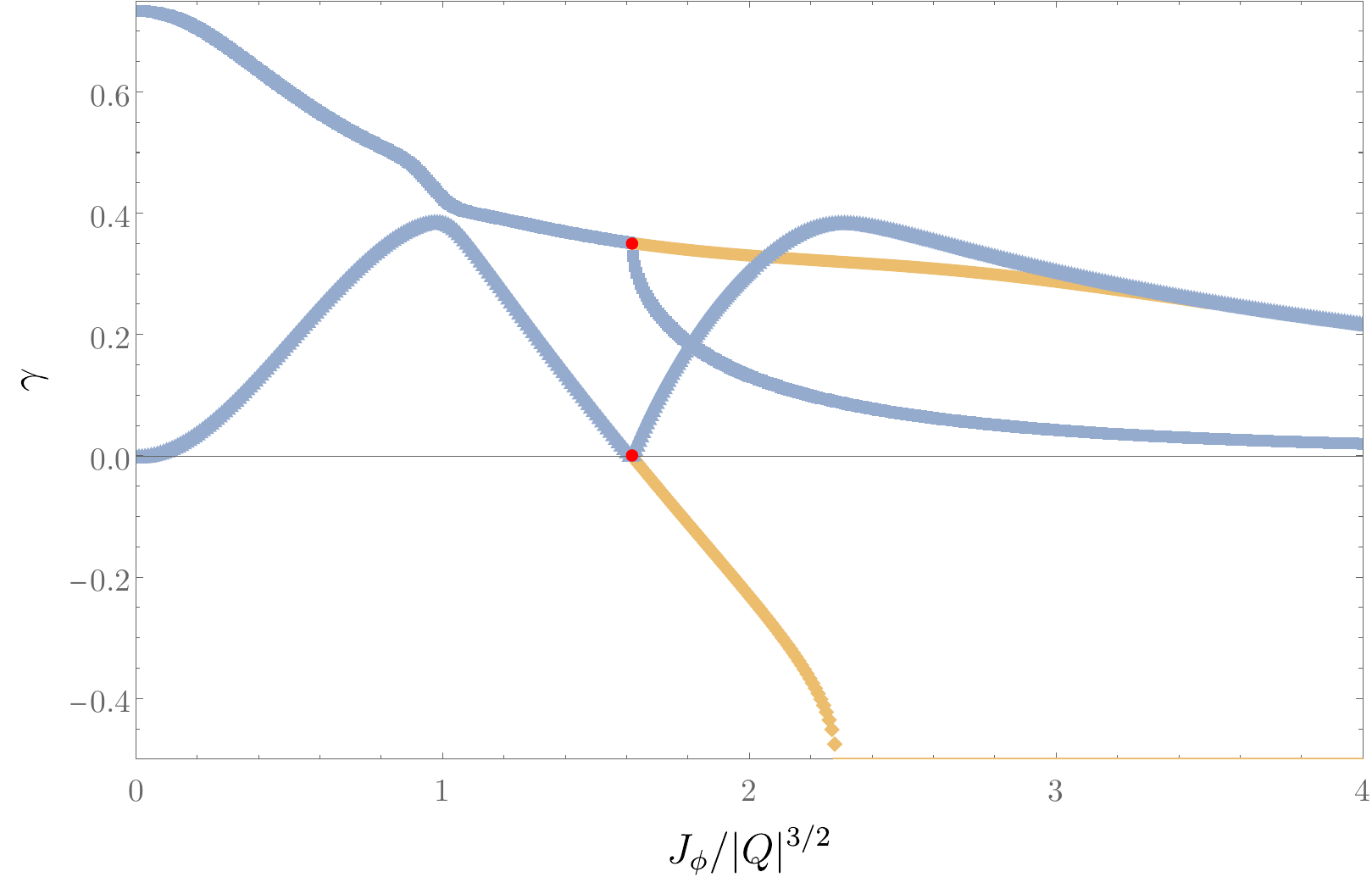}
\caption{The two lowest scaling dimensions $\gamma$ that we found numerically as a function of the dimensionless angular momentum $J_{\phi}/|Q|^{3/2}$, for fixed values of $\lambda=0$ and $J_{\psi}=3J_{\phi}/8$. Notice that the mode with smallest $\gamma$ changes as we increase $J_{\phi}/|Q|^{3/2}$.}
\label{fig:gamma_la_0}
\end{figure}

In Fig.~\ref{fig:gamma_la_0} we show the two smallest values of the scaling dimension $\gamma$ as a function of the normalised angular momentum $J_{\phi}/|Q|^{3/2}$ using the same choice of parameters as in Fig.~\ref{fig:ananum}. When $J_{\phi}=0$, we see two modes that are of interest. One of the modes starts at $\gamma = 0$ and was identified in Section \ref{sec:pcharge}. The second mode, with $\gamma \approx 0.733667(4)$ when $J_{\phi}=0$, belongs to the ${\rm G}_3$-type eigenvalues identified in the same section, calculated for  the squashed near-horizon geometry with $c_2/c_1 = \sqrt{3/8}$. This latter identification follows from Eq.~(\ref{eq:matchu1u1}). This figure also shows that for sufficiently large values of $J_{\phi}/|Q|^{3/2}$ the mode that dominates (with lowest scaling dimension) is not the one that dominates for small values of $J_{\phi}/|Q|^{3/2}$. {Indeed, we can read off the scaling of $\gamma$ at large $J_{\phi}/|Q|^{3/2}$ and, at finite $J_{\phi}/|Q|^{3/2}$, it appears to be consistent with $\left(|Q|^{3/2}/J_{\phi}\right)^{8/3}$, suggesting that this term can only be found by performing perturbation theory up to order $q^4$ about the near-horizon geometry of an extremal Myers-Perry black hole. This is consistent with the analysis of section \ref{sec:pcharge}.}

We now briefly discuss the $\lambda\neq0$ case. For $\lambda=1$, all the scaling dimensions become integers. This is to be expected, since a known solution exists and is analytic in $\rho$ when changed to the relevant gauge (see Eq.~(\ref{eqs:crazyNH}) for the corresponding near-horizon geometry). For $\lambda>1$ we always find negative scaling dimensions for at least one mode. These connect to the modes generated by ${\rm G}_3$ discussed in section \ref{sec:smalljs}. For $0<\lambda<1$ more interesting dynamics takes place. In particular, the orange and green families displayed in Fig.~\ref{fig:entropyla0.01} both have negative or complex scaling dimensions, showing that these are not suitable to connect to an asymptotically flat end.

\begin{figure}
\centering
\includegraphics[width=0.78\textwidth]{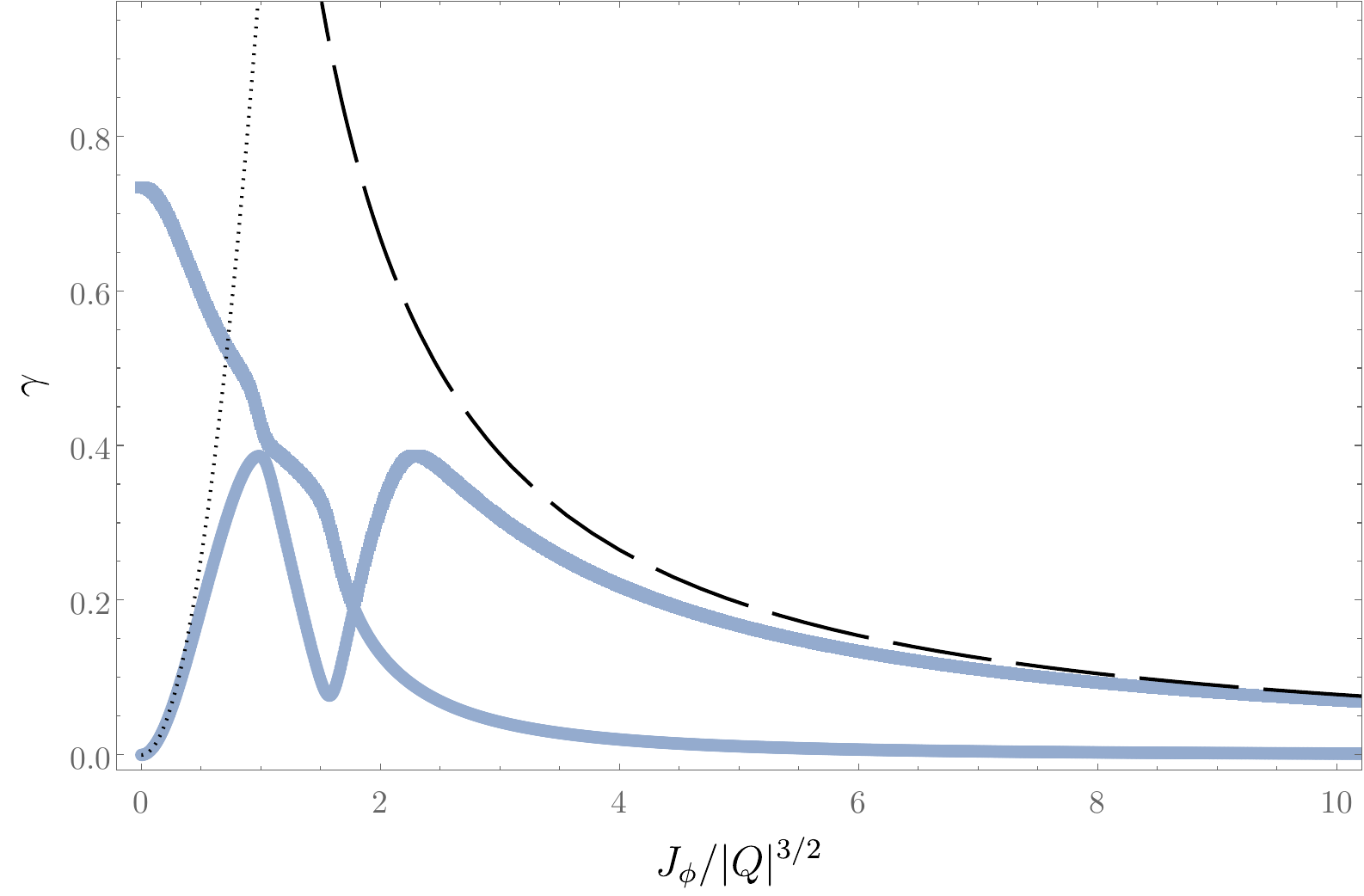}
\caption{The two lowest scaling dimensions $\gamma$ that we found numerically as a function of dimensionless angular momentum $J_{\phi}/|Q|^{3/2}$ for fixed values of $\lambda=10^{-2}$ and $J_{\psi}=3J_{\phi}/8$.}
\label{fig:gamma_la_not_0}
\end{figure}

Perhaps more interestingly, when $\lambda \neq 0$, the non-analytic behaviour observed in Fig.~\ref{fig:ananum} and Fig.~\ref{fig:gamma_la_0}  is smoothed out. This is most clearly seen in Fig.~\ref{fig:gamma_la_not_0}, where we plot the two lowest-lying scaling dimensions as a function of the normalised angular momentum $J_{\phi}/|Q|^{3/2}$, for fixed $\lambda = 10^{-2}$ and $J_{\psi} = 3J_{\phi}/8$. The colour coding is similar to that used in Fig.~\ref{fig:ananum}; in particular, the perturbative calculation valid at small angular momenta (Eq.~(\ref{eq:anasmalljs})) is shown by the black dotted line, while the perturbative prediction valid at small charge (Eq.~(\ref{eq:anasmallq})) is represented by the black dashed line.

\begin{figure}
\centering
\includegraphics[width=0.78\textwidth]{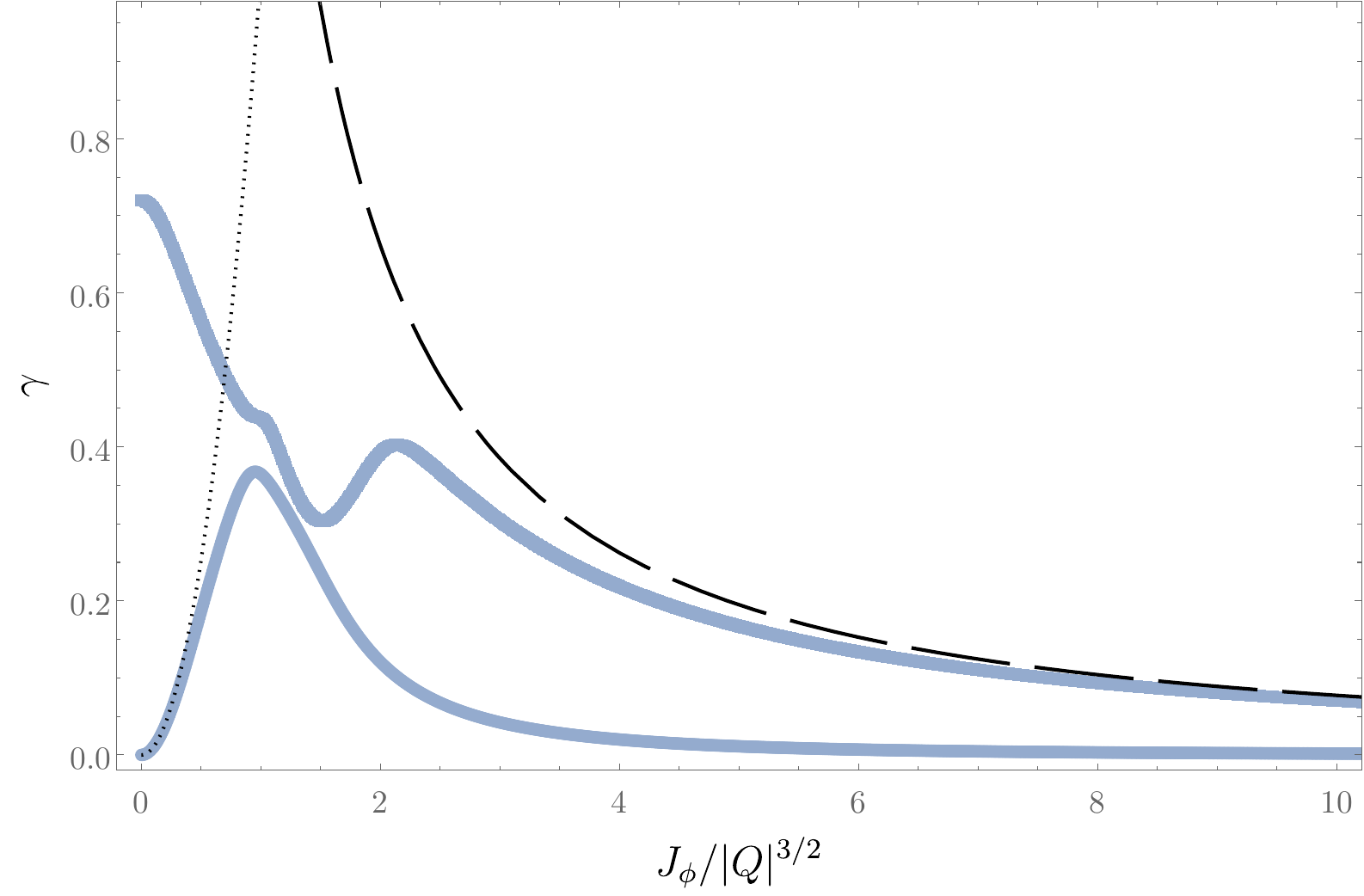}
\caption{The two lowest scaling dimensions $\gamma$ that we found numerically as a function of dimensionless angular momentum $J_{\phi}/|Q|^{3/2}$ for fixed values of $\lambda=0.1$ and $J_{\psi}=3J_{\phi}/8$. It is now clear that while the perturbative expressions (\ref{eq:anasmalljs}) and (\ref{eq:anasmallq}) are accurate approximations to the exact result in their respective domains, they apply to different modes. }
\label{fig:la_0.1}
\end{figure}

Another interesting phenomenon that occurs at finite $\lambda$ is level repulsion\footnote{The phenomenon of eigenvalue repulsion in the context of black holes was first observed while studying the quasinormal modes of Kerr-Newman black holes in \cite{Carullo:2021oxn,Dias:2021yju,Davey:2022vyx,Dias:2022oqm,Davey:2023fin,Davey:2024xvd}.}. Specifically, the two branches of scaling dimensions that cross in Fig.~\ref{fig:gamma_la_0} around $J_{\phi}/|Q|^{3/2}\approx1.80273(5)$ split apart and open a gap for any $\lambda\neq1$. This is difficult to visualize in Fig.~\ref{fig:gamma_la_not_0} since $\lambda=10^{-2}$. However, in Fig.~\ref{fig:la_0.1}, we plot the two lowest scaling dimensions for fixed $\lambda=0.1$ and $J_{\psi}=3J_{\phi}/8$, where this phenomenon can be clearly observed. So if the Chern-Simons term is included, the mode with lowest scaling dimension does not change as we increase the angular momentum.
\section{Near extremal black holes}
We now shift focus slightly and proceed to numerically construct the domain of outer communications of near-extremal, electrically charged and rotating black hole solutions in Einstein-Maxwell-Chern-Simons theory. The aim of this section is to demonstrate that modes with non-integer scaling dimensions are generically present in the near-horizon region of rotating near-extremal black holes in this theory for any $\lambda \neq 1$. This includes the Einstein-Maxwell case (where $\lambda = 0$) and likely explains why electrically charged and rotating black holes in Einstein-Maxwell theory are not known in closed form. These non-integer scaling dimensions are responsible for both divergent tidal forces measured by the Weyl tensor and divergent electric fields.

\subsection{Numerical method}

To construct these solutions we use the same strategy as in \cite{Horowitz:2024dch}, but adapted to five-dimensions. The idea is to use conformal coordinates, in which case the line element and gauge field ansatz reads
\begin{subequations}
\begin{multline}
{\rm d}s^2=-y^2 Q_1(x,y)\Delta(y){\rm d}t^2+\frac{Q_2(x,y)}{1-y^2}\left[\frac{{\rm d}y^2}{(1-y^2)\Delta(y)}+\frac{4{\rm d}x^2}{2-x^2}\right]
\\
+\frac{1}{1-y^2}\left[(1-x^2)^2Q_3(x,y)E_{\phi}^2+x^2(2-x^2)Q_5(x,y)E_{\psi}^2\right]
\end{multline}
and
\begin{equation}
A=Q_8(x,y)(1-y^2){\rm d}t+Q_9(x,y)(1-x^2)^2 E_{\phi}+Q_{10}(x,y)x^2(2-x^2)E_{\psi}\,,
\end{equation}
with
\begin{equation}
\begin{aligned}
&E_{\phi}\equiv {\rm d}\phi-(1-y^2)^2Q_4(x,y){\rm d}t\,,
\\
&E_{\psi}={\rm d}\psi-(1-y^2)^2Q_6(x,y){\rm d}t+(1-x^2)^2Q_7(x,y)E_{\phi}\,,
\\
& \Delta(y)=1-\frac{\tilde{Q}^2}{3}(1-y^2)^2\,.
\end{aligned}
\end{equation}
\end{subequations}%
In the above, $(x, y) \in [0, 1]^2$, $\phi \sim \phi + 2\pi$, $\psi \sim \psi + 2\pi$, and $t \in \mathbb{R}$.

There are a total of ten functions $Q_I$ of $x$ and $y$ to solve for. When $Q_1=Q_2=Q_3=Q_5=1$, $Q_8=\tilde{Q}$, $Q_4=Q_6=Q_7=Q_9=Q_{10}=0$ we recover the line elemement of an eletrically charged Reissner-Nordstr\"om black hole with the electric charge $\pi \tilde{Q}/(4 )$. For this particular case, the map between standard spherical polar coordinates and the $(x,y)$ coordinates reads
\begin{equation}
r=\frac{1}{1-y^2}\quad\text{and}\quad x\sqrt{2-x^2}=\cos \theta\,,
\end{equation}
where we used the rescaling symmetry $g\to\hat{\lambda}^2 g$ and $A\to \hat{\lambda}A$ with $\hat{\lambda}$ constant to set the black hole event horizon radius to unity and $r\in(1,+\infty)$ and $\theta\in[0,\pi/2]$.

Spatial infinity is located at $y = 1$, where the round $S^3$ diverges in size and $g_{tt} \to -1$, while $x = 0$ and $x = 1$ correspond to the axes of symmetry generated by $\partial/\partial \phi$ and $\partial/\partial \psi$, respectively. The hypersurface $y=0$ marks the location of the bifurcating Killing sphere. The Einstein equation demands that $Q_1(x,0)=A_0\,Q_2(x,0)$, $Q_4(x,0)=w_{\phi}$, $Q_6(x,0)=w_{\psi}$ and $Q_8(x,y)=w_t$ with $A_0$, $w_{\phi}$, $w_{\psi}$ and $w_t$ being constant. Together, these statements imply that $y=0$ is a Killing horizon generated by
\begin{equation}
K=\frac{\partial}{\partial t}+w_{\phi}\frac{\partial}{\partial \phi}+w_{\psi}\frac{\partial}{\partial \psi}
\end{equation}
that is to say, $\lVert K\rVert^2=0$ at $y=0$. The constants $w_{\phi}$ and $w_\psi$ are interpreted as the angular velocities of the horizon, while $w_t$ is the black hole chemical potential. The temperature can be computed via Eq.~(\ref{eq:temperature}) and turns out to be given by
\begin{equation}
T=\frac{\sqrt{A_0}\Delta(0)}{2\pi}=\frac{\sqrt{A_0}}{2\pi}\left(1-\frac{\tilde{Q}^2}{3}\right)\,,
\end{equation}
from which we conclude that as $\tilde{Q}\to\sqrt{3}^-$, $T\to0^+$ as long as $A_0$ remains finite in the limit.

Following \cite{Horowitz:2024dch}, we split the components of the Einstein equations into two sets: one that is solved directly on the grid, and another that is treated as constraints. Let
\begin{equation}
E_{ab}\equiv R_{ab}-\frac{1}{2}g_{ab}R-\frac{1}{2}\left(F_a^{\phantom{a}c}F_{bc}-\frac{g_{ab}}{4}F^{cd}F_{cd}\right)\quad\text{and}\quad J^{a}\equiv \nabla_b F^{ba}-\frac{\lambda}{4\sqrt{3}}\varepsilon^{abcde}F_{bc}F_{de}
\end{equation}
We solve on the grid the following set of partial differential equations
\begin{multline}
{\rm D}=\{E_{tt}=0,g^{xx}E_{xx}+g^{yy}E_{yy}=0,E_{t\phi}=0,E_{t\psi}=0,E_{\phi\phi}=0,
\\
E_{\phi\psi}=0,E_{\psi\psi}=0,J^t=0,J^{\phi}=0,J^{\psi}=0\}\,.
\end{multline}
and for constraints we take
\begin{equation}
C_1\equiv \frac{\sqrt{2-x^2}(1-y^2)\sqrt{\Delta}}{4}\sqrt{-g}(g^{xx}E_{xx}-g^{yy}E_{yy})\quad\text{and}\quad C_2\equiv (1-y^2)^2\Delta(y)\sqrt{-g}g^{xx}E_{xy}\,.
\end{equation}
When all the equations in group ${\rm D}$ are satisfied, one can show that the constraints above obey
\begin{subequations}
\begin{equation}
(1-y^2)\sqrt{\Delta(y)}\frac{\partial C_1}{\partial y}- {2}\sqrt{2-x^2}\frac{\partial C_2}{\partial x}=0\,,
\end{equation}
and
\begin{equation}
(1-y^2)\sqrt{\Delta(y)}\frac{\partial C_2}{\partial y}+  {2}\sqrt{2-x^2}\frac{\partial C_1}{\partial x}=0\,.
\end{equation}
\end{subequations}
With a suitable choice of boundary conditions, detailed below, the two sets of equations above are sufficient to show that $C_1=C_2=0$ throughout the integration domain if the equations in group ${\rm D}$ are satisfied. For additional details, we refer the reader to \cite{Horowitz:2024dch}. Furthermore, the equations in group ${\rm D}$ can be shown to be Elliptic, for the same choice of boundary conditions. This means that they can be readily solved using the numerical methods detailed in \cite{Dias:2015nua}.

In order to understand the asymptotic structure at spatial infinity, we expand the equations of motion in a power series about $y = 1$. At each order in $(1 - y)$, the equations decouple and can be readily solved after imposing regularity at $x = 0$ and $x = 1$. To the first order in $1-y$, we find:
\begin{equation}
\begin{aligned}
&Q_1(x,y)=1+(1-y) \mathcal{E}+\mathcal{O}[(1-y)^2]\,,
\\
&Q_2(x,y)=1+(1-y) \left[\lambda _1 \left(2 x^4-4 x^2+1\right)-\frac{\mathcal{E}}{2}\right]+\mathcal{O}[(1-y)^2]\,,
\\
&Q_3(x,y)=1+\lambda _0 (1-y)+\mathcal{O}[(1-y)^2]\,,
\\
&Q_4(x,y)=\mathcal{J}_{\phi }+(1-y) \left\{\lambda _3 \left(x^4-2 x^2+\frac{1}{3}\right)+\frac{1}{3} \left[\mathcal{Q} \mu _{\phi }+2 \mathcal{J}_{\phi } \left(\mathcal{E}-\lambda _0\right)\right]\right\}+\mathcal{O}[(1-y)^2]\,,
\\
&Q_5(x,y)=1-(1-y) \left(\lambda _0+\mathcal{E}\right)+\mathcal{O}[(1-y)^2]\,,
\\
&Q_6(x,y)=\mathcal{J}_{\psi }+(1-y) \left\{\lambda _4 \left(x^4-2 x^2+\frac{2}{3}\right)+\frac{1}{3} \left[\mathcal{Q} \mu _{\psi }+2 \mathcal{J}_{\psi } \left(\lambda _0+2 \mathcal{E}\right)\right]\right\}+\mathcal{O}[(1-y)^2]\,,
\\
&Q_7(x,y)=\mathcal{O}[(1-y)^2]
\\
&Q_8(x,y)=\mathcal{Q}+(1-y) \left[\lambda _5 \left(x^4-2 x^2+\frac{1}{2}\right)+\frac{\mathcal{Q} \mathcal{E}}{2}\right]+\mathcal{O}[(1-y)^2]\,,
\\
&Q_9(x,y)=(1-y) \mu_{\phi }+\mathcal{O}[(1-y)^2]\,,
\\
&Q_{10}(x,y)=(1-y) \mu_{\psi }+\mathcal{O}[(1-y)^2]\,.
\end{aligned}
\end{equation}

The constraints demand that
\begin{equation}
\lambda _0+\lambda _1+\frac{\mathcal{E}}{2}=0
\end{equation}
which we regard as a Robin-type boundary condition for $Q_2$, and thus impose at $y=1$ that
\begin{equation}
\left(\frac{1}{2} \left.\frac{\partial Q_1(x,y)}{\partial y}\right|_{y=1}+\left.\frac{\partial Q_5(x,y)}{\partial y}\right|_{y=1}\right) \left(2-8 x^2+4 x^4\right)-\left.\frac{\partial Q_1(x,y)}{\partial y}\right|_{y=1}-2 \left.\frac{\partial Q_2(x,y)}{\partial y}\right|_{y=1}=0\,.
\label{eq:BCy1}
\end{equation}

Using the asymptotic expansion above, we can readily compute the energy $M$, angular momenta $(J_{\phi},J_{\psi})$ and electric charge $Q$ of these black hole solutions, which turn out to be given by
\begin{equation}
M=\frac{\pi}{16 }(6+2\tilde{Q}^2-3 \mathcal{E})\,,\quad J_{\phi}=\frac{\pi \mathcal{J}_{\phi}}{4 }\,,\quad J_{\psi}=\frac{\pi \mathcal{J}_{\psi}}{4 }\quad\text{and}\quad Q=\frac{\pi \mathcal{Q}}{4 }\,.
\end{equation}

As indicated above, at spatial infinity ($y=1$), we  impose that 
\begin{multline}
Q_1(x,1)=Q_3(x,1)=Q_5(x,1)=1\,,\quad Q_4(x,1)=\mathcal{J}_{\phi}\,,\quad Q_6(x,1)=\mathcal{J}_{\psi}\,,
\\
Q_8(x,1)=\mathcal{Q}\quad\text{and}\quad Q_9(x,1)=Q_{10}(x,1)=0\,,
\end{multline}
together with Eq.~(\ref{eq:BCy1}). At the axis of symmetry generated by $\partial/\partial \psi$, i.e. $x=0$, we demand that
\begin{equation}
\left.\frac{\partial Q_I(x,y)}{\partial x}\right|_{x=0}=0,\quad\text{for}\quad I=1,\dots,10
\end{equation}
together with $Q_2(0,y)=Q_5(0,y)$, with the latter condition ensuring that $\psi$ has period $2\pi$. Similarly, at $x=1$, we demand that
\begin{equation}
\left.\frac{\partial Q_I(x,y)}{\partial x}\right|_{x=1}=0,\quad\text{for}\quad I=1,\dots,10
\end{equation}
together with $Q_2(1,y)=Q_3(1,y)$, thus ensuring that $\phi$ has period $2\pi$.

Finally, we discuss the boundary conditions at the bifurcating Killing surface located at $y=0$. The constraint equations impose constant temperature and constant angular velocity, but these are guaranteed to hold throughout the integration domain so long as the boundary conditions discussed above are imposed (see \cite{Horowitz:2024dch} for more details). As such, we impose Neumann boundary conditions on all variables at $y=0$, i.e.
\begin{equation}
\left.\frac{\partial Q_I(x,y)}{\partial y}\right|_{y=0}=0\,,\quad\text{for}\quad I=1,\ldots,10\,.
\end{equation}

To recap, once we specify $\tilde{Q}$, $\mathcal{Q}$, $\mathcal{J}_{\phi}$, and $\mathcal{J}_{\psi}$, we obtain a well-defined elliptic problem that can be readily solved using the methods detailed in \cite{Dias:2015nua}. Specifically, we discretize all the equations in group $\rm D$ in both the $x$ and $y$ directions using a spectral collocation approximation on two Gauss-Lobatto grids, with $N_x$ and $N_y$ points, respectively. The resulting discretised equations are then solved using a standard Newton-Raphson routine. To check for convergence, we monitor the two constraints $C_1$ and $C_2$, both of which approach zero exponentially fast as $N_x$ and $N_y$ increase, consistent with the expected convergence properties of spectral collocation methods. Note that up to factors of $\pi/(4 )$, $\mathcal{Q}$, $\mathcal{J}_{\phi}$, and $\mathcal{J}_{\psi}$ directly control the electric charge and angular momenta, respectively. As such, our integration scheme allows us to cool down the black hole (by tuning $\tilde{Q}$ close to $\sqrt{3}$) while keeping the electric charge and angular momenta constant.

\subsection{Results}

Without loss of generality, we will focus on solutions for which $J_{\phi} \geq J_{\psi}$. For such solutions, we expect the largest tidal forces to occur at $x = 0$. Rather than work with infalling timelike geodesics, we will study these effects using  ingoing null geodesics. Thus, the first step is to consider affinely parameterised null geodesics $U^a={\rm d}x^a/{\rm d}\lambda$, i.e., $U^a \nabla_a U^b = 0$ with $U^a U_a = 0$, which we now briefly discuss.

Our spacetime has three commuting Killing vector fields $\partial/\partial t$, $\partial /\partial \phi$ and $\partial/\partial \psi$. As such we expect three associated conserved charges. Let $\mathcal{L}=g_{ab}U^a U^b$, so that
\begin{equation}
E_n=-\frac{1}{2}\frac{\partial \mathcal{L}}{ \partial \dot{t}}\,,\quad L^{\phi}_n=\frac{1}{2}\frac{\partial \mathcal{L}}{ \partial \dot{\phi}}\quad\text{and}\quad L^{\psi}_n=\frac{1}{2}\frac{\partial \mathcal{L}}{ \partial \dot{\psi}}
\end{equation}
(where $\dot{}\equiv {\rm d}/{\rm d}\lambda$) are constants of motion. These conserved charges have no physical meaning (since they are not invariant under affine transformations), however, that is not the case for their ratio. Indeed, $L^{\phi}_n/E_n$ and $L^{\psi}_n/E_n$ can be interpreted as the impact parameters of incoming geodesics. Using that all functions have Neumann boundary conditions at $x=0$, it is a simple exercise to show null geodesics with $x(\lambda)=0$ exist. We focus on  geodesics along $x(\lambda)=0$ with vanishing $L^{\phi}_n$ and $L^{\psi}_n$ and using affine transformations we fix $E_n=1$. From $g_{ab}U^a U^b=0$ we can thus determine $\dot{y}$ for such geodesics up to an overall sign. To fix the overall sign, we introduce Kerr-like ingoing coordinates in the vicinity of the future event horizon
\begin{equation}
\begin{aligned}
&{\rm d}t={\rm d}v-\frac{3}{2\sqrt{A_0}(3-\tilde{Q}^2)}\frac{{\rm d}Y}{Y}
\\
&{\rm d}\phi={\rm d}\tilde{\phi}-\frac{3w_{\phi}}{2\sqrt{A_0}(3-\tilde{Q}^2)}\frac{{\rm d}Y}{Y}+w_{\phi}\,{\rm d}v
\\
&{\rm d}\psi={\rm d}\tilde{\psi}-\frac{3w_{\psi}}{2\sqrt{A_0}(3-\tilde{Q}^2)}\frac{{\rm d}Y}{Y}+w_{\psi}\,{\rm d}v\,.
\end{aligned}
\end{equation}
with $y=\sqrt{Y}$. Our black hole solutions are completely regular across $Y=0$ in such coordinates, and since constant $v$ hypersurfaces require ${\rm d}Y/{\rm d}t<0$, these coordinates are ingoing, and thus cover the future event horizon. It is in these coordinates that we want the ingoing geodesics to be regular, which in turn fixes the overall sign in $\dot{y}$. Once the dust settles we find that, in our original coordinates $\{t,x,y,\phi,\psi\}$,
\begin{equation}
U_{a}\mathrm{d}x^a=-{\rm d}t-\frac{\sqrt{Q_2(0,y)}}{y(1-y^2)^{3/2}\sqrt{Q_1(0,y)}\Delta(y)}{\rm d}y\,,
\end{equation}
with our choice $E_n=1$ also making $U^a$ future-directed.

Tidal effects are traditionally measured using the Riemann tensor since this is the tensor that naturally enters the geodesic deviation equation. However, in the present context, the singularity is easier to measure using the following scalar quantity
\begin{equation}
\Phi=\left.U^a \left(\frac{\partial }{\partial \phi}\right)^b F_{ab}\right|_{x=0}\,.
\label{eq:tidalF}
\end{equation}
 This is analogous to the $\phi$ component of the electric field that an observer would measure along a timelike geodesic. If $\Phi$ diverges, the curvature cannot remain smooth.
Since $\Phi$ is not invariant under rescalings $g\to\hat{\lambda}^2g$ and $A\to\hat{\lambda}A$, with $\hat{\lambda}$ constant, we will always plot the rescaled quantity $\Phi \sqrt{|Q|}$, which is dimensionless.

We can also compute this quantity using our near-horizon analysis. Since the Maxwell field involves one derivative of the vector potential which is scaling like $\rho^\gamma$, one expects 
\begin{equation}
\Phi \sqrt{|Q|}\propto \gamma \rho^{\gamma-1}\,,
\end{equation}
and indeed, this is what we find.
So any $0<\gamma<1$ will cause this quantity to diverge at the horizon of an extremal black hole.

\begin{figure}
\centering
\includegraphics[width=0.7\textwidth]{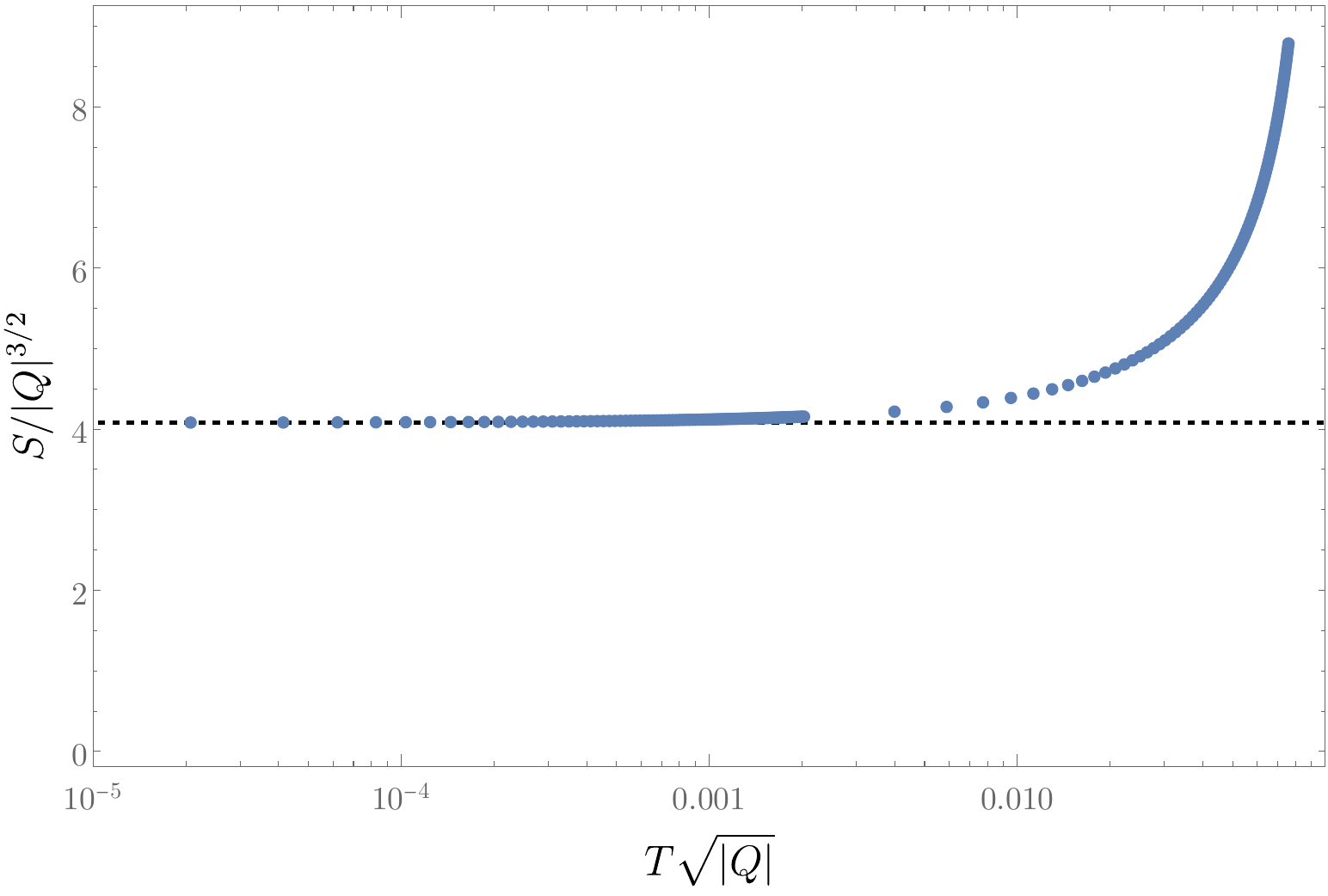}
\caption{The rescaled entropy $S/|Q|^{3/2}$ as a function of the rescaled temperature $T \sqrt{|Q|}$ for fixed $J_{\phi}/|Q|^{3/2}=8/(5 \sqrt{\pi })\approx0.902703$, $J_{\psi}/J_{\phi}=3/8$ and $\lambda=0$. The black horizontal dashed lined was obtained using the near-horizon geometries found in section \ref{subsec:fullback}. }
\label{fig:entropyfull}
\end{figure}

For the remainder of this section, we will focus the case with $J_{\psi}/J_{\phi}=3/8$, since this was used in previous sections. We have studied other linear combinations of angular momenta, and the qualitative results remain unchanged.  We start by describing results in Einstein-Maxwell theory ($\lambda = 0$) and later show what happens for nonzero $\lambda$. In Fig.~\ref{fig:entropyfull} we plot the rescaled entropy $S/|Q|^{3/2}$ as a function of the rescaled temperature $T\sqrt{|Q|}$ for fixed $J_{\phi}/|Q|^{3/2}=8/(5 \sqrt{\pi })\approx0.902703$. The black horizontal dashed lined was obtained using the 
near-horizon geometries found in section \ref{subsec:fullback}. We see that as we cool down the black hole at fixed $J_{\phi}/|Q|^{3/2}$ and $J_{\psi}/|Q|^{3/2}$, the entropy precisely approaches that of the near-horizon geometries found in section \ref{subsec:fullback}. The agreement between the near-horizon analysis and full extremal black hole solution indicates that we have correctly identified the corresponding near-horizon geometry.

\begin{figure}
\centering
\includegraphics[width=0.7\textwidth]{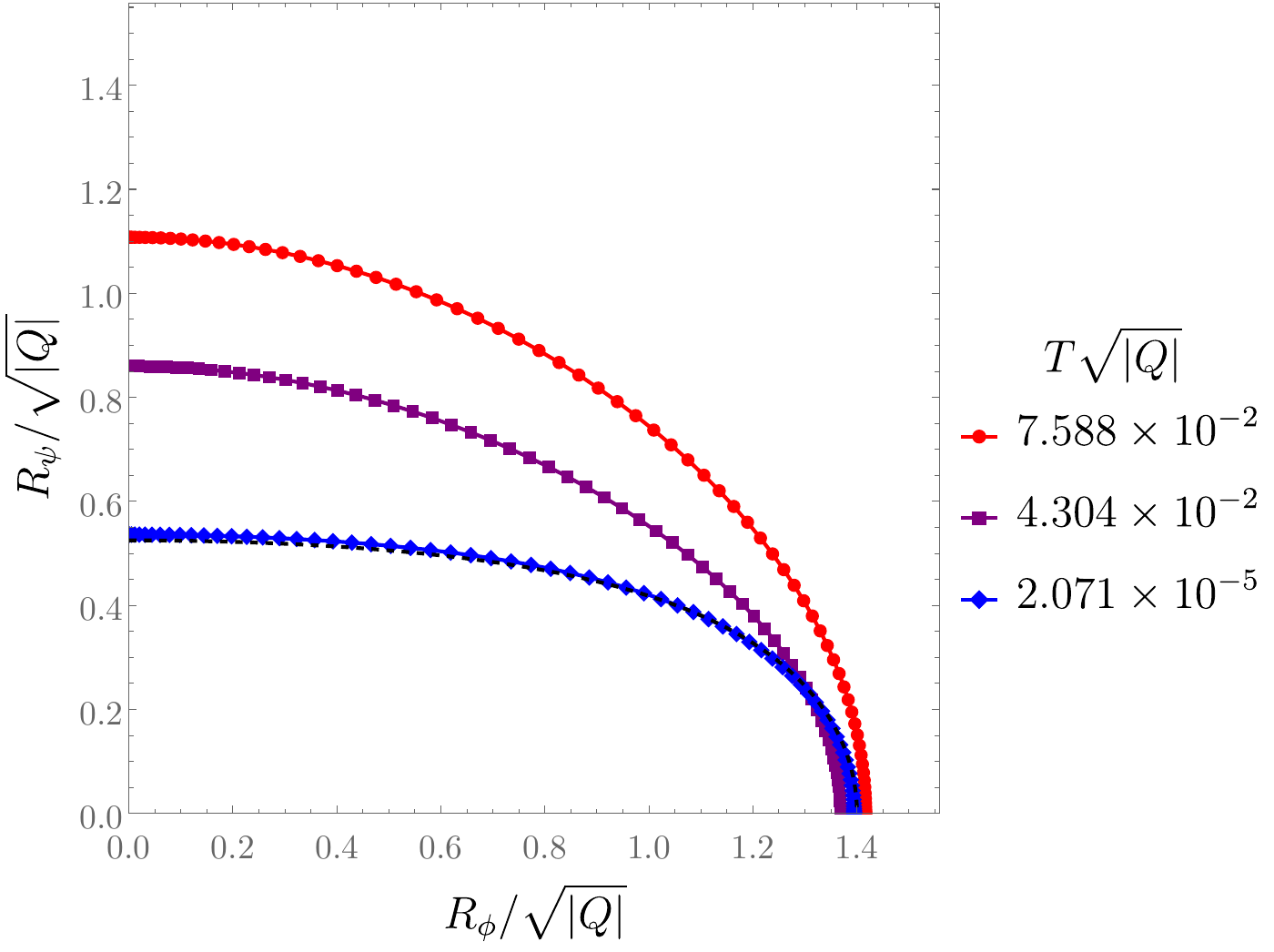}
\caption{A parametric plot of $\{R_{\phi}, R_{\psi}\}/\sqrt{|Q|}$ along a spatial cross section of the black hole event horizon, with $R_{\phi}\equiv\sqrt{g_{\phi\phi}}$ and $R_{\psi}\equiv\sqrt{g_{\psi\psi}}$. The black dashed line corresponds to the near-horizon geometry found in section \ref{subsec:fullback}. All plots were generated with  $J_{\phi}/|Q|^{3/2}=8/(5 \sqrt{\pi }), \ J_{\psi}/J_{\phi}=3/8$, and $\lambda=0$, and the different symbols correspond to the rescaled temperatures labelled on the right.}
\label{fig:cross}
\end{figure}

Not only can we match the entropy, but we can also match the local geometry of spatial cross-sections of the black hole event horizon. In Fig.~\ref{fig:cross}, we present a parametric plot of $\{R_{\phi}, R_{\psi}\}/\sqrt{|Q|}$, where $R_{\phi}\equiv\sqrt{g_{\phi\phi}}$ and $R_{\psi}\equiv\sqrt{g_{\psi\psi}}$, along spatial cross-sections of the horizon. For a round three-sphere, this would result in a perfect quarter circle, but with rotation, the local geometry becomes deformed. The black dashed line is obtained using the near-horizon geometries found in \ref{subsec:fullback}, while the remaining curves are derived from the full black hole solution at different temperatures, labeled on the right. The agreement between these curves as the temperature decreases is further evidence that that we have correctly identified the relevant near-horizon geometry. This plot was generated with $\lambda=0$ and $J_{\psi}/J_{\phi}=3/8$, but we find a similar behaviour whenever the scaling dimensions found in section \ref{sec:scalings} are positive.

\begin{figure}
\centering
\includegraphics[width=0.78\textwidth]{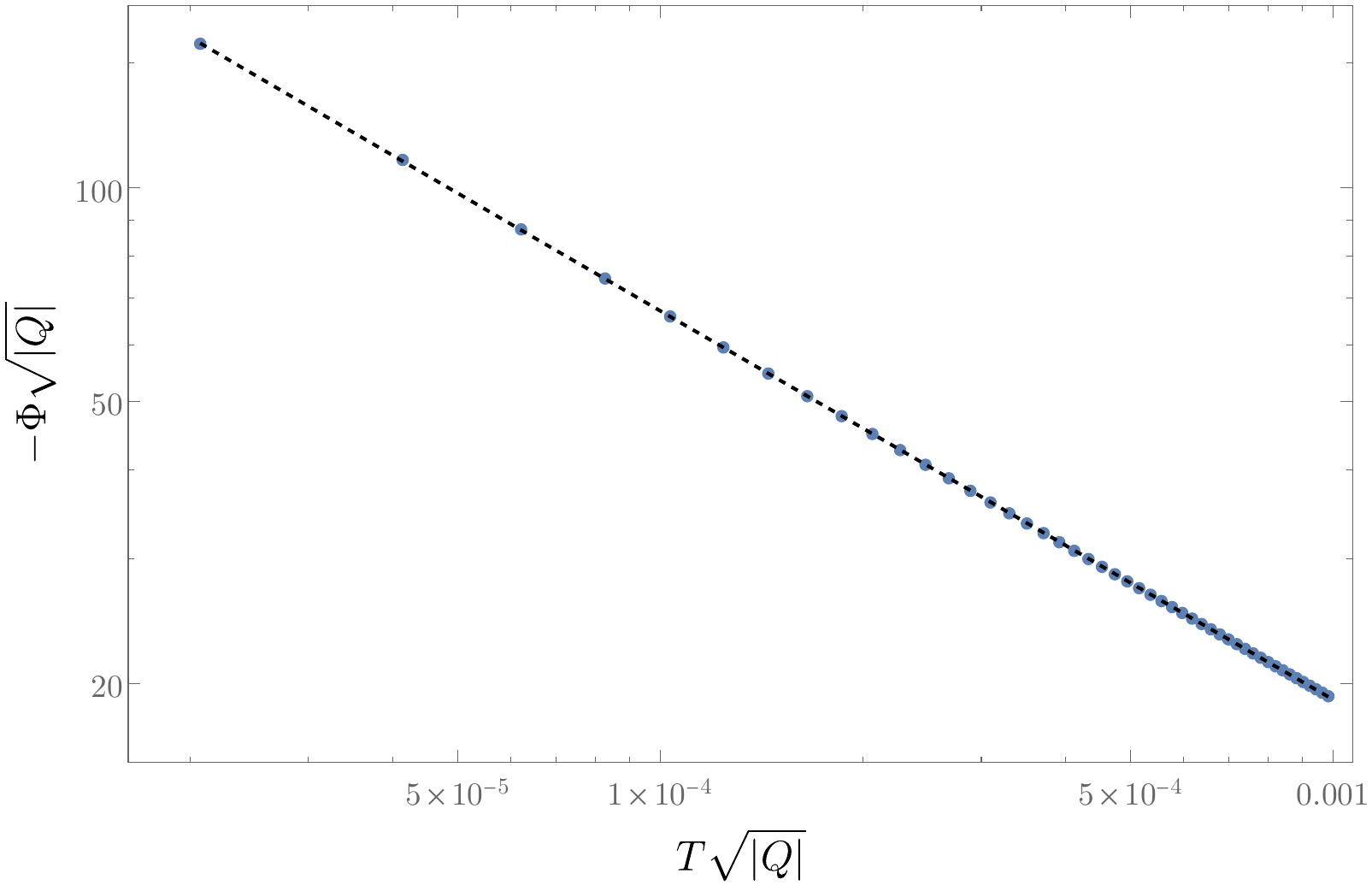}
\caption{A log-log plot of $-\Phi \sqrt{|Q|}$ (defined in Eq.~\eqref{eq:tidalF}) as a function of the rescaled temperature $T\sqrt{|Q|}$ for fixed $J_{\phi}/|Q|^{3/2}=8/(5 \sqrt{\pi }), \ J_{\psi}/J_{\phi}=3/8$, and $\lambda=0$. The black dashed line shows a fit to Eq.~(\ref{eq:fit}), with the predicted exponents computed in section \ref{sec:scalings}.}
\label{fig:tidal}
\end{figure}

To see the singularity, we now turn our attention to the Maxwell field on the horizon. Given what we have seen before, one can expect $\Phi$ (defined in Eq.~(\ref{eq:tidalF})) to become large as we lower the  temperature. In Fig.~\ref{fig:tidal} we plot $\Phi \sqrt{|Q|}$  as a function of the rescaled temperature, $T \sqrt{|Q|}$, in a log-log plot. This plot was again generated for $\lambda=0$, $J_{\psi}/J_{\phi}=3/8$ and $J_{\phi}/|Q|^{3/2}=8/(5 \sqrt{\pi })$. The divergence is very clearly seen across several decades.  What is more, we can compare this divergence to the scaling dimensions found in the near-horizon analysis. Using the scaling argument in \cite{Horowitz:2022mly} one can relate the dependence on $\rho$ of the extremal solution to the dependence on $T$ for the near extremal solution. The net result is that we expect $\Phi \sqrt{|Q|} \sim \gamma T^{\gamma-1}$ where $\gamma$ is the same scaling dimension computed in section 4. In fact we
computed the two lowest scaling dimensions for these values of the parameters from the near-horizon analysis, and these turn out to be given by $\gamma_0\approx0.37709(5)$ and $\gamma_1\approx0.47806(9)$. The black dashed line shows a  \emph{two-parameter} fit to a curve of the form 
\begin{equation}
f(\tau)\equiv a_0\,\tau^{\gamma_0-1}+a_1\,\tau^{\gamma_1-1}
\label{eq:fit}
\end{equation}
for which we find $a_0\approx 0.0614(5)$ and $a_1\approx 0.3920(5)$. The agreement is excellent. In fact, the error in the fit is estimated to only affect the last digit of each of the constants reported above. If we try to fit with only one of the two terms, the fit is not accurate. Based on the above numbers, we can estimate when the first term dominates over the second. This happens for $T \sqrt{|Q|}\lesssim 10^{-8}$, which is outside what our numerical scheme can handle without resorting to extended precision and forbidden grid sizes.

  Given this confirmation that the near-horizon analysis correctly describes the behavior of the full solution, it is clear that  tidal forces will also diverge in the extremal limit:
\begin{equation}
    R_{a\phi c\phi} U^a U^c \sim \gamma (\gamma-1) T^{\gamma -2}\,,
\end{equation}
{with $R_{abcd}$ denoting the  Riemann tensor.}

\begin{figure}
\centering
\includegraphics[width=0.78\textwidth]{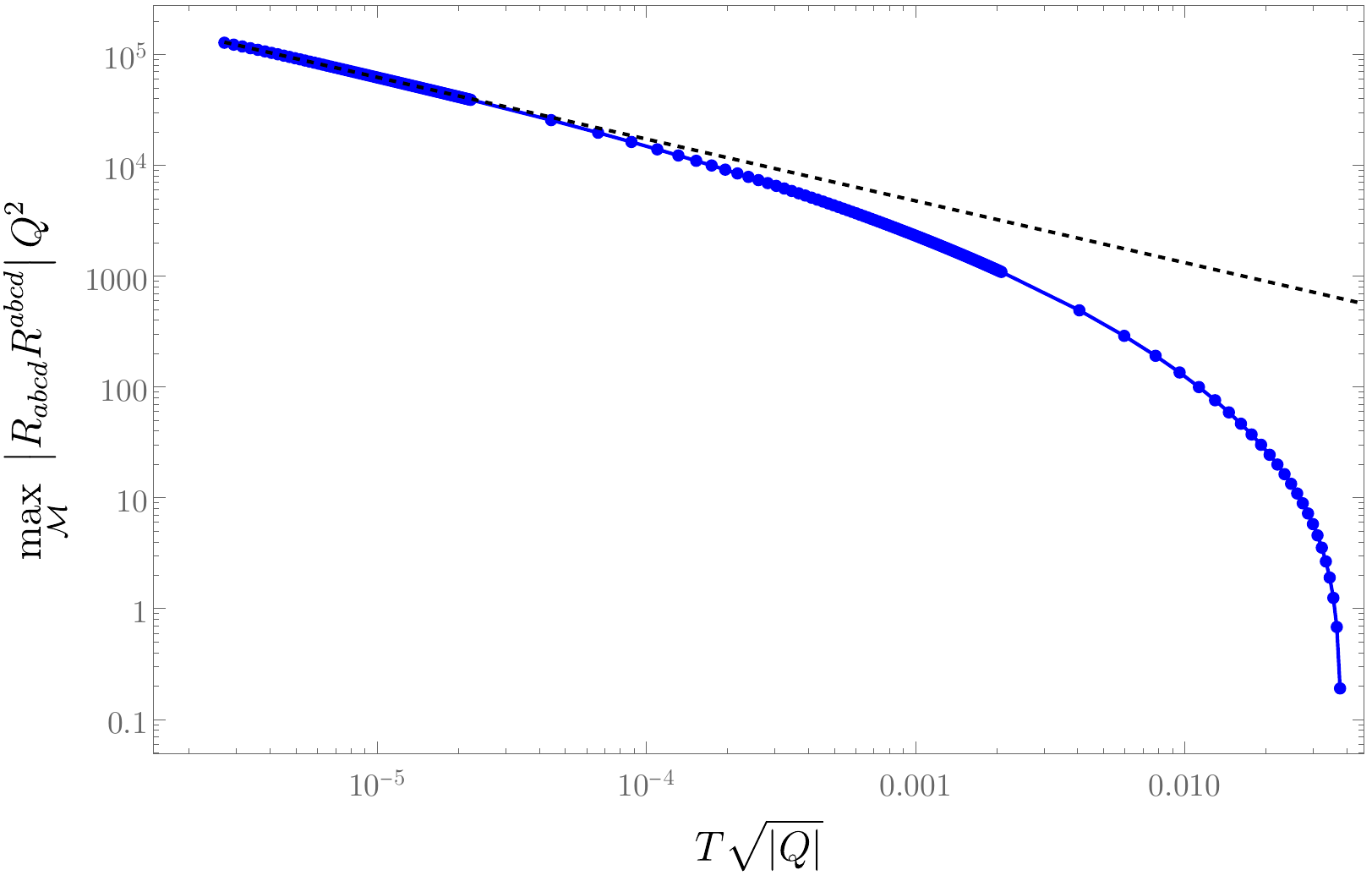}
\caption{A log-log plot of the maximum of the absolute value of the normalised Kretschmann scalar $Q^2R_{abcd}R^{abcd}$ as a function of the normalised temperature $T \sqrt{|Q|}$. This figure was generated for fixed $\lambda=3/2$, $J_{\psi}/J_{\phi}=3/8$ and $J_{\phi}/|Q|^{3/2}=8/(5 \sqrt{\pi })$. The black dashed line denotes the power law $T^{-0.557}$.}
\label{fig:kret}
\end{figure}

When $|\lambda| > 1$, the situation becomes more intriguing. In this case, our near-horizon analysis consistently predicts the existence of at least one mode (possibly more) with a negative scaling dimension. This strongly suggests that smooth near-horizon geometries may not exist as we cool the black holes while keeping the electric charge and angular momenta constant. Indeed, in Fig.~\ref{fig:kret} we plot the maximum of the absolute value of the normalised Kretschmann scalar $Q^2R_{abcd}R^{abcd}$ as a function of the normalised temperature $T \sqrt{|Q|}$. This figure was generated for fixed $\lambda=3/2$, $J_{\psi}/J_{\phi}=3/8$ and $J_{\phi}/|Q|^{3/2}=8/(5 \sqrt{\pi })$. The divergent curvature at small temperatures is consistent with an inverse power law scaling of the form $\max_{\mathcal{M}}\left|R_{abcd}R^{abcd}\right|Q^2\propto T^{-0.557}$ (shown as a black dashed curve in Fig.~\ref{fig:kret}).

\begin{figure}
\centering
\includegraphics[width=0.7\textwidth]{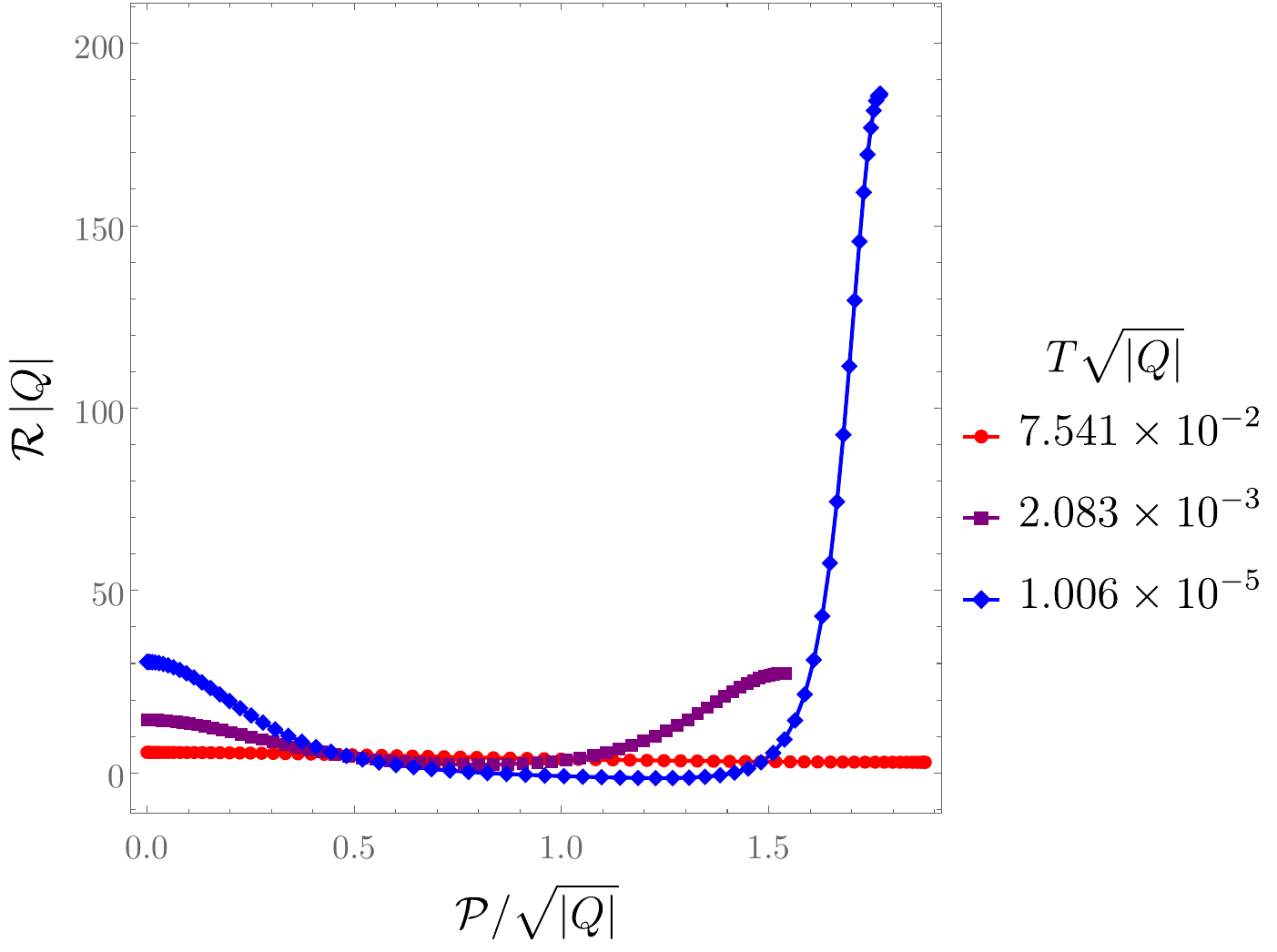}
\caption{The normalised Ricci scalar $\mathcal{R}\,|Q|$ as a function of the normalised proper distance from the $\psi$ axis $\mathcal{P}/\sqrt{|Q|}$ at fixed $\lambda=3/2$, $J_{\psi}/J_{\phi}=3/8$ and $J_{\phi}/|Q|^{3/2}=8/(5 \sqrt{\pi })$. The normalised temperatures are labelled on the right.}
\label{fig:curvax}
\end{figure}

The local geometry of spatial cross sections of the horizon also develops interesting features. In Fig.~\ref{fig:curvax} we plot the normalised Ricci scalar of the spatial cross sections of the horizon, i.e. $\mathcal{R} |Q|$, as a function of the normalised proper distance from the $\psi$ axis ($x=0$), $\mathcal{P}/\sqrt{|Q|}$, with
\begin{equation}
\mathcal{P}(x)\equiv \int_0^x\frac{2\sqrt{Q_2(\tilde{x},0)}}{\sqrt{2-\tilde{x}^2}}{\rm d}\tilde{x}\,.
\end{equation}
This figure was again generated at fixed $\lambda=3/2$, $J_{\psi}/J_{\phi}=3/8$ and $J_{\phi}/|Q|^{3/2}=8/(5 \sqrt{\pi })$. Fig.~\ref{fig:curvax} shows that the Ricci scalar becomes very large, possibly diverging in the strict $T\to 0$ limit, as we lower the normalised temperature $T\sqrt{|Q|}$ at the axis of rotation generated by orbits of $\partial/\partial \psi$. This turns out to be the case whenever $J_{\phi}>J_{\psi}$.

\begin{figure}
\centering
\includegraphics[width=\textwidth]{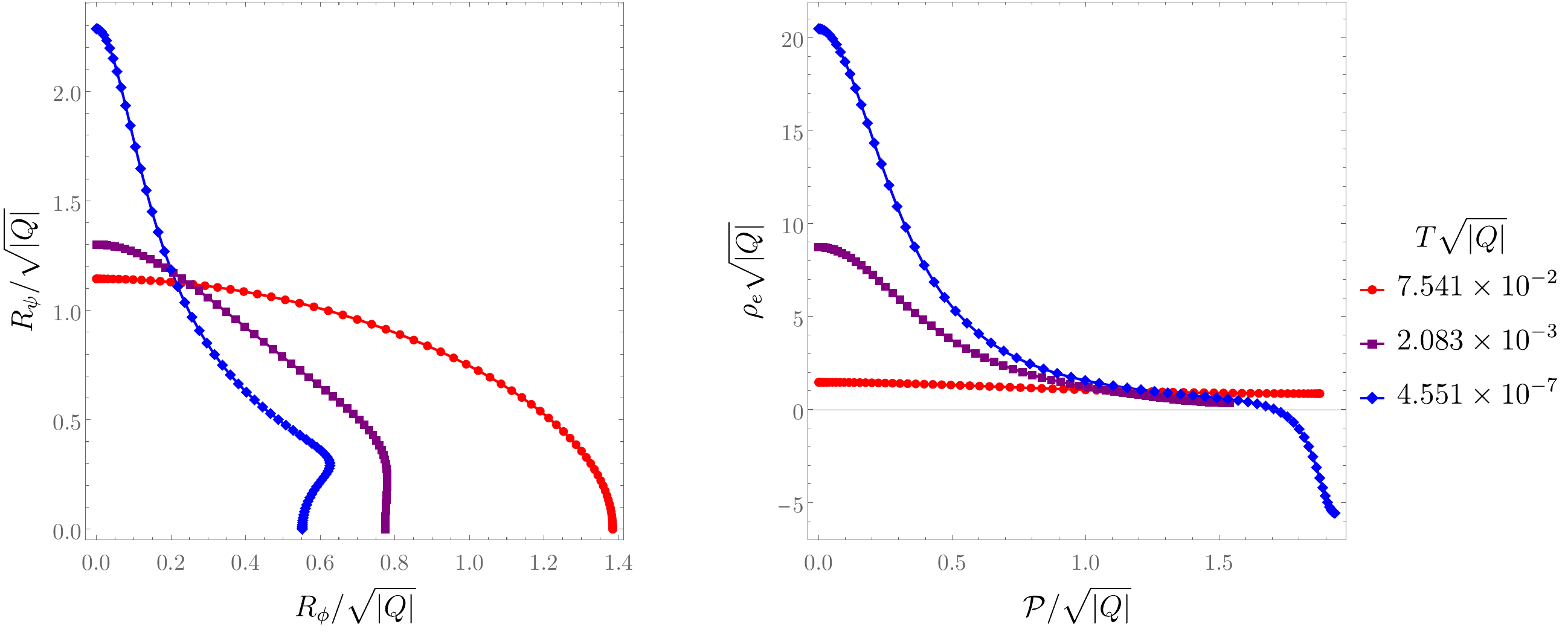}
\caption{{\bf Left panel.} Parametric plot of ${R_{\phi}, R_{\psi}}/\sqrt{|Q|}$ along spatial cross-sections of the horizon. {\bf Right panel.} Normalised charge density, $\rho_e \sqrt{|Q|}$ (as defined in the main text), plotted as a function of the normalised proper distance, $\mathcal{P}/\sqrt{|Q|}$. Both panels are generated with $\lambda = 3/2$, $J_{\psi} / J_{\phi} = 3/8$ and $J_{\phi}/|Q|^{3/2}=8/(5 \sqrt{\pi })$, with various temperatures indicated on the right.}
\label{fig:final}
\end{figure}

Finally, to gain further insight into the behavior of the local geometry of the spatial horizon cross-sections, we create a parametric plot of ${R_{\phi}, R_{\psi}}/\sqrt{|Q|}$ along spatial cross-sections of the horizon, with $\lambda=3/2$, $J_{\psi}/J_{\phi}=3/8$ and $J_{\phi}/|Q|^{3/2}=8/(5 \sqrt{\pi })$ held constant, across various temperatures. This is shown in the left panel of Fig.~\ref{fig:final}. As the temperature decreases, the horizon adopts a spindle shape, with high curvature concentrated at the spindle's tip. We also monitored the electric charge density, $\rho_e$, defined as the pullback of $\star F + \frac{\lambda}{\sqrt{3}} F \wedge A$ onto spatial 
cross-sections of the horizon, normalised by the volume element of the spatial cross-sections. Even though there is no charged matter, the Chern-Simons term provides an effective charge density as shown in Eq.~\eqref{eq:charges}. The right panel of Fig.~\ref{fig:final} shows the normalised charge density $\rho_e \sqrt{|Q|}$, revealing a region of negative charge density within the spindle, while near the axis generated by $\partial / \partial \psi$, a concentration of positive charge density is observed.
\section{Discussion}

We have explored the extremal limit of the generic 5D (asymptotically flat) rotating charged black hole, and found a number of surprises. Perhaps the most important is the fact that the horizon becomes singular. In Einstein-Maxwell theory, curvature scalars remain finite at the extremal horizon, but infalling observers see diverging electric fields and experience diverging tidal forces. We have also shown that ``general relativity knows about supergravity" in the sense that if one adds a Chern-Simons term and increases its coefficient $\lambda$, this singularity remains until one reaches the supergravity value $\lambda = 1$. The extremal horizon is smooth  for this one value of $\lambda$, but for $\lambda > 1$ the singularity becomes stronger and  even curvature scalars diverge. 

Another surprise concerns the limit of vanishing angular momenta. We found that this limit depends on the ratio $J_\phi/J_\psi$ as both angular momenta are scaled to zero. One approaches the 5D Reissner-Nordstr\"om solution only when $J_\phi/J_\psi =1$.  In general,  one approaches a new family of static, extremal black holes which are not spherically symmetric. However these limiting solutions might have a stronger singularity on the horizon than the ones with $J_i \ne 0$. Linear perturbations of the near horizon geometry no longer scale to zero like a power law $\rho^\gamma$ (with $\gamma$ given in Eq. \eqref{eq:anasmalljs}) but can now diverge like $\log \rho$. If these perturbations are present in the full asymptotically flat solution, the metric would no longer be continuous on the horizon, and would not have a well defined horizon geometry. This deserves further investigation.

We have also studied generic near-extremal black holes and shown that the tidal forces are anomously large but finite. They diverge only in the extremal limit.  If one starts with a near extremal black hole, the double limit of approaching extremality and turning off the angular momenta depends on the order of limits. If one first removes the angular momenta, the result is always the spherical Reissner-Nordstr\"om solution. As we just discussed, the opposite order yields a nonspherical solution.

The fact that extremal black holes have smooth horizons just for the Chern-Simons coefficient required for supergravity might lead one to wonder if extremal black holes in supergravity will always have  smooth horizons. It is easy to see that this is not the case. The Myers-Perry solution with only one nonzero angular momentum has a singular extremal limit, even though it is a solution of minimal 5D supergravity. Perhaps a better question is whether {\it supersymmetric} black holes in supergravity always have smooth extremal limits. However the answer is still no. The BMPV solution \cite{Breckenridge:1996is} is a well known supersymmetric black hole and has a  smooth horizon. But multi-BMPV solutions exist which are still supersymmetric and have tidal force singularities on the horizons \cite{Candlish:2007fh,Candlish:2009vy}.

We have focussed on five-dimensional black holes,  but we expect that generic extremal black holes in higher dimensions will also have singular horizons. (There is no extremal limit if some angular momenta vanish, but that is not the generic situation.) In fact the singularities are likely to be worse, since that is what happens in AdS for  extremal black holes  with perturbed boundary conditions \cite{Horowitz:2022mly, Horowitz:2022leb}. Recalling that even in 4D general relativity, extremal black holes have singular horizons as soon as any higher derivative correction is included \cite{Horowitz:2023xyl,Horowitz:2024dch},  the lesson is that smooth extremal horizons are the exception and not the rule.

\begin{acknowledgments}
It is a pleasure to thank Maciek Kolanowski for collaboration at an early stage of this work.  
G.H. was supported in part by NSF Grant PHY-2408110. 
J.E.S. has been partially supported by STFC consolidated grant ST/X000664/1. G.H. and J.E.S. were also supported in part by grant NSF PHY-2309135 to the Kavli Institute for Theoretical Physics (KITP) where this work was begun.
\end{acknowledgments}

\appendix

\section{\label{app:U1U1}Scaling dimensions of the  $U(1)\times U(1)$ symmetric near-horizon geometries}

In this appendix we compute the scaling dimensions of the static $U(1)\times U(1)$ symmetric near-horizon geometries given in Eq.~\eqref{eq:U1U1}. As discussed in section \ref{sec:smalljs}, the linear equations that determine the scaling dimenions decouple into three groups $G_1, G_2, G_3$ \eqref{eq:groups}.
Define
\begin{equation}
S_p(x)\equiv \, _2F_1\left(-p,p+1;1;x^2\right)\quad\text{with}\quad p\geq0\,.
\end{equation}

We find three families of modes in ${\rm G}_1$ with $\gamma\geq1$. The first family is given by
\begin{equation}
\begin{aligned}
&\gamma=p-1\,,\quad \text{with}\quad p\geq2\,,
\\
&q_1^{(0)}(x)=-2 p S_p(x)+\frac{x}{\sigma (x)}\left(1-x^2\right) (c_1-c_2) S_p^\prime(x)\,,
\\
&q_3^{(0)}(x)=-2 p (1+p) \left(1-2 x^2\right) S_p(x)+\frac{x}{2 \sigma (x)}\left[4 (1+p) \left(1-x^2\right) \sigma (x)-c_2\right] S_p'(x)\,,
\\
&q^{(0)}_4(x)=2 p (1+p) \left(1-2 x^2\right) S_p(x)+\frac{\left(1-x^2\right)}{2 x \sigma (x)}\left[c_1-4 (1+p) x^2 \sigma (x)\right] S_p'(x)\,,
\\
&q_8^{(0)}(x)=S_p(x)\,.
\end{aligned}
\end{equation}

The second family is given by
\begin{equation}
\begin{aligned}
&\gamma=p\,,\quad \text{with}\quad p\geq1\,,
\\
&q_1^{(0)}(x)=\frac{x}{\sigma (x)} \left(1-x^2\right) \left(c_1-c_2\right)^2 S_p'(x)\,,
\\
&q_3^{(0)}(x)=2 p (1+p) \left(c_1+c_2\right) S_p(x)-x \left[1+\frac{c_1-c_2}{2 \sigma (x)}\right] c_2 S_p'(x)\,,
\\
&q^{(0)}_4(x)=-2 p (1+p) \left(c_1+c_2\right) S_p(x)-\left(1-x^2\right) \left[1+\frac{c_2-c_1}{2 \sigma (x)}\right]c_1 \frac{S_p'(x)}{x}\,,
\\
& q_8^{(0)}(x)=S_p(x)\,.
\end{aligned}
\end{equation}

Finally, the third family is given by
\begin{equation}
\begin{aligned}
&\gamma=p+1\,,\quad \text{with}\quad p\geq0\,,
\\
&q_1^{(0)}(x)=2 (1+p) S_p(x)+\frac{x \left(1-x^2\right)}{\sigma (x)} \left(c_1-c_2\right) S_p'(x)\,,
\\
&q_3^{(0)}(x)=-2 p (1+p) \left(1-2 x^2\right) S_p(x)-x \left[\frac{c_2}{2 \sigma (x)}+2 p \left(1-x^2\right)\right] S_p'(x)\,,
\\
&
q^{(0)}_4(x)=2 p (1+p) \left(1-2 x^2\right) S_p(x)+\left(1-x^2\right) \left[\frac{c_1}{2 \sigma (x)}+2 p x^2\right] \frac{S_p'(x)}{x}\,,
\\
& q_8^{(0)}(x)=S_p(x)\,.
\end{aligned}
\end{equation}

For deformations in ${\rm G}_2$ we find
\begin{equation}
\begin{aligned}
q_5^{(0)}(x)=\, _2F_1\left(-p,p+3;2;x^2\right)\quad\text{with}\quad \gamma=p+1.\quad \text{where}\quad p\geq0\,.
\end{aligned}
\end{equation}
Note that the scaling dimensions $\gamma$ are integers for all of these modes.

The remaining sector, i.e. ${\rm G}_3$, is discussed in the main text as well as the modes with $\gamma=0$.

\section{\label{app:MP}The scaling dimensions of the near-horizon geometries of extremal Myers-Perry black holes}
In this appendix we compute the general scaling dimensions associated to the near-horizon limit of extremal Myers-Perry black holes. We will assume throughout that $\gamma^{(0)}\neq0$ (note that modes with $\gamma^{(0)}=0$ were discussed in section \ref{sec:pcharge}), and show that $\gamma$ is always a positive integer, consistent with earlier results in the literature \cite{Durkee:2010ea,Murata:2012ct}.  Unlike modes with $\gamma^{(0)}=0$, modes with $\gamma^{(0)}\neq0$ only decouple in two groups:
\begin{equation}
{\rm G}_1=\{z_1^{(0)},z_2^{(0)},z_3^{(0)},z_4^{(0)},z_5^{(0)},z_6^{(0)},z_7^{(0)}\}\quad\text{and}\quad {\rm G}_2=\{z_8^{(0)},z_9^{(0)},z_{10}^{(0)}\}\,.
\end{equation}

We will describe first the modes in ${\rm G}_2$, as these are substantially simpler to write. Set
\begin{equation}
H_1(x)\equiv x^2+4 \Omega_{\phi}^2,\quad \quad H_2(x)=x^2+4\Omega_{\phi}^2(1-x^2)\,.
\end{equation}
and
\begin{equation}
S_p(x)=\, _2F_1\left(-p,p+1;1;x^2\right)\,.
\end{equation}
We find three families of modes, given by
\begin{equation}
\begin{aligned}
&\gamma^{(0)}=p-1\quad\text{with}\quad p\geq2
\\
&z^{(0)}_8(x)=\frac{1}{(p-1) \Omega _{\phi } H_1(x)}\left\{-2 p H_2(x) S_p(x)+x \left[H_2(x)+2 \Omega _{\phi }^2-\frac{1}{2}\right] S_p'(x)\right\}\,,
\\
&z^{(0)}_9(x)=-\frac{4 \Omega _{\phi }}{(p-1) H_2(x)}\left\{2 p H_2(x) S_p(x)+\left(1-x^2\right) \left[H_2(x)-2 \Omega _{\phi }^2\right] \frac{S_p'(x)}{x}\right\}\,,
\\
&z^{(0)}_{10}(x)=S_p(x)\,,
\end{aligned}
\end{equation}
\begin{equation}
\begin{aligned}
&\gamma^{(0)}=p\quad\text{with}\quad p\geq1
\\
&z^{(0)}_8(x)=-\frac{x}{2 p \Omega _{\phi } H_1(x)} (1+4 \Omega _{\phi }^2) S_p'(x)
\\
&z^{(0)}_9(x)=\frac{8 \left(1-x^2\right) \Omega _{\phi }^3}{p H_2(x)}\frac{ S_p'(x)}{x}
\\
&z^{(0)}_{10}(x)=S_p(x)
\end{aligned}
\end{equation}
and
\begin{equation}
\begin{aligned}
&\gamma^{(0)}=p+1\quad\text{with}\quad p\geq0
\\
&z^{(0)}_8(x)=\frac{1}{\Omega _{\phi } H_1(x)}\left\{2 H_2(x) S_p(x)+\frac{x}{2 (p+1)} \left[2 H_2(x)+4 \Omega _{\phi }^2-1\right] S_p'(x)\right\}
\\
&z^{(0)}_9(x)=4 \Omega _{\phi } \left\{2 S_p(x)-\frac{\left(1-x^2\right)}{(p+1) H_2(x)} \left[H_2(x)-2 \Omega _{\phi }^2\right]\frac{ S_p'(x)}{ x}\right\})
\\
&z^{(0)}_{10}(x)=S_p(x)
\end{aligned}
\end{equation}

In ${\rm G_1}$, the equations are substantially harder to solve. Nevertheless, we identified five families of modes, matching the expected number of families based on this symmetry ansatz and the order of the differential equations being solved. Before presenting the modes, we will first describe a few generalities.

So long as $\gamma^{(0)}\neq0,1$, the Einstein equation demands
\begin{equation}
z^{(0)}_7(x)=-\frac{3}{2}z^{(0)}_1(x)-\frac{1}{2}z_2^{(0)}(x)-\frac{1}{2}z_4^{(0)}(x)\,.
\end{equation}
For $\gamma^{(0)}=1$ we impose the above as gauge condition, since modes with $\gamma^{(0)}=1$ exhibit a residual gauge symmetry similar to the one found in \cite{Horowitz:2024dch}.

An additional component of the Einstein equation implies
\begin{multline}
z^{(0)}_3(x)=\frac{H_1(x) \left[H_2(x)+4 \Omega _{\phi }^2\right]}{8 x \Omega _{\phi }^2 H_2(x)} \Bigg\{-\frac{3 H_3^+(x) H_3^-(x) H_2(x)}{x^3 \left(1-x^2\right) \left[H_2(x)+4 \Omega_{\phi }^2\right]} z^{(0)}_1(x)
   \\
   +\frac{H_2(x) H_4(x)}{x^3 \left(1-x^2\right) H_1(x) \left[H_2(x)+4 \Omega _{\phi }^2\right]} z^{(0)}_2(x)+\frac{H_2(x) \left(H_2(x)-2 x^4\right)}{x^3
   \left(1-x^2\right) \left(x^2+4 \Omega _{\phi }^2\right)} z^{(0)}_4(x)-\frac{1}{\gamma^{(0)} }{z^{(0)}_6}'(x)
   \\
   +\frac{H_2(x)^2}{x^2 \left[H_2(x)+4 \Omega_{\phi }^2\right]}\left[3 {z_1^{(0)}}'(x)+{z^{(0)}_2}'(x)+{z^{(0)}_4}'(x)\right]-\frac{4 \left(1-x^2\right) \Omega_\phi^2 \left[H_2(x)+1\right]}{x^2 \left[H_2(x)+4 \Omega _{\phi }^2\right] \gamma^{(0)} }{z^{(0)}_5}'(x)\Bigg\}\,,
\end{multline}
with
\begin{equation}
\begin{aligned}
&H_3^{\pm}(x)\equiv x^2\pm2\Omega_{\phi}(1-x^2)\,,
\\
&H_4(x)\equiv 16 \Omega _{\phi }^4-48 \Omega _{\phi }^4 x^2- (1+12 \Omega _{\phi }^2-32 \Omega _{\phi }^4)x^4- (1-4 \Omega _{\phi }^2)x^6\,.
\end{aligned}
\end{equation}
At this stage we are left with solving for $\{z_1^{(0)}(x),z_2^{(0)}(x),z_4^{(0)}(x),z_5^{(0)}(x),z_6^{(0)}(x)\}$. These all obey second order equations of motion, and we thus expect five degrees of freedom to emerge.

One of these degrees of freedom can be decoupled via the following change of variables
\begin{equation}
\begin{aligned}
&z^{(0)}_1(x)=\frac{1}{3}\left[h_T(x)-z^{(0)}_2(x)-z^{(0)}_4(x)\right]\,,
\\
&z_2^{(0)}(x)=r_1(x)-r_2(x)-x^4 h_T(x)+x^3(1-x^2)h_T'(x)\,,
\\
&z_4^{(0)}(x)=r_2(x)-(1-x^2+2x^4) h_T(x)-(1-2x^2)x (1-x^2)h_T'(x)
\\
&z_5^{(0)}(x)=r_3(x)+\frac{\Theta_1(x)}{8 (\gamma^{(0)}+1 ) (1+4 \Omega _{\phi }^2) \Omega _{\phi }^2}h_T(x)+\frac{\Theta_2(x)}{16 (\gamma^{(0)}+1 ) (1+4 \Omega _{\phi }^2) \Omega _{\phi }^2}x(1-x^2)h_T'(x)
\\
&z_6^{(0)}(x)=r_4(x)+\frac{\Theta_3(x)}{2(\gamma^{(0)}+1)(1+4\Omega_{\phi}^2)}h_T(x)+\frac{\Theta_4(x)}{4(\gamma^{(0)}+1)(1+4\Omega_{\phi}^2)}x(1-x^2)h_T'(x)
\end{aligned}
\end{equation}
with
\begin{equation}
\begin{aligned}
&\Theta_1(x)\equiv 4 \Omega _{\phi }^2-8 x^2 \Omega _{\phi }^2-x^4 \left[1-8 \left(2-\lambda ^{(0)}\right) \Omega _{\phi }^2-16 \left(3-2 \lambda^{(0)}\right) \Omega _{\phi }^4\right]
\\
&\Theta_2(x)\equiv4 \Omega_{\phi }^2+x^2 \left(1-16 \Omega_{\phi }^2-48 \Omega _{\phi }^4\right)
\\
&\Theta_3(x)\equiv3+(4+8 \lambda^{(0)} ) \Omega _{\phi }^2-16 (3-2 \lambda^{(0)} ) \Omega _{\phi }^4-x^2 \left[6+8 \Omega _{\phi }^2-96 \Omega _{\phi }^4-4 \lambda^{(0)}  \left(1-16 \Omega _{\phi }^4\right)\right]
\\
&\quad\quad\quad\quad\quad\quad\quad+x^4 \left[5+16 \Omega
   _{\phi }^2-48 \Omega _{\phi }^4-4 \lambda^{(0)}  \left(1+2 \Omega _{\phi }^2-8 \Omega _{\phi }^4\right)\right]
\\
& \Theta_4(x)\equiv 3+4 \Omega _{\phi }^2-48 \Omega _{\phi }^4-x^2 \left(5+16 \Omega _{\phi }^2-48 \Omega _{\phi }^4\right)
\end{aligned}
\end{equation}
where $\lambda^{(0)}\equiv \gamma^{(0)}(\gamma^{(0)}+1)$.

The equation for $h_T$ decouples from the remainder four equations for $r_1$, $r_2$, $r_3$ and $r_4$, and can be readily solved via
\begin{equation}
h_T(x)=\, _2F_1\left(-p,p+3;2;x^2\right)\quad\text{with}\quad \gamma^{(0)}=p+1\quad p\geq0\,.
\end{equation}
We are left with solving for $r_1$, $r_2$, $r_3$ and $r_4$, which all couple and obey linear second order differential equations with variable coefficients.

It turns out that these four equations can be solved exactly and solutions fall into four classes which we now detail.

In the first family
\begin{equation}
\begin{aligned}
&\gamma^{(0)}=p
\\
&r_1(x)=p S_1(x)\,,
\\
&r_2(x)=\frac{1+4\Omega_{\phi}^2}{H_1(x)}\left[p x^2S_1(x)+\frac{x(1-x^2)}{2(p+1)}S'_1(x)\right]\,,
\\
&r_3(x)=\frac{1}{8 (p+1) \Omega _{\phi }^2}\Bigg\{-p H_2(x) S_1(x)
\\
&\quad\quad\quad\quad\quad\quad\quad+\frac{x }{2 (p+1)}\left[\left(p^2+p+1\right) H_2(x)-1+4 p (1+p) \Omega _{\phi }^2\right] S_1'(x)\Bigg\}\,,
\\
&r_4(x)=\frac{1}{8 (p+1)}\Bigg\{-4 p H_2(x) S_1(x)
\\
&\quad\quad\quad\quad\quad\quad\quad+\frac{2}{p+1} \left[\left(p^2+p+1\right) H_2(x)-4 \Omega _{\phi }^2+p^2+p\right] \frac{(1-x^2)S_1'(x)}{x}\Bigg\}\,,
\end{aligned}
\end{equation}
with $S_1(x)\equiv \, _2F_1\left(-p,p+1;1;x^2\right)$, where $p\geq1$.

In the second family
\begin{equation}
\begin{aligned}
&\gamma^{(0)}=p
\\
&r_1(x)=x^2 S_2(x)\,,
\\
&r_2(x)=\frac{2 x^2 (x^2+2 \Omega _{\phi }^2)}{3 (x^2+4 \Omega _{\phi }^2)}S_2(x)\,,
\\
&r_3(x)=-\frac{\Omega _{\phi }^2}{3 (p+1) (1+4 \Omega _{\phi }^2)} \left[2 S_2(x)+x \left(1-2 x^2\right) S_2'(x)\right]\,,
\\
&r_4(x)=-\frac{1}{12 (p+1) (1+4 \Omega _{\phi }^2)}\left[2 S_2(x)+x (1-2 x^2) S_2'(x)\right]\,,
\end{aligned}
\end{equation}
with $S_2(x)\equiv \, _2F_1\left(-p,p+1;2;x^2\right)$, where $p\geq1$.

In the third family
\begin{equation}
\begin{aligned}
&\gamma^{(0)}=p+1
\\
&r_1(x)=\frac{x^6\gamma^{(0)}}{H_2(x)} S_3(x)\,,
\\
&r_2(x)=\frac{2}{3}\frac{x^6\gamma^{(0)}}{H_2(x)} S_3(x)
\\
&r_3(x)=-\frac{x^4 (p+1)}{12 (p+2) (1+4 \Omega
   _{\phi }^2)}\left[\frac{6 \Omega _{\phi }^2+x^2 (1-2 \Omega _{\phi }^2) }{\left(1-x^2\right) \Omega _{\phi }^2}S_p(x)+x S_p'(x)\right]\,,
\\
&r_4(x)=\frac{1}{3 (2+p)^2 (1+4 \Omega _{\phi }^2)}\left[\varpi_1(x)S_3(x)+\frac{x}{2}S_p'(x)\right]\,.
\end{aligned}
\end{equation}
with $S_3(x)\equiv \, _2F_1\left(-p,p+3;4;x^2\right)$, where $p\geq0$ and
\begin{equation}
\begin{aligned}
&\varpi_1(x)=4 x^4 (1+\Omega _{\phi }^2)-3 (1+4 \Omega _{\phi }^2)-p (p+3) x^2 \left[1+4 \Omega _{\phi }^2-2 x^2 \left(1+\Omega _{\phi }^2\right)\right]\,,
\\
&\varpi_2(x)=2 (p+1) (p+2) x^4 \left(1+2 \Omega _{\phi }^2\right)-p (p+3) x^2 \left(1+4 \Omega _{\phi }^2\right)-2-8 \Omega _{\phi }^2\,.
\end{aligned}
\end{equation}

Finally, for the fourth family we find
\begin{equation}
\begin{aligned}
&\gamma^{(0)}=p
\\
&r_1(x)=\frac{x^4}{H_2(x)} S_4(x)\,,
\\
&r_2(x)=\frac{2}{3}\frac{x^4}{H_2(x)} S_4(x)
\\
&r_3(x)=\frac{x^5}{48(1-x^2)(p+1)\Omega_{\phi}^2}S_4'(x)\,,
\\
&r_4(x)=-\frac{1}{3(p+1)}\left[S_4(x)+\frac{x(1-x^2)}{4}S_4'(x)\right]\,.
\end{aligned}
\end{equation}
with $S_4(x)\equiv \, _2F_1\left(-p,p+1;3;x^2\right)$, where $p\geq1$.

\bibliography{extremal}{}

\providecommand{\href}[2]{#2}\begingroup\raggedright\begin{thebibliography}{10}

\bibitem{Tangherlini:1963bw}
F.~R. Tangherlini, ``{Schwarzschild field in n dimensions and the
  dimensionality of space problem},''
  \href{http://dx.doi.org/10.1007/BF02784569}{{\em Nuovo Cim.} {\bfseries 27}
  (1963) 636--651}.

\bibitem{Myers:1986un}
R.~C. Myers and M.~J. Perry, ``{Black Holes in Higher Dimensional
  Space-Times},'' \href{http://dx.doi.org/10.1016/0003-4916(86)90186-7}{{\em
  Annals Phys.} {\bfseries 172} (1986) 304}.

\bibitem{Chong:2005hr}
Z.~W. Chong, M.~Cvetic, H.~Lu, and C.~N. Pope, ``{General non-extremal rotating
  black holes in minimal five-dimensional gauged supergravity},''
  \href{http://dx.doi.org/10.1103/PhysRevLett.95.161301}{{\em Phys. Rev. Lett.}
  {\bfseries 95} (2005) 161301},
  \href{http://arxiv.org/abs/hep-th/0506029}{{\ttfamily arXiv:hep-th/0506029}}.

\bibitem{Deshpande:2024vbn}
R.~Deshpande and O.~Lunin, ``{Rotating Einstein-Maxwell black holes in odd
  dimensions},'' \href{http://arxiv.org/abs/2411.01795}{{\ttfamily
  arXiv:2411.01795 [hep-th]}}.

\bibitem{Horowitz:2023xyl}
G.~T. Horowitz, M.~Kolanowski, G.~N. Remmen, and J.~E. Santos, ``{Extremal Kerr
  Black Holes as Amplifiers of New Physics},''
  \href{http://dx.doi.org/10.1103/PhysRevLett.131.091402}{{\em Phys. Rev.
  Lett.} {\bfseries 131} no.~9, (2023) 091402},
  \href{http://arxiv.org/abs/2303.07358}{{\ttfamily arXiv:2303.07358
  [hep-th]}}.

\bibitem{Horowitz:2024dch}
G.~T. Horowitz, M.~Kolanowski, G.~N. Remmen, and J.~E. Santos, ``{Sudden
  breakdown of effective field theory near cool Kerr-Newman black holes},''
  \href{http://dx.doi.org/10.1007/JHEP05(2024)122}{{\em JHEP} {\bfseries 05}
  (2024) 122}, \href{http://arxiv.org/abs/2403.00051}{{\ttfamily
  arXiv:2403.00051 [hep-th]}}.

\bibitem{Horowitz:2022mly}
G.~T. Horowitz, M.~Kolanowski, and J.~E. Santos, ``{Almost all extremal black
  holes in AdS are singular},''
  \href{http://dx.doi.org/10.1007/JHEP01(2023)162}{{\em JHEP} {\bfseries 01}
  (2023) 162}, \href{http://arxiv.org/abs/2210.02473}{{\ttfamily
  arXiv:2210.02473 [hep-th]}}.

\bibitem{Kunduri:2009ud}
H.~K. Kunduri and J.~Lucietti, ``{Static near-horizon geometries in five
  dimensions},'' \href{http://dx.doi.org/10.1088/0264-9381/26/24/245010}{{\em
  Class. Quant. Grav.} {\bfseries 26} (2009) 245010},
  \href{http://arxiv.org/abs/0907.0410}{{\ttfamily arXiv:0907.0410 [hep-th]}}.

\bibitem{Blazquez-Salcedo:2013yba}
J.~L. Bl\'azquez-Salcedo, J.~Kunz, and F.~Navarro-Lerida, ``{Angular momentum -
  area - proportionality of extremal charged black holes in odd dimensions},''
  \href{http://dx.doi.org/10.1016/j.physletb.2013.10.046}{{\em Phys. Lett. B}
  {\bfseries 727} (2013) 340--344},
  \href{http://arxiv.org/abs/1309.2088}{{\ttfamily arXiv:1309.2088 [gr-qc]}}.

\bibitem{Komar:1958wp}
A.~Komar, ``{Covariant conservation laws in general relativity},''
  \href{http://dx.doi.org/10.1103/PhysRev.113.934}{{\em Phys. Rev.} {\bfseries
  113} (1959) 934--936}.

\bibitem{Page:1983mke}
D.~N. Page, ``{Classical Stability of Round and Squashed Seven Spheres in
  Eleven-dimensional Supergravity},''
  \href{http://dx.doi.org/10.1103/PhysRevD.28.2976}{{\em Phys. Rev. D}
  {\bfseries 28} (1983) 2976}.

\bibitem{Hanaki:2007mb}
K.~Hanaki, K.~Ohashi, and Y.~Tachikawa, ``{Comments on charges and near-horizon
  data of black rings},''
  \href{http://dx.doi.org/10.1088/1126-6708/2007/12/057}{{\em JHEP} {\bfseries
  12} (2007) 057}, \href{http://arxiv.org/abs/0704.1819}{{\ttfamily
  arXiv:0704.1819 [hep-th]}}.

\bibitem{Kunduri:2013gce}
H.~K. Kunduri and J.~Lucietti, ``{Classification of near-horizon geometries of
  extremal black holes},'' \href{http://dx.doi.org/10.12942/lrr-2013-8}{{\em
  Living Rev. Rel.} {\bfseries 16} (2013) 8},
  \href{http://arxiv.org/abs/1306.2517}{{\ttfamily arXiv:1306.2517 [hep-th]}}.

\bibitem{Bardeen:1973gs}
J.~M. Bardeen, B.~Carter, and S.~W. Hawking, ``{The Four laws of black hole
  mechanics},'' \href{http://dx.doi.org/10.1007/BF01645742}{{\em Commun. Math.
  Phys.} {\bfseries 31} (1973) 161--170}.

\bibitem{Carter:1969zz}
B.~Carter, ``{Killing horizons and orthogonally transitive groups in
  space-time},'' \href{http://dx.doi.org/10.1063/1.1664763}{{\em J. Math.
  Phys.} {\bfseries 10} (1969) 70--81}.

\bibitem{Hawking:1974rv}
S.~W. Hawking, ``{Black hole explosions},''
  \href{http://dx.doi.org/10.1038/248030a0}{{\em Nature} {\bfseries 248} (1974)
  30--31}.

\bibitem{Kunduri:2007vf}
H.~K. Kunduri, J.~Lucietti, and H.~S. Reall, ``{Near-horizon symmetries of
  extremal black holes},''
  \href{http://dx.doi.org/10.1088/0264-9381/24/16/012}{{\em Class. Quant.
  Grav.} {\bfseries 24} (2007) 4169--4190},
  \href{http://arxiv.org/abs/0705.4214}{{\ttfamily arXiv:0705.4214 [hep-th]}}.

\bibitem{Cvetic:1996xz}
M.~Cvetic and D.~Youm, ``{General rotating five-dimensional black holes of
  toroidally compactified heterotic string},''
  \href{http://dx.doi.org/10.1016/0550-3213(96)00355-0}{{\em Nucl. Phys. B}
  {\bfseries 476} (1996) 118--132},
  \href{http://arxiv.org/abs/hep-th/9603100}{{\ttfamily arXiv:hep-th/9603100}}.

\bibitem{Headrick:2009pv}
M.~Headrick, S.~Kitchen, and T.~Wiseman, ``{A New approach to static numerical
  relativity, and its application to Kaluza-Klein black holes},''
  \href{http://dx.doi.org/10.1088/0264-9381/27/3/035002}{{\em Class. Quant.
  Grav.} {\bfseries 27} (2010) 035002},
  \href{http://arxiv.org/abs/0905.1822}{{\ttfamily arXiv:0905.1822 [gr-qc]}}.

\bibitem{Wiseman:2011by}
T.~Wiseman, {\em {Numerical construction of static and stationary black
  holes}}, pp.~233--270.
\newblock {Cambridge University Press}, 2012.
\newblock \href{http://arxiv.org/abs/1107.5513}{{\ttfamily arXiv:1107.5513
  [gr-qc]}}.

\bibitem{Dias:2015nua}
O.~J.~C. Dias, J.~E. Santos, and B.~Way, ``{Numerical Methods for Finding
  Stationary Gravitational Solutions},''
  \href{http://dx.doi.org/10.1088/0264-9381/33/13/133001}{{\em Class. Quant.
  Grav.} {\bfseries 33} no.~13, (2016) 133001},
  \href{http://arxiv.org/abs/1510.02804}{{\ttfamily arXiv:1510.02804
  [hep-th]}}.

\bibitem{Figueras:2016nmo}
P.~Figueras and T.~Wiseman, ``{On the existence of stationary Ricci
  solitons},'' \href{http://dx.doi.org/10.1088/1361-6382/aa764a}{{\em Class.
  Quant. Grav.} {\bfseries 34} no.~14, (2017) 145007},
  \href{http://arxiv.org/abs/1610.06178}{{\ttfamily arXiv:1610.06178 [gr-qc]}}.

\bibitem{Durkee:2010ea}
M.~Durkee and H.~S. Reall, ``{Perturbations of near-horizon geometries and
  instabilities of Myers-Perry black holes},''
  \href{http://dx.doi.org/10.1103/PhysRevD.83.104044}{{\em Phys. Rev. D}
  {\bfseries 83} (2011) 104044},
  \href{http://arxiv.org/abs/1012.4805}{{\ttfamily arXiv:1012.4805 [hep-th]}}.

\bibitem{Kunz:2005ei}
J.~Kunz and F.~Navarro-Lerida, ``{D=5 Einstein-Maxwell-Chern-Simons black
  holes},'' \href{http://dx.doi.org/10.1103/PhysRevLett.96.081101}{{\em Phys.
  Rev. Lett.} {\bfseries 96} (2006) 081101},
  \href{http://arxiv.org/abs/hep-th/0510250}{{\ttfamily arXiv:hep-th/0510250}}.

\bibitem{Murata:2012ct}
K.~Murata, ``{Instability of higher dimensional extreme black holes},''
  \href{http://dx.doi.org/10.1088/0264-9381/30/7/075002}{{\em Class. Quant.
  Grav.} {\bfseries 30} (2013) 075002},
  \href{http://arxiv.org/abs/1211.6903}{{\ttfamily arXiv:1211.6903 [gr-qc]}}.

\bibitem{Carullo:2021oxn}
G.~Carullo, D.~Laghi, N.~K. Johnson-McDaniel, W.~Del~Pozzo, O.~J.~C. Dias,
  M.~Godazgar, and J.~E. Santos, ``{Constraints on Kerr-Newman black holes from
  merger-ringdown gravitational-wave observations},''
  \href{http://dx.doi.org/10.1103/PhysRevD.105.062009}{{\em Phys. Rev. D}
  {\bfseries 105} no.~6, (2022) 062009},
  \href{http://arxiv.org/abs/2109.13961}{{\ttfamily arXiv:2109.13961 [gr-qc]}}.

\bibitem{Dias:2021yju}
O.~J.~C. Dias, M.~Godazgar, J.~E. Santos, G.~Carullo, W.~Del~Pozzo, and
  D.~Laghi, ``{Eigenvalue repulsions in the quasinormal spectra of the
  Kerr-Newman black hole},''
  \href{http://dx.doi.org/10.1103/PhysRevD.105.084044}{{\em Phys. Rev. D}
  {\bfseries 105} no.~8, (2022) 084044},
  \href{http://arxiv.org/abs/2109.13949}{{\ttfamily arXiv:2109.13949 [gr-qc]}}.

\bibitem{Davey:2022vyx}
A.~Davey, O.~J.~C. Dias, P.~Rodgers, and J.~E. Santos, ``{Strong Cosmic
  Censorship and eigenvalue repulsions for rotating de Sitter black holes in
  higher-dimensions},'' \href{http://dx.doi.org/10.1007/JHEP07(2022)086}{{\em
  JHEP} {\bfseries 07} (2022) 086},
  \href{http://arxiv.org/abs/2203.13830}{{\ttfamily arXiv:2203.13830 [gr-qc]}}.

\bibitem{Dias:2022oqm}
O.~J.~C. Dias, M.~Godazgar, and J.~E. Santos, ``{Eigenvalue repulsions and
  quasinormal mode spectra of Kerr-Newman: an extended study},''
  \href{http://dx.doi.org/10.1007/JHEP07(2022)076}{{\em JHEP} {\bfseries 07}
  (2022) 076}, \href{http://arxiv.org/abs/2205.13072}{{\ttfamily
  arXiv:2205.13072 [gr-qc]}}.

\bibitem{Davey:2023fin}
A.~Davey, O.~J.~C. Dias, and J.~E. Santos, ``{Scalar QNM spectra of Kerr and
  Reissner-Nordstr\"om revealed by eigenvalue repulsions in Kerr-Newman},''
  \href{http://dx.doi.org/10.1007/JHEP12(2023)101}{{\em JHEP} {\bfseries 12}
  (2023) 101}, \href{http://arxiv.org/abs/2305.11216}{{\ttfamily
  arXiv:2305.11216 [gr-qc]}}.

\bibitem{Davey:2024xvd}
A.~Davey, O.~J.~C. Dias, and D.~S. Gil, ``{Strong Cosmic Censorship in
  Kerr-Newman-de Sitter},''
  \href{http://dx.doi.org/10.1007/JHEP07(2024)113}{{\em JHEP} {\bfseries 07}
  (2024) 113}, \href{http://arxiv.org/abs/2404.03724}{{\ttfamily
  arXiv:2404.03724 [gr-qc]}}.

\bibitem{Breckenridge:1996is}
J.~C. Breckenridge, R.~C. Myers, A.~W. Peet, and C.~Vafa, ``{D-branes and
  spinning black holes},''
  \href{http://dx.doi.org/10.1016/S0370-2693(96)01460-8}{{\em Phys. Lett. B}
  {\bfseries 391} (1997) 93--98},
  \href{http://arxiv.org/abs/hep-th/9602065}{{\ttfamily arXiv:hep-th/9602065}}.

\bibitem{Candlish:2007fh}
G.~N. Candlish and H.~S. Reall, ``{On the smoothness of static multi-black hole
  solutions of higher-dimensional Einstein-Maxwell theory},''
  \href{http://dx.doi.org/10.1088/0264-9381/24/23/022}{{\em Class. Quant.
  Grav.} {\bfseries 24} (2007) 6025--6040},
  \href{http://arxiv.org/abs/0707.4420}{{\ttfamily arXiv:0707.4420 [gr-qc]}}.

\bibitem{Candlish:2009vy}
G.~N. Candlish, ``{On the smoothness of the multi-BMPV black hole spacetime},''
  \href{http://dx.doi.org/10.1088/0264-9381/27/6/065005}{{\em Class. Quant.
  Grav.} {\bfseries 27} (2010) 065005},
  \href{http://arxiv.org/abs/0904.3885}{{\ttfamily arXiv:0904.3885 [hep-th]}}.

\bibitem{Horowitz:2022leb}
G.~T. Horowitz, M.~Kolanowski, and J.~E. Santos, ``{A deformed IR: a new IR
  fixed point for four-dimensional holographic theories},''
  \href{http://dx.doi.org/10.1007/JHEP02(2023)152}{{\em JHEP} {\bfseries 02}
  (2023) 152}, \href{http://arxiv.org/abs/2211.01385}{{\ttfamily
  arXiv:2211.01385 [hep-th]}}.

\end{thebibliography}\endgroup
\bibliographystyle{utphys-modified}

\end{document}